\documentclass[aps,prl,showpacs,superscriptaddress,twocolumn,final,numerical,10pt]{revtex4-1}
\usepackage{hyperref}
\usepackage{verbatim}
\usepackage{amsmath}
\usepackage{latexsym}
\usepackage{revsymb}
\usepackage{ifthen}
\usepackage{color}
\usepackage{natbib}
\usepackage{amsfonts}
\usepackage{amsmath}
\usepackage{amssymb}
\usepackage{amsthm}
\usepackage{graphicx}
\usepackage{bm}
\usepackage{float}
\usepackage{framed}
\definecolor{shadecolor}{rgb}{0.92,0.92,0.92}

\begin{document}

\title{Practical challenges in quantum key distribution}

\author{Eleni Diamanti}
\affiliation{Laboratoire Traitement et Communication de l'Information, \\ CNRS, T\'{e}l\'{e}com ParisTech, Universit\'e Paris-Saclay, Paris 75103, France}

\author{Hoi-Kwong Lo}
\email{hklo@ece.utoronto.ca}
\affiliation{Center for Quantum Information and Quantum Control, Department of Physics and Department of Electrical \& Computer Engineering, University of Toronto, M5S 3G4 Toronto, Canada}

\author{Bing Qi}
\affiliation{Quantum Information Science Group, Computational Sciences and Engineering Division, Oak Ridge National Laboratory, Oak Ridge, Tennessee 37831-6418, USA}
\affiliation{Department of Physics and Astronomy, University of Tennessee, Knoxville, Tennessee 37996, USA}

\author{Zhiliang Yuan}
\affiliation{Toshiba Research Europe Limited, 208 Cambridge Science Park, Cambridge CB4 0GZ, United Kingdom}
\affiliation{Corporate Research \& Development Center, Toshiba Corporation, 1 Komukai-Toshiba-Cho, Saiwai-ku, Kawasaki 212-8582, Japan}

\date{\today}

\begin{abstract}

Quantum key distribution (QKD) promises unconditional security in data communication and is currently being deployed in commercial applications. Nonetheless, before QKD can be widely adopted, it faces a number of important challenges such as secret key rate, distance, size, cost and practical security. Here, we survey those key challenges and the approaches that are currently being taken to address them.

\end{abstract}

%\pacs{}
\maketitle

\noindent {\large \bf Introduction.}

\noindent {\bf Why quantum key distribution?}
For thousands of years, human beings have been using codes to keep secrets. With the rise of the Internet and recent trends to the Internet of Things, our sensitive personal financial and health data as well as commercial and national secrets are routinely being transmitted through the Internet. In this context, communication security is of utmost importance. In conventional symmetric cryptographic algorithms, communication security relies solely on the secrecy of an encryption key. If two users, Alice and Bob, share a long random string of secret bits -- the key -- then they can achieve unconditional security by encrypting their message using the standard one-time-pad encryption scheme. The central question then is: how do Alice and Bob share a secure key in the first place? This is called the key distribution problem. Unfortunately, all classical methods to distribute a secure key are fundamentally insecure because in classical physics there is nothing preventing an eavesdropper, Eve, from copying the key during its transit from Alice to Bob. On the other hand, standard asymmetric or public-key cryptography solves the key distribution problem by relying on computational assumptions such as the hardness of factoring. Therefore, such schemes do not provide information-theoretic security because they are vulnerable to unexpected future advances in hardware and algorithms, including the construction of a large-scale quantum computer \cite{Shor:focs94}.

We remark that some secrets, such as, for instance, census data, need to be kept secret for decades (e.g. 92 years in Canada \cite{CanadianCensus}). Currently, however, data transmitted in 2016 is vulnerable to technological advances made in the future as Eve might simply save the transcripts of communication in her memory and wait for the construction, for example, of a quantum computer some time before 2108 (92 years from 2016). This is highly probable. Recall that ENIAC, the first general purpose electronics computer \cite{ENIAC}, which was largely inferior to modern computers, was invented only 70 years ago. The US National Security Agency is taking the threat of quantum computing seriously and has recently announced transition plans to quantum-resistant classical algorithms \cite{NSA,notePQC}.

Quantum cryptography, or more specifically, quantum key distribution (QKD) \cite{BB84,Ekert:prl91,SBC:rmp09,LCT:natphoton14}, promises in principle unconditional security -- the Holy Grail of communication security -- based on the laws of physics only \cite{Mayers:jacm01,LC:science99,SP:prl00}. QKD has the advantage of being future-proof \cite{Unruh:iacr12}: unlike classical key distribution, it is not possible for an eavesdropper to keep a transcript of quantum signals sent in a QKD process, owing to the quantum non-cloning theorem \cite{WZ:nature82,Dieks:pl82}. For this reason, QKD is an essential element of the future quantum-safe infrastructure, which will include both quantum-resistant classical algorithms and quantum cryptographic solutions. In the bigger context of quantum information, there has been tremendous scientific and engineering effort towards the long-term vision of a global quantum internet \cite{Kimble:nature08}. Imagine a world where only a few large-scale quantum computers are available (just like the early days of classical computing when only a few classical computers were available and in line with the current trend towards cloud computing); users will have to access those powerful quantum computers at long distances via a quantum internet. QKD will play a central role in securing data communication links in such a quantum internet.

The potential applications of QKD include securing critical infrastructures (for instance, the Smart Grid), financial institutions and national defense. Experimental QKD has been performed over distances on the order of 100 km in standard telecom fibers as well as in free space, while the secure key rate has now reached a few Mbits per second. QKD has leaped out of the lab \cite{Qiu:natnews14}. In China, the deployment of a 2000 km QKD network between Shanghai and Beijing is underway; in Europe, after the SECOQC network demonstration in 2008 \cite{PPA:njp09}, the UK is now creating a quantum network facilitating device and system trials, and the integration of quantum and conventional communications; in Japan, QKD technologies will be put into test to secure transmission of sensitive genome data; and the US has also started installing its own QKD network.

\noindent {\bf Why practical challenges in QKD?}
In this review, we will focus on practical issues in QKD. We remark that, historically, practical considerations in QKD have led to ground-breaking inventions. For example, the need to counter the photon-number-splitting attack \cite{HIG:pra95} triggered the invention of the decoy-state protocol \cite{Hwang:prl03,LMC:prl05,Wang:prl05}, which allows efficient distillation of secure keys using weak coherent pulse based QKD systems that once were vulnerable. As another example, the need to counter detector side-channel attacks has led to the discovery of measurement device independent (MDI) QKD \cite{LCQ:prl12}. New theory that is due to practical advances in QKD also includes, for instance, the quantum de Finetti theorem \cite{CKM:commmathphys07}, while security loopholes in QKD are closely related to loopholes in Bell inequality tests \cite{HBD:nature15} -- a key subject in the foundations of quantum mechanics. These issues are therefore of great interest to mathematicians and theoretical physicists.

QKD is clearly of interest to engineers too. For instance, practical QKD is closely linked to the development of new single-photon detection technologies such as superconducting nanowire single-photon detectors (SNSPDs) \cite{GOC:apl01}, superconducting transition-edge sensors (TES) \cite{LMN:opex08}, frequency up-conversion single photo detectors \cite{AW:ol04,LDR:ol05}, and self-differential InGaAs avalanche photodiodes (APDs) \cite{YKS:apl07}, as well as of high performance homodyne detection techniques \cite{HAH:ol01}. It is also the motivation for high-speed quantum random number generators (QRNG) \cite{JAW:rsi00} and broadband entangled photon sources \cite{ZTQ:ol13}.

Practical QKD has steered innovation and is a precursor in the field of Quantum Information Processing.

\noindent {\bf Outline of the review.}
Despite the important theoretical and experimental achievements, a number of key challenges remain for QKD to be widely used for securing everyday interactions. For instance, much effort is being put into increasing the communication rate and range of QKD and making QKD systems low cost, compact and robust. New hardware such as chip-based QKD and new software such as novel protocols are being studied and developed. The security of practical QKD systems is another important challenge. In order to foil quantum hackers, protocols such as MDI-QKD and loss-tolerant QKD \cite{TCK:pra14} have been developed and are currently being experimentally implemented. Yet, a comprehensive theory of the model of a QKD source remains to be constructed. To further extend the reach of QKD, two different approaches -- quantum repeaters and ground-to-satellite QKD -- are being pursued. In view of the proliferation of mobile computing devices including smart phones, mobile QKD applications have also attracted recent attention. Furthermore, the standardization of QKD components is currently being pursued in ETSI (European Telecommunications Standards Institute) \cite{ADM:GC14}. In what follows, we will highlight some of the above challenges and the various approaches that are being taken to tackle them.\\

\noindent {\large \bf  Main protocols and implementations.}

\noindent We begin our discussion with a brief overview of the main QKD protocols currently studied and the state-of-the-art in their practical implementations. As our main focus here is the current challenges in the field, we refer the reader to a recent review \cite{LCT:natphoton14} for the necessary background on the rigorous information-theoretic (or, unconditional) security definition of QKD in the composable framework, secure communication schemes including the one-time pad, the standard BB84 QKD protocol, and basic QKD components.

\begin{figure*}[htb]
\centerline{\includegraphics[width=\textwidth]{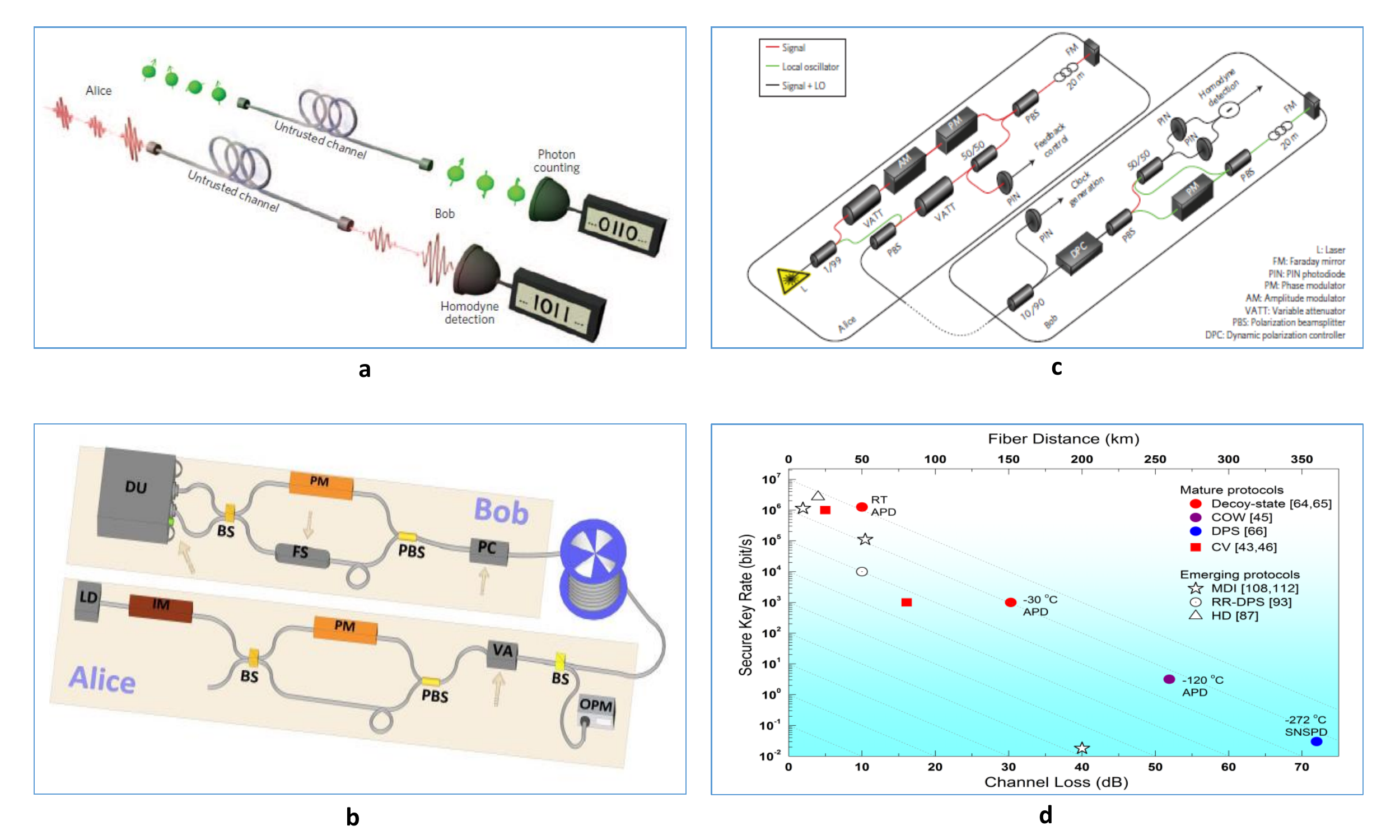}} \caption{a. Quantum key distribution systems use discrete-variable (DV) single-photon state encoding and single-photon detection techniques or continuous-variable (CV) quadrature field amplitude encoding and homodyne (or heterodyne) detection techniques. b. State-of-the-art experimental setup for the implementation of the decoy-state BB84 QKD protocol \cite{LPD:opex13}. c. State-of-the-art experimental setup for the implementation of the coherent state CV-QKD protocol \cite{JKL:natphoton13}. d. Secret key generation rates demonstrated in some representative recent QKD experiments. Note that this figure is not meant to provide an exhaustive list of QKD implementations. Furthermore, protocol performance cannot be directly compared as different security assumptions are considered; for instance, decoy-state BB84 is secure against general coherent attacks while COW and CV-QKD are secure against collective attacks. QKD is a subject of active ongoing research and so further developments are likely to occur in the near future. The loss coefficient of 0.2 dB/km in standard single-mode fibers at telecom wavelengths is assumed in this figure. Figures adapted with permission from: a. ref. \cite{LR:natphoton13}, \copyright 2013 NPG, courtesy of Ping Koy Lam; b. ref. \cite{LPD:opex13}, \copyright 2013 OSA; c. ref. \cite{JKL:natphoton13}, \copyright 2013 NPG.}
\label{fig:dv-cv-qkd}
\end{figure*}

QKD protocols can be in essence divided with respect to the detection technique required to recover the key information encoded in the properties of light (Fig. \ref{fig:dv-cv-qkd}a). In  discrete-variable (DV) protocols information is typically encoded in the polarization or phase of weak coherent pulses simulating true single-photon states; hence the corresponding implementations employ single-photon detection techniques. The previously mentioned BB84 and decoy-state protocols are prominent examples in this category. Single-photon detection techniques are also necessary for the so-called distributed-phase-reference protocols, such as the coherent-one-way (COW) \cite{SBG:apl05} and differential-phase-shift (DPS) \cite{IWY:prl02} protocols, where the key information is encoded in photon arrival times or in the phase between adjacent weak coherent pulses. On the other hand, in continuous-variable (CV) QKD protocols information is encoded in the quadratures of the quantized electromagnetic field, such as those of coherent states \cite{GG:prl02,GVW:nat03}, and homodyne or heterodyne detection techniques are used in this case. Such detectors are routinely deployed in classical optical communications, hence the CV approach offers the possibility for implementations based only on mature telecom components. All these protocols are prepare-and-measure in the sense that the transmitter, Alice, sends the encoded pulses to the receiver, Bob, who decodes as required by the specific protocol. On the contrary, in entanglement-based protocols \cite{Ekert:prl91}, both parties receive parts of an entangled state and perform suitable measurements. More details on all protocols can be found in refs. \cite{SBC:rmp09,LCT:natphoton14,DL:entropy15,MFL:pra07}.

When it comes to practical demonstrations, performance of point-to-point links is assessed by the distance over which secret keys can be distributed and the rate of their distribution for a given security level. The security level is determined by the type of attacks considered in the corresponding security proof; demonstrating security against the so-called collective attacks \cite{SBC:rmp09} is an important challenge for an implementation, however information-theoretic security is achieved only when security against the most general (or coherent) attacks is proven. Hence, the ultimate goal is to provide this level of security at a speed and a distance that are compatible with practical applications. Some recent implementations have provided high levels of security: several QKD protocols have been demonstrated to provide composable security against collective attacks using reasonable data block sizes and practical setups, including decoy-state BB84 \cite{LPD:opex13}, COW \cite{KLH:natphoton15}, and CV-QKD \cite{JKL:natphoton13,HLW:opex15}. Among those protocols, the security of decoy-state BB84 QKD has been extended to cover coherent attacks, for realistic block sizes and with a minimal sacrifice in the secret key rate \cite{LCW:pra14,LDF:JSTQE15}. Unfortunately, for COW, the best security proof against coherent attacks currently gives a secret key rate that only scales quadratically with the loss \cite{MCL:prl12}. For CV-QKD with coherent states and heterodyne detection, a composable security proof against the most general attacks has recently been provided \cite{Lev:prl15}, but the current proof techniques do not allow a positive key rate for realistic block sizes in this case. Extending the security proofs for the latter protocols is therefore a pressing task in the theoretical study of QKD.

Figures \ref{fig:dv-cv-qkd}b,c show examples of advanced fiber-optic QKD systems allowing for real-time secret key generation over distances of 50 km with Mbit/s rates. In Fig. \ref{fig:dv-cv-qkd}d we summarize some important experimental achievements from both established and emerging QKD protocols (discussed in the following sections). Although the security assumptions and technological maturity vary in these implementations, these results illustrate the diversity of protocols and experimental solutions that the research community has invented to push the performance of QKD technology. Indeed, tremendous progress has been achieved in recent years, and avenues for further progress will be discussed in the next section. We remark, however, that there are fundamental limitations on what can be ultimately achieved. Over optical fiber networks, the attenuation of light in standard fibers at the telecom wavelength of 1550 nm is 0.2 dB/km (or 0.16 dB/km in newly developed ultra low loss fibers). This unavoidable loss will not allow the range of point-to-point QKD links to exceed a few hundreds of kilometers as with overly excessive channel loss it would take several years to generate just one bit even using perfect light sources and detectors. Furthermore, with a practical lossy channel, the ultimate key rate is upper bounded by the so-called TGW bound \cite{TGW:natcomm14} (see also \cite{PLO:arxiv15} for a more recent result). The TGW bound provides a useful benchmark for the performance of all QKD protocol implementations.\\

\noindent {\large \bf Major challenges in performance and cost.}

\noindent In the quest for high performance and low-cost QKD systems, both hardware and software solutions are currently being pursued.

\noindent {\bf Hardware development.}

\noindent {\it Key rate.}
Encryption keys generated by QKD can be used in a symmetric cipher scheme, such as AES (Advanced Encryption Standard), which is quantum-resistant, for enhanced security, or they can be combined with the one-time-pad encryption scheme for unconditional security. In both cases, the secure key rate achieved by the underlying QKD layer in a typical application scenario is crucial. Higher secure rates allow for a more frequent update of encryption keys in symmetric ciphers, and for a proportional increase in the one-time-pad communication bandwidth as this scheme requires the key to be as long as the message.

%If a key generated by QKD is subsequently used in a symmetric encryption scheme such as AES, then, as AES itself is not unconditionally secure, such a usage of QKD key does not provide
%unconditional security for the encrypted message. In summary, if the goal is the unconditional security of the encrypted message, then the one-time-pad encryption scheme should be used.
%Now, if the one-time-pad encryption scheme is used, then a key has to be as long as the message.
%The secret key rate achievable by the underlying QKD layer in a typical application scenario is crucial for the speed of the targeted secure communication tasks.

Presently, strong disparity exists between the classical and QKD communication rates. Classical optical communications delivering speeds of 100 Gbit/s per wavelength channel are currently being deployed \cite{Winzer:OPN15}, and a field trial featuring 54.2 Tbit/s aggregated data rate has recently been performed \cite{HTI:jlt14}. On the other hand, the Mbit/s rates achieved by QKD systems today are sufficient, for instance, for video transmission; however, it is clear that if we want in the longer term to encrypt high volumes of classical network traffic using the one-time-pad, major developments on the secure key rate generated by QKD will be required.

\begin{figure}[htb]
\centerline{\includegraphics[width=9cm]{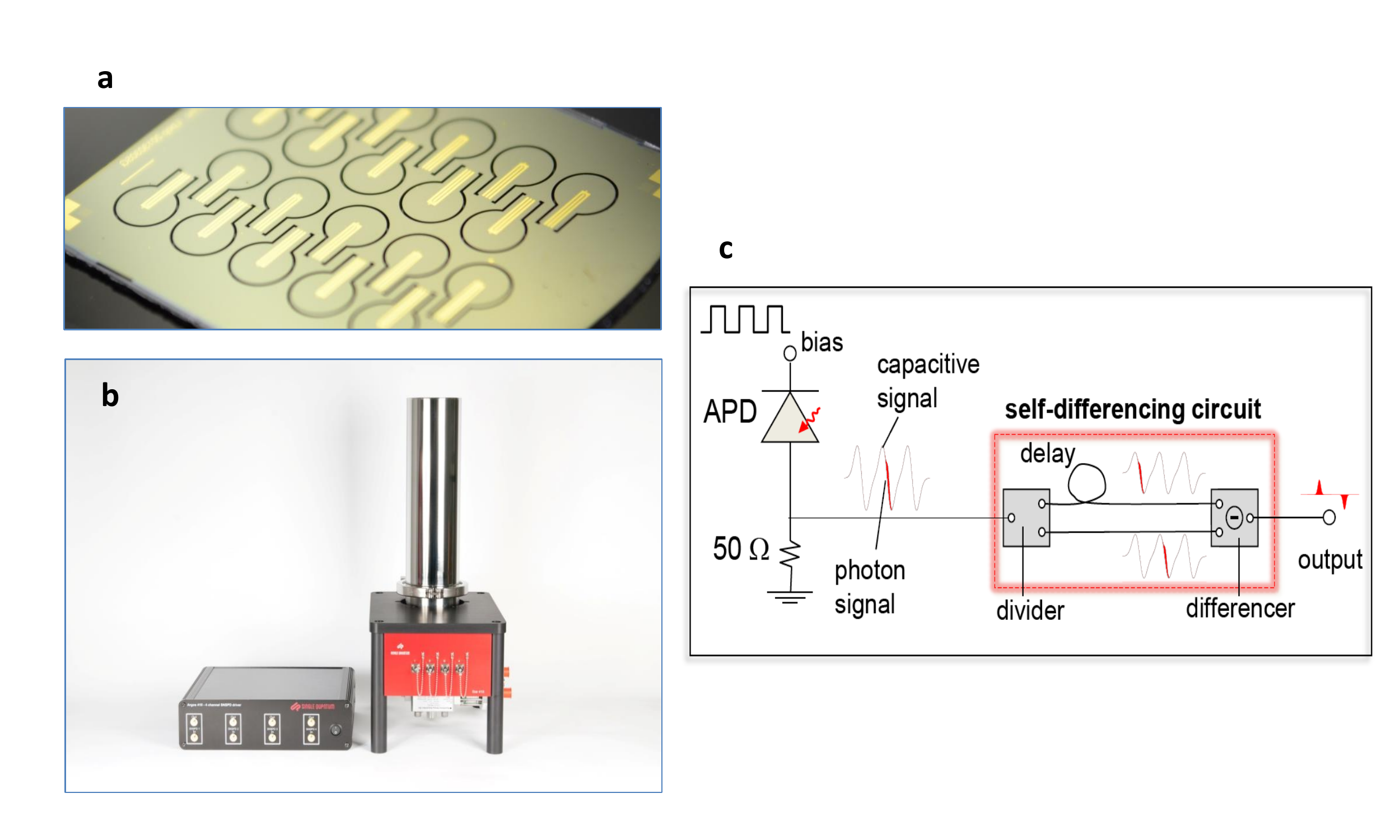}} \caption{a. Superconducting nanowire chips. b. Commercial SNSPDs with high detection efficiency. c. Characterization circuit for self-differencing InGaAs avalanche photodiodes \cite{CFL:apl14}. Figures adapted with permission from: a. http://www.photonspot.com/, courtesy of Vikas Anant; b. http://www.singlequantum.com/products, courtesy of Jessie Qin-Dregely.}
\label{fig:detectors}
\end{figure}

The obtained key rate depends crucially on the performance of the detectors used. For QKD systems employing single-photon detection techniques, high efficiency and short dead time of the detectors are essential for reaching a high bit rate. The latest developments on high efficiency detectors \cite{PSM:natcomm12,MVS:natphoton13,CFD:jap15} are extremely promising; quantum efficiencies as high as 93\% at telecom wavelengths have been reported for SNSPDs \cite{MVS:natphoton13}, and devices based on this technology with short dead time, low dark count, low time jitter and high detection efficiency are commercially available \cite{SNSPD} (Figs. \ref{fig:detectors}a,b). These results may allow for as much as a four-fold increase in the secret key rate, which currently stands at 1~Mbit/s over a 50~km fiber (or 10~dB loss) achieved using self-differential InGaAs APDs with an ultrashort dead time \cite{LPD:opex13} (Fig. \ref{fig:detectors}c). Further key rate increase is possible using wavelength or spatial mode multiplexing technologies which have been routinely used for increasing the bandwidth in data communications \cite{Winzer:OPN15,BRS:jlt15,DKT:opex16}. For CV-QKD systems, increasing the bandwidth of the homodyne or heterodyne detectors, while keeping at the same time the electronic noise low, is a necessary step for increasing the key rate beyond the 1 Mbit/s over 25~km that has been achieved \cite{HLW:opex15}. Further progress continues to be pursued, targeting also higher efficiency, which is currently around 60\% for fiber-coupled detectors at telecom wavelengths \cite{JKL:natphoton13}. Furthermore, as shown in Fig. \ref{fig:dv-cv-qkd}c, a practical issue in these systems is that the strong phase reference pulse (or, local oscillator) needs to be transmitted together with the signal at high clock rates; recent proposals  that avoid this and use instead a local oscillator generated at Bob's site \cite{QLP:prx15,SBC:prx15,HHL:ol15} are promising and will lead to more practical, high performance implementations.
%(note however that the electronic noise of these detectors has to be kept low at the same time).

\noindent {\it Distance.}
Extending the communication range of QKD systems is a major driving factor for technological developments in view of future network applications. QKD systems based on single-photon detection champion the point-to-point communication distance (or channel loss). Here, the low noise of single-photon detectors is the key enabling  factor; in particular, the attainable range depends on the type and operation temperature of the detectors. InGaAs APDs can tolerate losses of 30~dB and 52~dB when cooled to --30$^\circ$C and --120$^\circ$C \cite{FDL:scirep15,KLH:natphoton15}, respectively, while SNSPDs cooled to cryogenic temperatures have been demonstrated to withstand a record loss of 72~dB \cite{SHS:ol14}. This loss is equivalent to 360~km of standard single mode fiber or about 450~km of ultra low loss fiber. Although technologically possible, further extending the point-to-point distance is increasingly unappealing because the channel loss will inevitably reduce the key rate to a level of little practical relevance. This is also true for CV-QKD systems, which are in general more sensitive to losses. Here, it is crucial to keep the excess noise -- the noise exceeding the fundamental shot noise of coherent states -- low and especially to be able to estimate the noise value precisely, which becomes increasingly difficult with the distance \cite{JKL:natphoton13,DL:entropy15}.

We remark that advances towards high performance QKD systems in terms of key rate and distance are coupled with the security guarantees offered by these systems. For instance,
achieving composable security against general attacks requires in practice being able to perform efficient post-processing, including parameter estimation, over large data blocks with stable setups. Particularly for CV-QKD, performing efficient error correction and precise parameter estimation is of utmost importance \cite{DL:entropy15,JEK:pra14}.

\noindent {\it Cost and robustness.}
For QKD systems to be used in real world applications, low cost and robustness are indispensable features alongside high performance. Several avenues are currently being pursued. First, QKD systems have been shown to coexist with intense data traffic in the same fiber \cite{PDL:apl14,CZD:opex14,QZQ:njp10,KQA:njp15}, thus eliminating the need for dark fibers that are not only expensive but also often unavailable. Access network architecture allows simultaneous access by a multitude of QKD users, and importantly they are compatible with full power GPON (Gigabit Passive Optical Network) traffic in the same network \cite{FDL:nature13,FDL:scirep15}. Room-temperature single-photon detectors have been shown to be suitable for DV-QKD over up to 100~km fiber, thus removing cooling requirements for the entire QKD system \cite{LCW:pra14,CFL:apl14}; for CV-QKD cooling is unnecessary. All these developments help reduce deployment cost as well as system complexity, footprint and power consumption.

\begin{figure}[htb]
\centerline{\includegraphics[width=9cm]{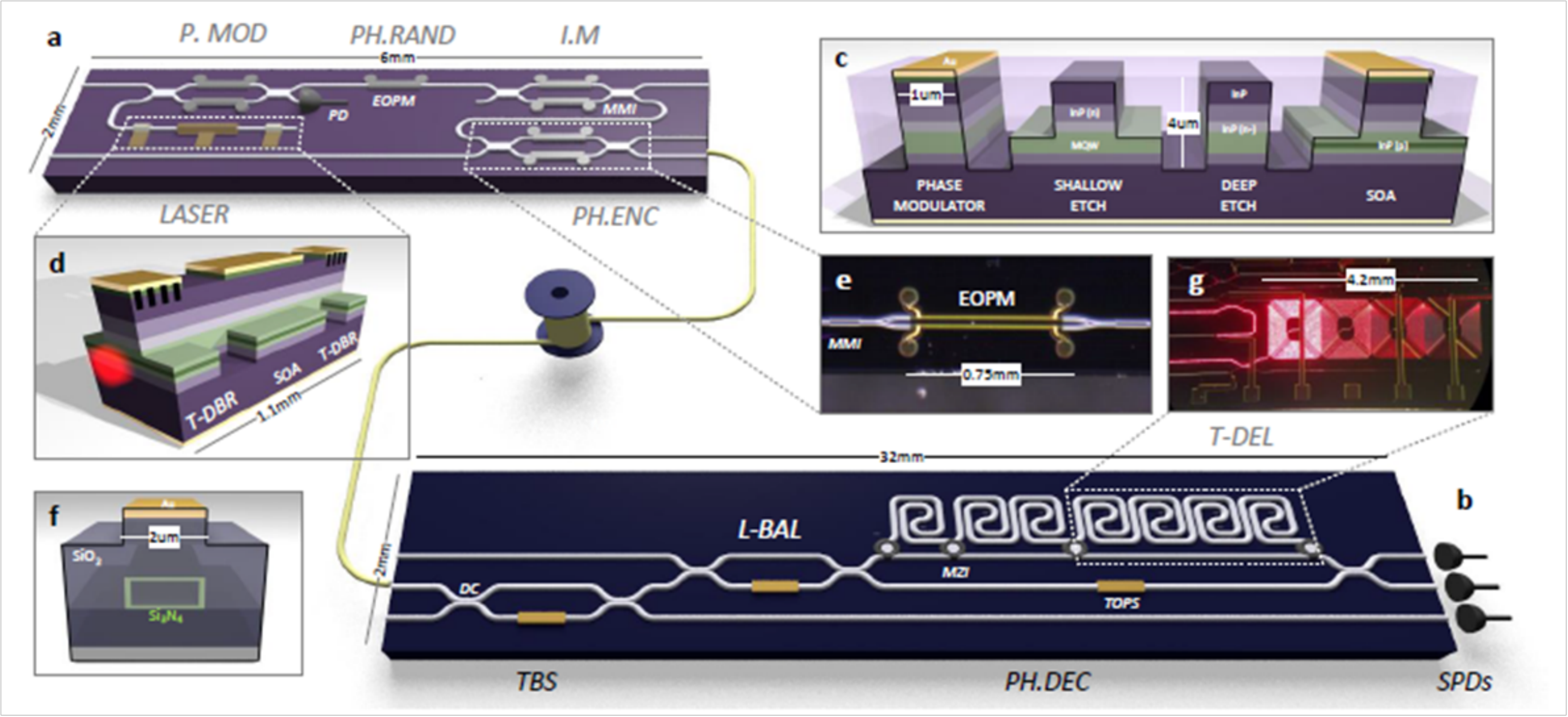}} \caption{Chip architecture combining several integrated photonic devices for the implementation of DV-QKD. Figure adapted with permission from ref. \cite{SEG:arxiv15}, courtesy of Chris Erven.} 
\label{fig:chip}
\end{figure}

Another important avenue to address the issue of cost and robustness is photonic integration \cite{HNM:arxiv13}. Chip-scale integration will bring high level of miniaturization, leading to compact and light-weight QKD modules that can be mass-manufactured at low cost. Two main integration platforms are currently being explored, namely silicon (Si) \cite{LSF:JSTQE14} and indium phosphide (InP) \cite{SLA:SST14}, while alternative techniques include lithium niobate (LiNbO$_3$) integration and glass waveguide technologies. For QKD protocols employing single-photon detection the main difficulty comes from the receiver side so initial experiments have focused on transmitter integration. A LiNbO$_3$ integrated polarization controller was used for state preparation in a QKD implementation \cite{ZAM:prl14}, while several techniques were combined to construct a handheld QKD sender module in ref. \cite{VRF:jstqe14}. More recently, a QKD transmitter chip that is reconfigurable to accommodate the state preparation for several QKD protocols, including decoy-state BB84, COW and DPS, has been developed on InP \cite{SEG:arxiv15} (Fig. \ref{fig:chip}).

Chip-scale QKD receivers are also progressing. Low-loss planar-lightwave-circuits (PLCs) based on silica-on-silicon technology have been routinely used to replace fiber-based asymmetric Mach-Zehnder interferometers \cite{TDH:njp05,NYT:jmo08,SEG:arxiv15}, a key enabling component for phase-based QKD protocols. Research efforts are currently focused on the integration of single-photon detectors using the aforementioned techniques, which will be essential for developing complete integrated systems. CV-QKD systems are particularly well suited for this objective because they only require the use of standard components. Indeed, Si photonic chips integrating many functionalities of a CV-QKD setup, including active elements such as amplitude and phase modulators and homodyne/heterodyne detectors based on germanium (Ge) photodiodes, have been developed \cite{ZPH:cleoeurope15}.

Development of chip-scale QKD is still at its early stages. Further research in this direction will help bring the QKD technology closer to its wide adoption.

\noindent {\bf New QKD protocols.}

\noindent In parallel to hardware development, much effort has also been devoted to novel QKD protocols aiming to outperform the established ones. Encouragingly, this line of research has led to protocols that may exhibit advantages when certain technical constraints are in place.  Below, we discuss two protocols featuring high photon information capacity or noise tolerance.

\noindent{\it HD-QKD.}
High dimension (HD) QKD allows retrieval of more than 1 bit from each detected photon, thus offering an advantage in the photon information capacity when the photon rate is restrained \cite{BP:pra00,BKB:pra01,CBK:prl02}. The choice for encoding is to use the arrival times of time-energy entangled photon pairs \cite{ZSW:prl08}, whose continuous nature permits encoding of extremely large alphabets. A security proof against collective attacks has been developed \cite{ZME:prl14}, which was followed by a laboratory experiment demonstrating a photon information capacity of up to 6.9 bits per coincidence and a key rate of 2.7~Mbit/s over a 20~km fiber \cite{ZZH:njp15}. While this development has narrowed the key rate gap between entanglement based and prepare-and-measure QKD systems, its potential in a field environment will face a challenge to maintain the near unity interference visibility which was key to the obtained information capacity.  HD-QKD without entanglement is also possible by exploiting the spatial degree of freedom, but its potential is restricted by the availability of high speed modulators \cite{MMO:njp15,ECG:scirep13}.

\begin{figure*}[htb]
\centerline{\includegraphics[width=15cm]{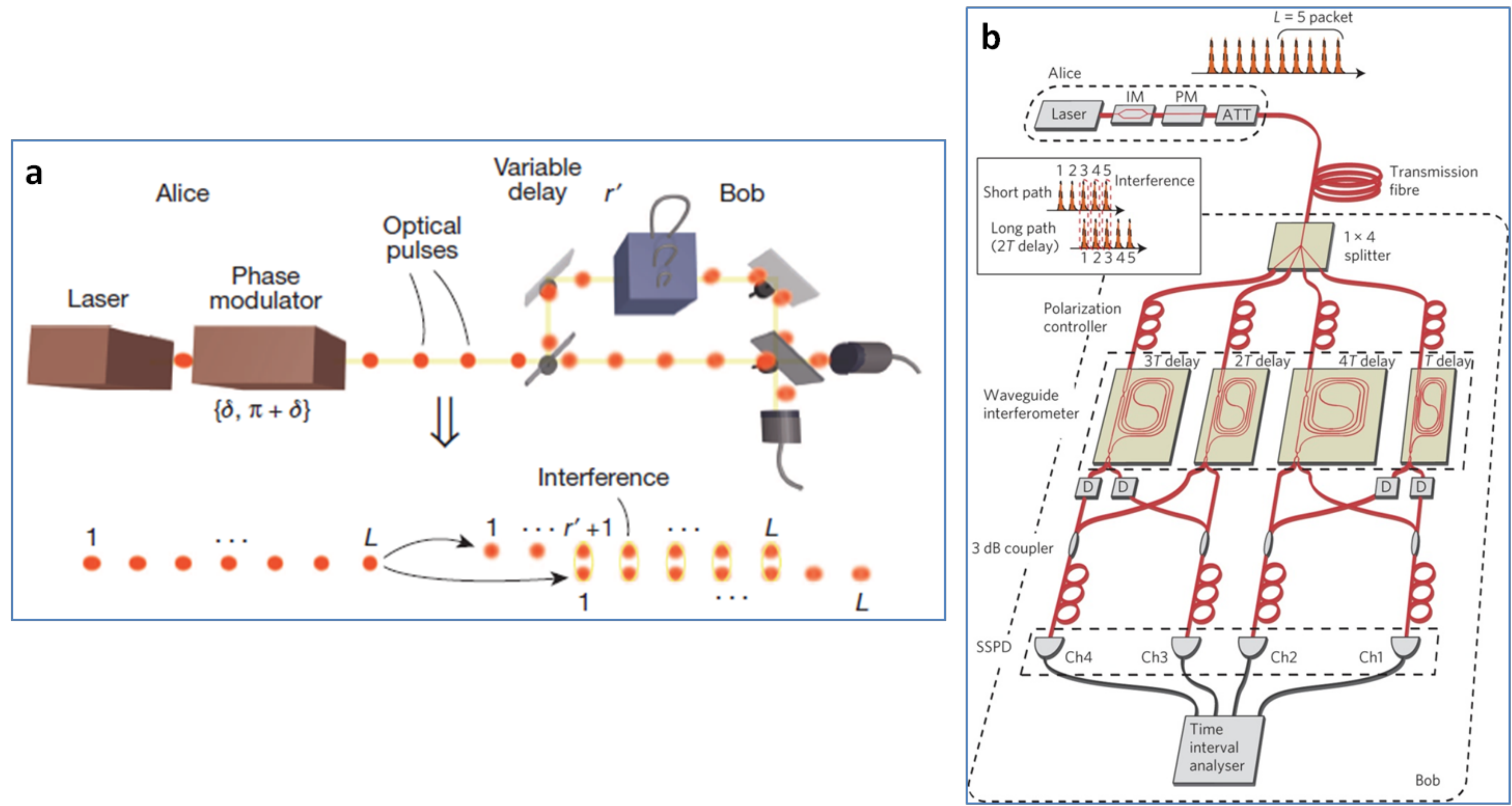}} \caption{a. Basic principle of RR-DPS QKD protocol \cite{SWK:nature14}. b. Example of experimental implementation of the RR-DPS QKD protocol \cite{TST:natphoton15}. Figures adapted with permission from: a. ref. \cite{SWK:nature14}, \copyright 2014 NPG; b. ref. \cite{TST:natphoton15}, \copyright 2015 NPG.  Courtesy of Masato Koashi.} 
\label{fig:rrdps}
\end{figure*}

\noindent{\it RR-DPS-QKD.}
The Round-Robin (RR) DPS protocol, which was proposed in 2014 \cite{SWK:nature14}, removes the need for monitoring the channel disturbance to establish security, in stark contrast with conventional QKD protocols (see Fig. \ref{fig:rrdps}a for the principle). Instead, Eve's information can be tightly set, even to an arbitrarily low level, by just choosing experimental parameters. In theory, a positive key rate is possible for any quantum bit error rate (QBER) lower than 50\%. This extraordinary QBER tolerance makes it attractive for deployment when large systematic errors cannot be avoided. Shortly after its introduction the protocol has stimulated a number of experimental demonstrations \cite{GCL:prl15,TST:natphoton15,WYC:natphoton15,LCD:pra16}. The RR-DPS-QKD protocol uses a transmitter identical to that found in a conventional DPS system \cite{IWY:prl02}, but requires a receiver that is capable of measuring the differential phase between any two pulses within a pulse group sent by Alice. Two different approaches are adopted. In the first, direct approach, a combination of optical switches and delay lines is used to bring the intended pulses into temporal overlap and then let them interfere \cite{TST:natphoton15,WYC:natphoton15,LCD:pra16} (see for example Fig. \ref{fig:rrdps}b). A more ingenious approach is to let a common phase reference interfere with all pulses sent by Alice, and then determine the differential phase between those pulses whose interference with the common reference produces a photon click \cite{GCL:prl15}. This approach avoids many problems associated with the direct one, such as loss and phase instability caused by optical delay lines and switches, but it will require remote optical phase locking for optimal performance.  As it currently stands, the best key rate for RR-DPS-QKD is around 10~kbit/s for a 50 km distance in fiber \cite{WYC:natphoton15} and cannot compete with the more mature decoy-state BB84 protocol. RR-DPS-QKD has the advantage of being robust against encoding errors \cite{MIT:pra15}, but it is vulnerable to attacks on detectors, which will be discussed in the next section.\\

\noindent {\large \bf Major challenges in practical security.}

\noindent While the security of a QKD protocol can be proven rigorously, its real-life implementation often contains imperfections that may be overlooked in the corresponding security proof. By exploiting such imperfections, various attacks, targeting either the source or the detectors, have been proposed; some of them have even been demonstrated to be effective against commercial systems \cite{ZFQ:pra08,LWW:natphoton10,XQL:njp10}. We refer the reader to a recent review \cite{LCT:natphoton14} for more details on quantum hacking and also countermeasures. To regain security in practical QKD, several solutions, including QKD based on testable assumptions \cite{LCT:natphoton14}, device independent (DI) QKD \cite{MY:focs98,ABG:prl07} (see also \cite{BP:prl12}), and MDI-QKD \cite{LCQ:prl12}, have been proposed. In the following, we discuss some important recent developments in this direction.

\begin{figure*}[htb]
\centerline{\includegraphics[width=12cm]{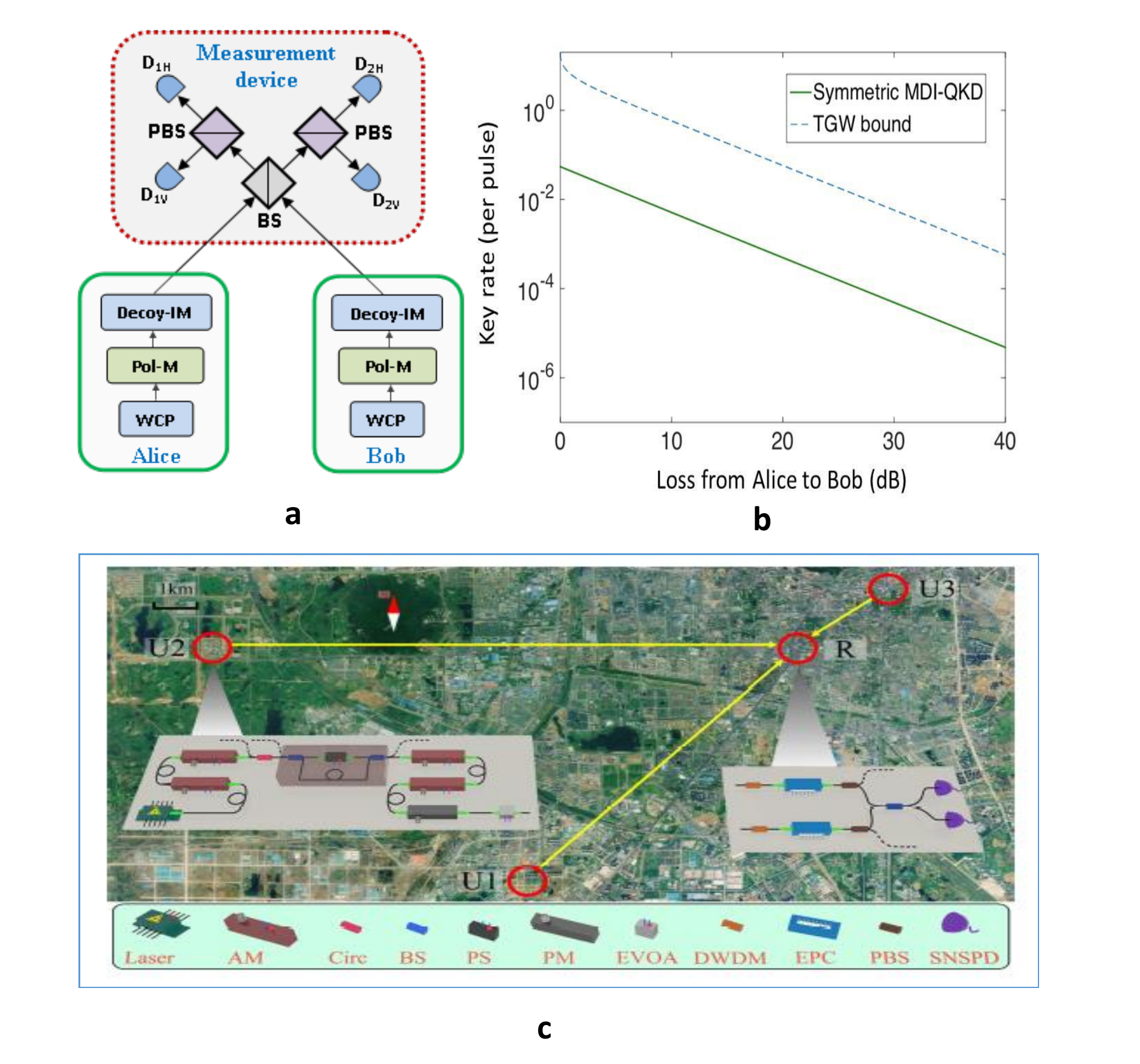}} \caption{a. The schematic diagram of DV MDI-QKD proposed in \cite{LCQ:prl12}. b. Simulation results of MDI-QKD and TGW bound
\cite{XCQ:natphoton15}. DV MDI-QKD has a high key rate and is suitable for metropolitan networks. The achievable key rate is about 0.01 bit/pulse at a channel loss of 5 dB (which corresponds to 25 km telecom fiber). The key rate of DV MDI-QKD is only about 2 orders of magnitude away from the TGW bound at a practical distance. The simulation corresponds to the symmetric MDI-QKD case where the channels between Alice and Charlie and Charlie and Bob have the same amount of losses. It assumes high-efficiency SNSPDs with detection efficiency of $93\%$ and dark count probability of $10^{-6}$ (per pulse) \cite{MVS:natphoton13}, and an intrinsic error rate of $0.1\%$ \cite{TYC:prl14}. The efficiency of error correction is assumed to be $1.16$. Note that if the detection efficiency is reduced, for instance, to 50\%, this induces a drop of the key rate of about a factor of 4. This means that for the metropolitan applications of DV MDI-QKD, the requirement on detector efficiency is not stringent. c. MDI-QKD metropolitan area network experimental field test with untrusted relays \cite{TYZ:prx15}. Figures adapted with permission from: a. ref. \cite{LCQ:prl12}, \copyright 2012 APS; b. ref. \cite{XCQ:natphoton15}, courtesy of Feihu Xu; c. ref. \cite{TYZ:prx15}, courtesy of Qiang Zhang.}%courtesy of XX.
\label{fig:mdi-qkd}
\end{figure*}

\noindent{\bf MDI-QKD.}
One promising long-term solution to side-channel attacks is DI-QKD, where the security relies on the violation of a Bell inequality and can be proven without knowing the implementation details. While recent loophole-free Bell experiments \cite{HBD:nature15,SMC:prl15,GVW:prl15} imply that DI-QKD could be implemented, the expected secure key rate is nevertheless impractically low even at short distances. A more practical solution is MDI-QKD, which is inherently immune to all side-channel attacks targeting the measurement device, usually the most vulnerable part in a QKD system. In fact, the measurement device in MDI-QKD can be treated as a ``black box'' which could even be manufactured and operated by Eve. Building upon \cite{BHM:pra96,Ina:algo02}, ref. \cite{LCQ:prl12} proposed a practical scheme with weak coherent pulses and decoy states (Fig. \ref{fig:mdi-qkd}a), whose security against the most general coherent attacks, taking into account the finite data size effect, has been proved in \cite{CXC:natcomm14} (see also ref. \cite{BP:prl12} which studied an entanglement-based representation with general finite-dimensional systems, and ref. \cite{LPT:prx13} which proposed a DI-QKD protocol with local Bell test).

MDI-QKD \cite{LCQ:prl12} is a natural building block for multi-user QKD networks, since the most expensive and complicated measurement device can be placed in an untrusted relay and shared among many QKD users \cite{FDL:nature13}. Several groups have demonstrated its feasibility. In particular, discrete-variable (DV) MDI-QKD was demonstrated over 200 km telecom fiber in lab conditions \cite{TYC:prl14} and over 30 km deployed fibers \cite{TYC:jstqe14}. With highly efficient single-photon detectors, the tolerable channel loss can be as high as 60 dB, which corresponds to 300 km telecom fiber \cite{VLC:jmo15}. A real-life fiber based multi-user MDI-QKD network was also implemented recently \cite{TYZ:prx15} (Fig. \ref{fig:mdi-qkd}c). Moreover, a 1 Mbit/s proof-of-principle MDI-QKD experiment was performed \cite{CLF:natphoton16}, thus illustrating the high key rate potential of DV MDI-QKD. This was also studied in \cite{XCQ:natphoton15} for MDI-QKD employing state-of-the-art SNSPDs; in Fig. \ref{fig:mdi-qkd}b, simulation results of the secret key rate in this case show an achievable key rate of 0.01 bit/pulse over 25 km. With a transmission rate of 1 GHz, this corresponds to a secret key rate of 10 Mbit/s, which is sufficient for many cryptographic applications. As a comparison, we also present in Fig. \ref{fig:mdi-qkd}b the previously mentioned fundamental upper bound per optical mode (TGW bound) \cite{TGW:natcomm14}. We see that the key rate of DV MDI-QKD is only about 2 orders of magnitude away from the TGW bound at a practical distance, hence this protocol is suitable for high speed communications in metropolitan area networks.

It is important to emphasize that one fundamental assumption in MDI-QKD is that Eve cannot interfere with Alice and Bob's state preparation processes. To prevent Eve from having access to quantum signals entering Alice's or Bob's labs and interfering with the state preparation process, MDI-QKD is commonly implemented using {\it independent} laser sources for Alice and Bob. Recently, gigahertz-clocked, phase-randomized pulses from independent gain-switched lasers have been demonstrated to interfere with high visibility, by control of the frequency chirp and/or emission jitter \cite{CLF:natphoton16,YLD:prapp14}.

\noindent\textit{DDI-QKD.} One drawback of MDI-QKD is that its key rate scales quadratically with the detector efficiency. This is because in most of existing MDI-QKD protocols (except for \cite{TLF:pra12}), secure keys are distilled from two-fold coincidence detection events \cite{noteMDI}. Recently, the detector-device-independent (DDI) QKD protocol, designed to bridge the strong security of MDI-QKD with the high efficiency of conventional QKD, was proposed \cite{GRF:pra15,LKM:apl14,CZZ:arxiv14}. In this protocol, the legitimate receiver employs a trusted linear optics network to decode information on photons received from an insecure quantum channel, and then performs a Bell state measurement (BSM) using uncharacterized detectors. One important advantage of this approach is that its key rate scales linearly with the detector efficiency. This is achieved by replacing the two-photon BSM scheme in the original MDI-QKD protocol (see Fig. \ref{fig:mdi-qkd}a) by a single-photon BSM scheme \cite{Kim:pra03}. However, its ability to completely remove detector side-channel attacks has yet to be proven. Either countermeasures to Trojan horse attacks \cite{GFK:pra06} or some trustworthiness to the BSM device is still required to establish the security of DDI-QKD \cite{Qi:pra15}. In fact, mathematically the standard BB84 QKD protocol based on a four-state modulation scheme can be formulated into a DDI-QKD protocol \cite{LLY:pra15}. This highlights the underlying connection between DDI-QKD and the BB84 protocol. Finally, we remark that the advantage of DDI-QKD compared to MDI-QKD becomes insignificant if high detection efficiency detectors are used in both schemes.

\noindent\textit{CV MDI-QKD.} The MDI-QKD scheme has been extended recently to the CV framework \cite{POS:natphoton15} (see also \cite{LZX:pra14,MSJ:pra14} for a more restricted security analysis). In the CV framework, both Alice and Bob prepare Gaussian-modulated coherent states and send them to an untrusted third party, Charlie, who measures the correlation between the incoming quantum states.
The CV MDI-QKD system requires high efficiency ($>85\%$) homodyne detectors for a positive key rate \cite{XCQ:natphoton15}. This efficiency requirement has been met in recent proof-of-principle laboratory free-space experiments \cite{POS:natphoton15,GHD:natcomm15}. However, achieving the required efficiencies in a fiber-based optical network setting is more challenging, owing to the detector coupling loss and losses by fiber network interconnects and components \cite{TYZ:prx15} (see also \cite{POSW:natphoton15} for a different perspective). When high efficiency detectors are in place, CV MDI-QKD would require an asymmetric configuration, where Charlie needs to be located close to one of the users. Even in this case, the expected key rate of the state-of-the-art CV MDI-QKD system drops to zero when the channel loss is above 6 dB (corresponding to 30 km standard telecom fiber) \cite{XCQ:natphoton15,POS:natphoton15}. Therefore, for long distance ($>$ 30 km) applications, DV MDI-QKD is currently the only option available for MDI-QKD. A reliable phase reference between Alice and Bob also needs to be established in CV MDI-QKD, and may be possible to realize using recently proposed techniques for standard CV-QKD \cite{QLP:prx15,SBC:prx15,HHL:ol15}. Despite these challenges, CV MDI-QKD has the potential for very high key rates, within one order of magnitude from the TGW bound, at relatively short communication distances.

\noindent{\bf QKD with imperfect sources.}
Given that the security loopholes associated with the measurement device can be closed by MDI-QKD, an important remaining question is how to justify the assumption of trustable quantum state preparation, including single-mode operation, perfect global phase randomization, no side channels, etc. On one hand, the imperfections in quantum state preparation need to be carefully quantified and taken into account in the security proof; on the other hand, practical countermeasures are required to prevent Trojan horse attacks \cite{GFK:pra06} on the source.

To address imperfections in quantum state preparation in QKD, a loss-tolerant protocol was proposed in ref. \cite{TCK:pra14}, which makes QKD tolerable to channel loss in the presence of source flaws (see also studies in \cite{YFM:pra13,YFM:pra14}). Based on the assumption that the single-photon components of the states prepared by Alice remain inside a two-dimensional Hilbert space, it was shown that Eve cannot enhance state preparation flaws by exploiting the channel loss and Eve's information can be bounded by the rejected data analysis \cite{BHP:jmo93}. The intuition for the security of loss-tolerant QKD protocol can be understood in the following manner. By assuming that the state prepared by Alice is a qubit, it becomes impossible for Eve to perform an unambiguous state discrimination (USD) attack \cite{DJL:pra00}. Indeed, in order for Eve to perform a USD attack, the states prepared by Alice must be linearly independent; but by having three or more states in a two-dimensional space, in general the set of states prepared by Alice is linearly dependent, thus making USD impossible.

The above loss-tolerant protocol has been further developed and demonstrated experimentally in ref. \cite{XWS:pra15}, where the authors implemented decoy-state QKD with imperfect state preparation and employed tight finite-key security bounds with composable security against coherent attacks. The work in \cite{TCK:pra14} has also been extended to the finite-key regime in \cite{MCL:njp15}, where a wide range of imperfections in the laser source, such as the intensity fluctuations, have been taken into account. In \cite{CZL:njp15}, a rigorous security proof of QKD systems using discrete-phase-randomized coherent states was given, thus removing the requirement for perfect phase randomization. With respect to this, we note that gain-switched laser diodes are presently the de facto QKD light source, capable of naturally providing phase-randomized coherent pulses at a clock rate of up to 2.5 GHz \cite{YLD:apl14}.

Progress has also been made on enhancing the security of QKD by carefully examining source imperfections in implementations. Refs. \cite{JAK:njp14,JSK:jstqe15}, studied the risk of Trojan horse attacks due to back reflections from commonly used optical components in QKD. Similar research was also conducted for CV-QKD \cite{SKJ:cleo15}. In \cite{LCW:prx15}, by using laser-induced damage threshold of single-mode optical fiber to bound the photon numbers in Eve's Trojan horse pulses, the authors provided quantitative security bounds and a purely passive solution against a general Trojan horse attack.

All the above advances strongly suggest the feasibility of long distance secure quantum communication with imperfect sources. A promising research direction is to apply the above techniques for QKD with imperfect sources to MDI-QKD leading to practical side-channel-free QKD. To achieve this goal, it is necessary to establish a comprehensive list of assumptions on the sources, and verify them one by one. In a recent experimental demonstration \cite{TWB:pra16}, the loss-tolerant protocol is applied to a MDI-QKD setting. Such an experiment thus addresses source and detector flaws at the same time.

We end our discussion on practical security by noting that in both classical and quantum cryptography, it is also important to carefully address the risks of side-channel attacks on the electronics and post-processing layers. Various side-channel attacks discovered in classical cryptography, such as the timing attack \cite{Kocher:crypto96}, the power-monitoring attack \cite{KJJ:crypto99}, and acoustic cryptanalysis \cite{GST:crypto14}, can also pose threats to quantum cryptography. Closing these side channels requires substantial future efforts.\\

\noindent {\large \bf Network QKD.}

\noindent So far, our discussion has been largely limited to point-to-point QKD links. While these links are useful for some applications, QKD network structures must be considered in order to enable access by a greater many users and also to extend the reach and geographical coverage. Additionally, the incorporation of mobile QKD nodes for key transports will add to network connection flexibility and allow even greater geographical coverage. In the following, we discuss approaches for building a QKD network and possibilities for future mobile QKD deployment.

\noindent {\bf Building QKD networks.}
An important issue in a network setting is the topology that allows for multiple users to access the network. A star topology is suitable for this purpose for relatively short distance (up to 400 km). Imagine a star network where there is at most one intermediate node between any two users, allowing for secure quantum communication among all users without the need for the relay to be trusted. In fact, this approach has already been demonstrated based on the MDI-QKD protocol \cite{TYZ:prx15}. The long-term vision is for each user to use a simple and cheap transmitter and outsource all the complicated devices for network control and measurement to an untrusted network operator. Since only one set of measurement devices will be needed for such a network that is shared by many users, the cost per user could be kept relatively low. The network provider would then be in a favorable position to deploy state-of-the-art technologies including high detection efficiency SNSPDs to enhance the performance of the network and to perform all network management tasks. The important advantage is that the network operator can be completely untrusted without compromising security. Experimental demonstrations of network MDI-QKD, either in optical fibers \cite{TYZ:prx15} or in free space, are a major step towards such QKD networks with untrusted relays.

Nonetheless, MDI-QKD is limited in distance, hence in order to address the great challenge of extending the distance of secure QKD, three further approaches are possible.
The first and the simplest approach is to use trusted relays. This is already feasible with current technology and indeed has been used as the standard in existing QKD networks \cite{PPA:njp09,SFI:oe11}. By setting up trusted nodes, for instance, every 50 km, to relay secrets, it is possible to achieve secure communication over arbitrarily long distances. The QKD network currently under development between Shanghai and Beijing is based on this approach.

The second approach is quantum repeaters, which remove the need for the users to trust the relay nodes. Quantum repeaters are beyond current technology, but have been a subject of intense research efforts in recent years. The long-term vision here is to construct a global quantum internet as described, for example, in ref. \cite{Kimble:nature08}. Research efforts on quantum repeaters have focused on matter quantum memories and their interface with photonic flying qubits \cite{NB:natphoton14,BCT:natphoton14}. However, new recent approaches manage to reduce the need for a quantum memory \cite{MSD:natphoton12} or to completely remove it by using all-photonic quantum repeaters \cite{ATL:natcomm15}.

Finally, the third approach is ground-to-satellite QKD. By using one or a few trusted satellites as relay stations, it is possible to extend the distance of secure QKD to the global scale. To this end, several free-space studies, including experiments with low earth orbit (LEO) satellites, have been conducted \cite{BHL:prl00,NMR:natphoton13,WYL:natphoton13,VBD:prl15,Meyers:fso15,EGK:icsos15,BHZ:opex15}. China, the EU and Canada are all currently exploring experimental ground-to-satellite QKD in ambitious long-term projects involving LEO satellites.

\noindent {\bf Mobile QKD.}
The studies in free-space QKD may also open the door to mobile QKD networks, which can be useful in many applications, such as ship-to-ship communication, airport traffic control, communication between autonomous vehicles, etc. In such a network, the mobility of QKD platforms requires the network to be highly reconfigurable -- the QKD users should be able to automatically determine the optimal QKD route in real time based on their locations. Fast beam tracking systems are indispensable. Furthermore, due to the strong ambient light, an effective filtering scheme is required to selectively detect quantum signals. A recent study shows that CV-QKD based on coherent detection could be robust against ambient noise photons due to the intrinsic filtering function of the local oscillator \cite{HPK:njp14}. We also note that preliminary studies suggest that QKD at microwave wavelengths, which are widely used in wireless communications, might be feasible over short distances \cite{WPL:prl10,WPR:pra12}. Driven by various potential applications, we expect that mobile QKD will become an active research topic in the coming years.\\

\noindent {\large \bf Conclusion.}

\noindent In this review, we have discussed important challenges in practical QKD. These range from extending security proofs to the most general attacks allowed by quantum mechanics to developing photonic chips as well as side-channel-free systems and global-scale QKD networks. Addressing these challenges using some of the approaches that we have presented will open the way to the use of QKD technology for securing everyday interactions.

As the lead application of the field of Quantum Information Processing, advances in QKD will have important implications in many other applications too. For example, a great range of quantum communication protocols beyond QKD have been studied in recent years and their development has directly benefited from research in QKD. These include, for instance, quantum bit commitment \cite{Mayers:prl97,LC:prl97,LKB:prl15}, quantum secret sharing \cite{CGL:prl99,HBB:pra99,BMH:natcommun14}, quantum coin flipping \cite{BBB:natcomm11,PJL:natcommun14}, quantum fingerprinting \cite{BCW:prl01,XAW:natcommun15}, quantum digital signatures \cite{GC:arxiv01,DCK:pra16}, blind quantum computing \cite{BFK:focs09,BKB:science12}, and position-based quantum cryptography \cite{LL:pra11,BCF:siam14,CL:pra15}. It is known that some of those protocols, such as quantum bit commitment and position-based quantum cryptography, cannot be perfectly achieved with unconditional security. However, other security models exist, such as, for instance, those based on relativistic constraints or on noisy storage assumptions \cite{WCS:pra10}, where by assuming that it is impossible for an eavesdropper to store quantum information for a long time, one can retrieve security for such protocols.

Determining the exact power and limitations of quantum communication is the subject of intense research efforts worldwide. The formidable developments that can be expected in the next few years will mark important milestones towards the quantum internet of the future.

\section*{Acknowledgments}
We acknowledge helpful comments from many colleagues including Romain All\'eaume, Hoi-Fung Chau, Marcos Curty, Philippe Grangier, Anthony Leverrier, Charles Ci Wen Lim, Marco Lucamarini, Li Qian, Xiongfeng Ma, Kiyoshi Tamaki and Feihu Xu. We thank colleagues including Ping Koy Lam, Vikas Anant, Jessie Qin-Dregely, Chris Erven, Masato Koashi and Qiang Zhang for allowing us to reproduce some of their figures. We thank Warren Raye of Nature Partner Journals for securing the permission for reproductions of figures from various publishers. We acknowledge financial support from NSERC, CFI, ORF, the U.S. Office of Naval Research (ONR), the Laboratory Directed Research and Development (LDRD) Program of Oak Ridge National Laboratory (managed by UT-Battelle LLC for the U.S. Department of Energy), the City of Paris, the French National Research Agency, the Ile-de-France Region, the France-USA Partner University Fund, and the Commissioned Research of National Institute of Information and Communications Technology (NICT), Japan.

\section*{Additional information}
Correspondence and requests for materials should be addressed to H.-K.L.

\bibliography{NQI_review}

\begin{thebibliography}{177}
\expandafter\ifx\csname natexlab\endcsname\relax\def\natexlab#1{#1}\fi
\expandafter\ifx\csname bibnamefont\endcsname\relax
  \def\bibnamefont#1{#1}\fi
\expandafter\ifx\csname bibfnamefont\endcsname\relax
  \def\bibfnamefont#1{#1}\fi
\expandafter\ifx\csname citenamefont\endcsname\relax
  \def\citenamefont#1{#1}\fi
\expandafter\ifx\csname url\endcsname\relax
  \def\url#1{\texttt{#1}}\fi
\expandafter\ifx\csname urlprefix\endcsname\relax\def\urlprefix{URL }\fi
\providecommand{\bibinfo}[2]{#2}
\providecommand{\eprint}[2][]{\url{#2}}

\bibitem[{\citenamefont{Shor}(1994)}]{Shor:focs94}
\bibinfo{author}{\bibfnamefont{P.~W.} \bibnamefont{Shor}}, in
  \emph{\bibinfo{booktitle}{Proc. 35th Annual Symposium on Foundations of
  Computer Science}}, edited by
  \bibinfo{editor}{\bibfnamefont{S.}~\bibnamefont{Goldwasser}}
  (\bibinfo{publisher}{IEEE Computer Society Press}, \bibinfo{year}{1994}), pp.
  \bibinfo{pages}{124--134}.

\bibitem[{Can()}]{CanadianCensus}
\bibinfo{note}{Statistical Canada webpage, ``Release of personal data after 92
  years'', see
  http://www12.statcan.gc.ca/census-recensement/2011/ref/about-apropos/personal-personnels-eng.cfm}.

\bibitem[{ENI()}]{ENIAC}
\bibinfo{note}{See http://www.britannica.com/EBchecked/topic/183842/ ENIAC}.

\bibitem[{NSA()}]{NSA}
\bibinfo{note}{See https://www.nsa.gov/ia/programs/suiteb\_cryptography/}.

\bibitem[{not({\natexlab{a}})}]{notePQC}
\bibinfo{note}{These algorithms are typically based on hard computational
  problems involving for instance the structure of elliptic curves or of some
  specific lattices. Despite important progress in the development of such
  algorithms, it is still an open question whether they are secure against a
  quantum computer.}

\bibitem[{\citenamefont{Bennett and Brassard}(1984)}]{BB84}
\bibinfo{author}{\bibfnamefont{C.~H.} \bibnamefont{Bennett}} \bibnamefont{and}
  \bibinfo{author}{\bibfnamefont{G.}~\bibnamefont{Brassard}}, in
  \emph{\bibinfo{booktitle}{Proc. IEEE International Conference on Computers,
  Systems, and Signal Processing}}, edited by
  \bibinfo{editor}{\bibfnamefont{S.}~\bibnamefont{Goldwasser}}
  (\bibinfo{publisher}{IEEE Press}, \bibinfo{year}{1984}), pp.
  \bibinfo{pages}{175--179}.

\bibitem[{\citenamefont{Ekert}(1991)}]{Ekert:prl91}
\bibinfo{author}{\bibfnamefont{A.~K.} \bibnamefont{Ekert}},
  \bibinfo{journal}{Phys. Rev. Lett.} \textbf{\bibinfo{volume}{67}},
  \bibinfo{pages}{661} (\bibinfo{year}{1991}).

\bibitem[{\citenamefont{Scarani et~al.}(2009)\citenamefont{Scarani,
  Bechmann-Pasquinucci, Cerf, Du{\v{s}}ek, L{\"u}tkenhaus, and
  Peev}}]{SBC:rmp09}
\bibinfo{author}{\bibfnamefont{V.}~\bibnamefont{Scarani}},
  \bibinfo{author}{\bibfnamefont{H.}~\bibnamefont{Bechmann-Pasquinucci}},
  \bibinfo{author}{\bibfnamefont{N.~J.} \bibnamefont{Cerf}},
  \bibinfo{author}{\bibfnamefont{M.}~\bibnamefont{Du{\v{s}}ek}},
  \bibinfo{author}{\bibfnamefont{N.}~\bibnamefont{L{\"u}tkenhaus}},
  \bibnamefont{and} \bibinfo{author}{\bibfnamefont{M.}~\bibnamefont{Peev}},
  \bibinfo{journal}{Rev. Mod. Phys.} \textbf{\bibinfo{volume}{81}},
  \bibinfo{pages}{1301} (\bibinfo{year}{2009}).

\bibitem[{\citenamefont{Lo et~al.}(2014)\citenamefont{Lo, Curty, and
  Tamaki}}]{LCT:natphoton14}
\bibinfo{author}{\bibfnamefont{H.-K.} \bibnamefont{Lo}},
  \bibinfo{author}{\bibfnamefont{M.}~\bibnamefont{Curty}}, \bibnamefont{and}
  \bibinfo{author}{\bibfnamefont{K.}~\bibnamefont{Tamaki}},
  \bibinfo{journal}{Nature Photon.} \textbf{\bibinfo{volume}{8}},
  \bibinfo{pages}{595} (\bibinfo{year}{2014}).

\bibitem[{\citenamefont{Mayers}(2001)}]{Mayers:jacm01}
\bibinfo{author}{\bibfnamefont{D.}~\bibnamefont{Mayers}},
  \bibinfo{journal}{Journal of the ACM} \textbf{\bibinfo{volume}{48}},
  \bibinfo{pages}{351} (\bibinfo{year}{2001}).

\bibitem[{\citenamefont{Lo and Chau}(1999)}]{LC:science99}
\bibinfo{author}{\bibfnamefont{H.-K.} \bibnamefont{Lo}} \bibnamefont{and}
  \bibinfo{author}{\bibfnamefont{H.~F.} \bibnamefont{Chau}},
  \bibinfo{journal}{Science} \textbf{\bibinfo{volume}{283}},
  \bibinfo{pages}{2050} (\bibinfo{year}{1999}).

\bibitem[{\citenamefont{Shor and Preskill}(2000)}]{SP:prl00}
\bibinfo{author}{\bibfnamefont{P.~W.} \bibnamefont{Shor}} \bibnamefont{and}
  \bibinfo{author}{\bibfnamefont{J.}~\bibnamefont{Preskill}},
  \bibinfo{journal}{Phys. Rev. Lett.} \textbf{\bibinfo{volume}{85}},
  \bibinfo{pages}{441} (\bibinfo{year}{2000}).

\bibitem[{\citenamefont{Unruh}(2012)}]{Unruh:iacr12}
\bibinfo{author}{\bibfnamefont{D.}~\bibnamefont{Unruh}}, \bibinfo{journal}{IACR
  Cryptology ePrint Archive} p. \bibinfo{pages}{177} (\bibinfo{year}{2012}).

\bibitem[{\citenamefont{Wootters and Zurek}(1982)}]{WZ:nature82}
\bibinfo{author}{\bibfnamefont{W.~K.} \bibnamefont{Wootters}} \bibnamefont{and}
  \bibinfo{author}{\bibfnamefont{W.~H.} \bibnamefont{Zurek}},
  \bibinfo{journal}{Nature} \textbf{\bibinfo{volume}{299}},
  \bibinfo{pages}{802} (\bibinfo{year}{1982}).

\bibitem[{\citenamefont{Dieks}(1982)}]{Dieks:pl82}
\bibinfo{author}{\bibfnamefont{D.}~\bibnamefont{Dieks}},
  \bibinfo{journal}{Phys. Lett.} \textbf{\bibinfo{volume}{92A}},
  \bibinfo{pages}{271} (\bibinfo{year}{1982}).

\bibitem[{\citenamefont{Kimble}(2008)}]{Kimble:nature08}
\bibinfo{author}{\bibfnamefont{H.~J.} \bibnamefont{Kimble}},
  \bibinfo{journal}{Nature} \textbf{\bibinfo{volume}{453}},
  \bibinfo{pages}{1023} (\bibinfo{year}{2008}).

\bibitem[{\citenamefont{Qiu}(2014)}]{Qiu:natnews14}
\bibinfo{author}{\bibfnamefont{J.}~\bibnamefont{Qiu}},
  \bibinfo{journal}{Nature} \textbf{\bibinfo{volume}{508}},
  \bibinfo{pages}{441} (\bibinfo{year}{2014}).

\bibitem[{\citenamefont{Peev et~al.}(2009)\citenamefont{Peev, Pacher,
  All\'eaume, Barreiro, Bouda, Boxleitner, Debuisschert, Diamanti, Dianati,
  Dynes et~al.}}]{PPA:njp09}
\bibinfo{author}{\bibfnamefont{M.}~\bibnamefont{Peev}},
  \bibinfo{author}{\bibfnamefont{C.}~\bibnamefont{Pacher}},
  \bibinfo{author}{\bibfnamefont{R.}~\bibnamefont{All\'eaume}},
  \bibinfo{author}{\bibfnamefont{C.}~\bibnamefont{Barreiro}},
  \bibinfo{author}{\bibfnamefont{J.}~\bibnamefont{Bouda}},
  \bibinfo{author}{\bibfnamefont{W.}~\bibnamefont{Boxleitner}},
  \bibinfo{author}{\bibfnamefont{T.}~\bibnamefont{Debuisschert}},
  \bibinfo{author}{\bibfnamefont{E.}~\bibnamefont{Diamanti}},
  \bibinfo{author}{\bibfnamefont{M.}~\bibnamefont{Dianati}},
  \bibinfo{author}{\bibfnamefont{J.~F.} \bibnamefont{Dynes}},
  \bibnamefont{et~al.}, \bibinfo{journal}{New J. Phys.}
  \textbf{\bibinfo{volume}{11}}, \bibinfo{pages}{075001}
  (\bibinfo{year}{2009}).

\bibitem[{\citenamefont{Huttner et~al.}(1995)\citenamefont{Huttner, Imoto,
  Gisin, and Mor}}]{HIG:pra95}
\bibinfo{author}{\bibfnamefont{B.}~\bibnamefont{Huttner}},
  \bibinfo{author}{\bibfnamefont{N.}~\bibnamefont{Imoto}},
  \bibinfo{author}{\bibfnamefont{N.}~\bibnamefont{Gisin}}, \bibnamefont{and}
  \bibinfo{author}{\bibfnamefont{T.}~\bibnamefont{Mor}},
  \bibinfo{journal}{Phys. Rev. A} \textbf{\bibinfo{volume}{51}},
  \bibinfo{pages}{1863} (\bibinfo{year}{1995}).

\bibitem[{\citenamefont{Hwang}(2003)}]{Hwang:prl03}
\bibinfo{author}{\bibfnamefont{W.-Y.} \bibnamefont{Hwang}},
  \bibinfo{journal}{Phys. Rev. Lett.} \textbf{\bibinfo{volume}{91}},
  \bibinfo{pages}{057901} (\bibinfo{year}{2003}).

\bibitem[{\citenamefont{Lo et~al.}(2005)\citenamefont{Lo, Ma, and
  Chen}}]{LMC:prl05}
\bibinfo{author}{\bibfnamefont{H.-K.} \bibnamefont{Lo}},
  \bibinfo{author}{\bibfnamefont{X.}~\bibnamefont{Ma}}, \bibnamefont{and}
  \bibinfo{author}{\bibfnamefont{K.}~\bibnamefont{Chen}},
  \bibinfo{journal}{Phys. Rev. Lett.} \textbf{\bibinfo{volume}{94}},
  \bibinfo{pages}{230504} (\bibinfo{year}{2005}).

\bibitem[{\citenamefont{Wang}(2005)}]{Wang:prl05}
\bibinfo{author}{\bibfnamefont{X.-B.} \bibnamefont{Wang}},
  \bibinfo{journal}{Phys. Rev. Lett.} \textbf{\bibinfo{volume}{94}},
  \bibinfo{pages}{230503} (\bibinfo{year}{2005}).

\bibitem[{\citenamefont{Lo et~al.}(2012)\citenamefont{Lo, Curty, and
  Qi}}]{LCQ:prl12}
\bibinfo{author}{\bibfnamefont{H.-K.} \bibnamefont{Lo}},
  \bibinfo{author}{\bibfnamefont{M.}~\bibnamefont{Curty}}, \bibnamefont{and}
  \bibinfo{author}{\bibfnamefont{B.}~\bibnamefont{Qi}}, \bibinfo{journal}{Phys.
  Rev. Lett.} \textbf{\bibinfo{volume}{108}}, \bibinfo{pages}{130503}
  (\bibinfo{year}{2012}).

\bibitem[{\citenamefont{Christandl et~al.}(2007)\citenamefont{Christandl,
  Koenig, Mitchison, and Renner}}]{CKM:commmathphys07}
\bibinfo{author}{\bibfnamefont{M.}~\bibnamefont{Christandl}},
  \bibinfo{author}{\bibfnamefont{R.}~\bibnamefont{Koenig}},
  \bibinfo{author}{\bibfnamefont{G.}~\bibnamefont{Mitchison}},
  \bibnamefont{and} \bibinfo{author}{\bibfnamefont{R.}~\bibnamefont{Renner}},
  \bibinfo{journal}{Comm. Math. Phys.} \textbf{\bibinfo{volume}{273}},
  \bibinfo{pages}{473} (\bibinfo{year}{2007}).

\bibitem[{\citenamefont{Hensen et~al.}(2015)\citenamefont{Hensen, Bernien,
  Dr\'{e}au, Reiserer, Kalb, Blok, Ruitenberg, Vermeulen, Schouten, Abell\'an
  et~al.}}]{HBD:nature15}
\bibinfo{author}{\bibfnamefont{B.}~\bibnamefont{Hensen}},
  \bibinfo{author}{\bibfnamefont{H.}~\bibnamefont{Bernien}},
  \bibinfo{author}{\bibfnamefont{A.~E.} \bibnamefont{Dr\'{e}au}},
  \bibinfo{author}{\bibfnamefont{A.}~\bibnamefont{Reiserer}},
  \bibinfo{author}{\bibfnamefont{N.}~\bibnamefont{Kalb}},
  \bibinfo{author}{\bibfnamefont{M.~S.} \bibnamefont{Blok}},
  \bibinfo{author}{\bibfnamefont{J.}~\bibnamefont{Ruitenberg}},
  \bibinfo{author}{\bibfnamefont{R.~F.~L.} \bibnamefont{Vermeulen}},
  \bibinfo{author}{\bibfnamefont{R.~N.} \bibnamefont{Schouten}},
  \bibinfo{author}{\bibfnamefont{C.}~\bibnamefont{Abell\'an}},
  \bibnamefont{et~al.}, \bibinfo{journal}{Nature}
  \textbf{\bibinfo{volume}{526}}, \bibinfo{pages}{682} (\bibinfo{year}{2015}).

\bibitem[{\citenamefont{Gol'Tsman et~al.}(2001)\citenamefont{Gol'Tsman, Okunev,
  Chulkova, Lipatov, Semenov, Smirnov, Voronov, Dzardanov, Williams, and
  Sobolewski}}]{GOC:apl01}
\bibinfo{author}{\bibfnamefont{G.~N.} \bibnamefont{Gol'Tsman}},
  \bibinfo{author}{\bibfnamefont{O.}~\bibnamefont{Okunev}},
  \bibinfo{author}{\bibfnamefont{G.}~\bibnamefont{Chulkova}},
  \bibinfo{author}{\bibfnamefont{A.}~\bibnamefont{Lipatov}},
  \bibinfo{author}{\bibfnamefont{A.}~\bibnamefont{Semenov}},
  \bibinfo{author}{\bibfnamefont{K.}~\bibnamefont{Smirnov}},
  \bibinfo{author}{\bibfnamefont{B.}~\bibnamefont{Voronov}},
  \bibinfo{author}{\bibfnamefont{A.}~\bibnamefont{Dzardanov}},
  \bibinfo{author}{\bibfnamefont{C.}~\bibnamefont{Williams}}, \bibnamefont{and}
  \bibinfo{author}{\bibfnamefont{R.}~\bibnamefont{Sobolewski}},
  \bibinfo{journal}{Appl. Phys. Lett.} \textbf{\bibinfo{volume}{79}},
  \bibinfo{pages}{705} (\bibinfo{year}{2001}).

\bibitem[{\citenamefont{Lita et~al.}(2008)\citenamefont{Lita, Miller, and
  Nam}}]{LMN:opex08}
\bibinfo{author}{\bibfnamefont{A.~E.} \bibnamefont{Lita}},
  \bibinfo{author}{\bibfnamefont{A.~J.} \bibnamefont{Miller}},
  \bibnamefont{and} \bibinfo{author}{\bibfnamefont{S.~W.} \bibnamefont{Nam}},
  \bibinfo{journal}{Opt. Express} \textbf{\bibinfo{volume}{16}},
  \bibinfo{pages}{3032} (\bibinfo{year}{2008}).

\bibitem[{\citenamefont{Albota and Wong}(2004)}]{AW:ol04}
\bibinfo{author}{\bibfnamefont{M.~A.} \bibnamefont{Albota}} \bibnamefont{and}
  \bibinfo{author}{\bibfnamefont{F.~N.} \bibnamefont{Wong}},
  \bibinfo{journal}{Opt. Lett.} \textbf{\bibinfo{volume}{29}},
  \bibinfo{pages}{1449} (\bibinfo{year}{2004}).

\bibitem[{\citenamefont{Langrock et~al.}(2005)\citenamefont{Langrock, Diamanti,
  Roussev, Yamamoto, Fejer, and Takesue}}]{LDR:ol05}
\bibinfo{author}{\bibfnamefont{C.}~\bibnamefont{Langrock}},
  \bibinfo{author}{\bibfnamefont{E.}~\bibnamefont{Diamanti}},
  \bibinfo{author}{\bibfnamefont{R.~V.} \bibnamefont{Roussev}},
  \bibinfo{author}{\bibfnamefont{Y.}~\bibnamefont{Yamamoto}},
  \bibinfo{author}{\bibfnamefont{M.~M.} \bibnamefont{Fejer}}, \bibnamefont{and}
  \bibinfo{author}{\bibfnamefont{H.}~\bibnamefont{Takesue}},
  \bibinfo{journal}{Opt. Lett.} \textbf{\bibinfo{volume}{30}},
  \bibinfo{pages}{1725} (\bibinfo{year}{2005}).

\bibitem[{\citenamefont{Yuan et~al.}(2007)\citenamefont{Yuan, Kardynal, Sharpe,
  and Shields}}]{YKS:apl07}
\bibinfo{author}{\bibfnamefont{Z.~L.} \bibnamefont{Yuan}},
  \bibinfo{author}{\bibfnamefont{B.~E.} \bibnamefont{Kardynal}},
  \bibinfo{author}{\bibfnamefont{A.~W.} \bibnamefont{Sharpe}},
  \bibnamefont{and} \bibinfo{author}{\bibfnamefont{A.~J.}
  \bibnamefont{Shields}}, \bibinfo{journal}{Appl. Phys. Lett.}
  \textbf{\bibinfo{volume}{91}}, \bibinfo{pages}{041114}
  (\bibinfo{year}{2007}).

\bibitem[{\citenamefont{Hansen et~al.}(2001)\citenamefont{Hansen, Aichele,
  Hettich, Lodahl, Lvovsky, Mlynek, and Schiller}}]{HAH:ol01}
\bibinfo{author}{\bibfnamefont{H.}~\bibnamefont{Hansen}},
  \bibinfo{author}{\bibfnamefont{T.}~\bibnamefont{Aichele}},
  \bibinfo{author}{\bibfnamefont{C.}~\bibnamefont{Hettich}},
  \bibinfo{author}{\bibfnamefont{P.}~\bibnamefont{Lodahl}},
  \bibinfo{author}{\bibfnamefont{A.~I.} \bibnamefont{Lvovsky}},
  \bibinfo{author}{\bibfnamefont{J.}~\bibnamefont{Mlynek}}, \bibnamefont{and}
  \bibinfo{author}{\bibfnamefont{S.}~\bibnamefont{Schiller}},
  \bibinfo{journal}{Opt. Lett.} \textbf{\bibinfo{volume}{26}},
  \bibinfo{pages}{1714} (\bibinfo{year}{2001}).

\bibitem[{\citenamefont{Jennewein et~al.}(2000)\citenamefont{Jennewein,
  Achleitner, Weihs, Weinfurter, and Zeilinger}}]{JAW:rsi00}
\bibinfo{author}{\bibfnamefont{T.}~\bibnamefont{Jennewein}},
  \bibinfo{author}{\bibfnamefont{U.}~\bibnamefont{Achleitner}},
  \bibinfo{author}{\bibfnamefont{G.}~\bibnamefont{Weihs}},
  \bibinfo{author}{\bibfnamefont{H.}~\bibnamefont{Weinfurter}},
  \bibnamefont{and}
  \bibinfo{author}{\bibfnamefont{A.}~\bibnamefont{Zeilinger}},
  \bibinfo{journal}{Rev. Sci. Instrum.} \textbf{\bibinfo{volume}{71}},
  \bibinfo{pages}{1675} (\bibinfo{year}{2000}).

\bibitem[{\citenamefont{Zhu et~al.}(2013)\citenamefont{Zhu, Tang, Qian, Helt,
  Liscidini, Sipe, Corbari, Canagasabey, Ibsen, and Kazansky}}]{ZTQ:ol13}
\bibinfo{author}{\bibfnamefont{E.~Y.} \bibnamefont{Zhu}},
  \bibinfo{author}{\bibfnamefont{Z.}~\bibnamefont{Tang}},
  \bibinfo{author}{\bibfnamefont{L.}~\bibnamefont{Qian}},
  \bibinfo{author}{\bibfnamefont{L.~G.} \bibnamefont{Helt}},
  \bibinfo{author}{\bibfnamefont{M.}~\bibnamefont{Liscidini}},
  \bibinfo{author}{\bibfnamefont{J.~E.} \bibnamefont{Sipe}},
  \bibinfo{author}{\bibfnamefont{C.}~\bibnamefont{Corbari}},
  \bibinfo{author}{\bibfnamefont{A.}~\bibnamefont{Canagasabey}},
  \bibinfo{author}{\bibfnamefont{M.}~\bibnamefont{Ibsen}}, \bibnamefont{and}
  \bibinfo{author}{\bibfnamefont{P.~G.} \bibnamefont{Kazansky}},
  \bibinfo{journal}{Opt. Lett.} \textbf{\bibinfo{volume}{38}},
  \bibinfo{pages}{4397} (\bibinfo{year}{2013}).

\bibitem[{\citenamefont{Tamaki et~al.}(2014)\citenamefont{Tamaki, Curty, Kato,
  Lo, and Azuma}}]{TCK:pra14}
\bibinfo{author}{\bibfnamefont{K.}~\bibnamefont{Tamaki}},
  \bibinfo{author}{\bibfnamefont{M.}~\bibnamefont{Curty}},
  \bibinfo{author}{\bibfnamefont{G.}~\bibnamefont{Kato}},
  \bibinfo{author}{\bibfnamefont{H.-K.} \bibnamefont{Lo}}, \bibnamefont{and}
  \bibinfo{author}{\bibfnamefont{K.}~\bibnamefont{Azuma}},
  \bibinfo{journal}{Phys. Rev. A} \textbf{\bibinfo{volume}{90}},
  \bibinfo{pages}{052314} (\bibinfo{year}{2014}).

\bibitem[{\citenamefont{All\'eaume et~al.}(2014)\citenamefont{All\'eaume,
  Degiovanni, Mink, Chapuran, L\"utkenhaus, Peev, Chunnilall, Martín~Ayuso,
  Lucamarini, Ward et~al.}}]{ADM:GC14}
\bibinfo{author}{\bibfnamefont{R.}~\bibnamefont{All\'eaume}},
  \bibinfo{author}{\bibfnamefont{I.~P.} \bibnamefont{Degiovanni}},
  \bibinfo{author}{\bibfnamefont{A.}~\bibnamefont{Mink}},
  \bibinfo{author}{\bibfnamefont{T.~E.} \bibnamefont{Chapuran}},
  \bibinfo{author}{\bibfnamefont{N.}~\bibnamefont{L\"utkenhaus}},
  \bibinfo{author}{\bibfnamefont{M.}~\bibnamefont{Peev}},
  \bibinfo{author}{\bibfnamefont{C.~J.} \bibnamefont{Chunnilall}},
  \bibinfo{author}{\bibfnamefont{V.}~\bibnamefont{Martín~Ayuso}},
  \bibinfo{author}{\bibfnamefont{M.}~\bibnamefont{Lucamarini}},
  \bibinfo{author}{\bibfnamefont{M.}~\bibnamefont{Ward}}, \bibnamefont{et~al.},
  in \emph{\bibinfo{booktitle}{Proc. IEEE Globecom Workshops (GC Wkshps)}}
  (\bibinfo{year}{2014}), pp. \bibinfo{pages}{656--551}.

\bibitem[{\citenamefont{Stucki et~al.}(2005)\citenamefont{Stucki, Brunner,
  Gisin, Scarani, and Zbinden}}]{SBG:apl05}
\bibinfo{author}{\bibfnamefont{D.}~\bibnamefont{Stucki}},
  \bibinfo{author}{\bibfnamefont{N.}~\bibnamefont{Brunner}},
  \bibinfo{author}{\bibfnamefont{N.}~\bibnamefont{Gisin}},
  \bibinfo{author}{\bibfnamefont{V.}~\bibnamefont{Scarani}}, \bibnamefont{and}
  \bibinfo{author}{\bibfnamefont{H.}~\bibnamefont{Zbinden}},
  \bibinfo{journal}{Appl. Phys. Lett.} \textbf{\bibinfo{volume}{87}},
  \bibinfo{pages}{194108} (\bibinfo{year}{2005}).

\bibitem[{\citenamefont{Inoue et~al.}(2002)\citenamefont{Inoue, Waks, and
  Yamamoto}}]{IWY:prl02}
\bibinfo{author}{\bibfnamefont{K.}~\bibnamefont{Inoue}},
  \bibinfo{author}{\bibfnamefont{E.}~\bibnamefont{Waks}}, \bibnamefont{and}
  \bibinfo{author}{\bibfnamefont{Y.}~\bibnamefont{Yamamoto}},
  \bibinfo{journal}{Phys. Rev. Lett.} \textbf{\bibinfo{volume}{89}},
  \bibinfo{pages}{037902} (\bibinfo{year}{2002}).

\bibitem[{\citenamefont{Grosshans and Grangier}(2002)}]{GG:prl02}
\bibinfo{author}{\bibfnamefont{F.}~\bibnamefont{Grosshans}} \bibnamefont{and}
  \bibinfo{author}{\bibfnamefont{P.}~\bibnamefont{Grangier}},
  \bibinfo{journal}{Phys. Rev. Lett.} \textbf{\bibinfo{volume}{88}},
  \bibinfo{pages}{057902} (\bibinfo{year}{2002}).

\bibitem[{\citenamefont{Grosshans et~al.}(2003)\citenamefont{Grosshans, Assche,
  Wenger, Brouri, Cerf, and Grangier}}]{GVW:nat03}
\bibinfo{author}{\bibfnamefont{F.}~\bibnamefont{Grosshans}},
  \bibinfo{author}{\bibfnamefont{G.~V.} \bibnamefont{Assche}},
  \bibinfo{author}{\bibfnamefont{J.}~\bibnamefont{Wenger}},
  \bibinfo{author}{\bibfnamefont{R.}~\bibnamefont{Brouri}},
  \bibinfo{author}{\bibfnamefont{N.~J.} \bibnamefont{Cerf}}, \bibnamefont{and}
  \bibinfo{author}{\bibfnamefont{P.}~\bibnamefont{Grangier}},
  \bibinfo{journal}{Nature} \textbf{\bibinfo{volume}{421}},
  \bibinfo{pages}{238} (\bibinfo{year}{2003}).

\bibitem[{\citenamefont{Diamanti and Leverrier}(2015)}]{DL:entropy15}
\bibinfo{author}{\bibfnamefont{E.}~\bibnamefont{Diamanti}} \bibnamefont{and}
  \bibinfo{author}{\bibfnamefont{A.}~\bibnamefont{Leverrier}},
  \bibinfo{journal}{Entropy} \textbf{\bibinfo{volume}{17}},
  \bibinfo{pages}{6072} (\bibinfo{year}{2015}).

\bibitem[{\citenamefont{Ma et~al.}(2007)\citenamefont{Ma, Fung, and
  Lo}}]{MFL:pra07}
\bibinfo{author}{\bibfnamefont{X.}~\bibnamefont{Ma}},
  \bibinfo{author}{\bibfnamefont{C.-H.~F.} \bibnamefont{Fung}},
  \bibnamefont{and} \bibinfo{author}{\bibfnamefont{H.-K.} \bibnamefont{Lo}},
  \bibinfo{journal}{Phys. Rev. A} \textbf{\bibinfo{volume}{76}},
  \bibinfo{pages}{012307} (\bibinfo{year}{2007}).

\bibitem[{\citenamefont{Lucamarini et~al.}(2013)\citenamefont{Lucamarini,
  Patel, Dynes, Fr\"ohlich, Sharpe, Dixon, Yuan, Penty, and
  Shields}}]{LPD:opex13}
\bibinfo{author}{\bibfnamefont{M.}~\bibnamefont{Lucamarini}},
  \bibinfo{author}{\bibfnamefont{K.~A.} \bibnamefont{Patel}},
  \bibinfo{author}{\bibfnamefont{J.~F.} \bibnamefont{Dynes}},
  \bibinfo{author}{\bibfnamefont{B.}~\bibnamefont{Fr\"ohlich}},
  \bibinfo{author}{\bibfnamefont{A.~W.} \bibnamefont{Sharpe}},
  \bibinfo{author}{\bibfnamefont{A.~R.} \bibnamefont{Dixon}},
  \bibinfo{author}{\bibfnamefont{Z.~L.} \bibnamefont{Yuan}},
  \bibinfo{author}{\bibfnamefont{R.~V.} \bibnamefont{Penty}}, \bibnamefont{and}
  \bibinfo{author}{\bibfnamefont{A.~J.} \bibnamefont{Shields}},
  \bibinfo{journal}{Opt. Express} \textbf{\bibinfo{volume}{21}},
  \bibinfo{pages}{24550} (\bibinfo{year}{2013}).

\bibitem[{\citenamefont{Jouguet et~al.}(2013)\citenamefont{Jouguet,
  Kunz-Jacques, Leverrier, Grangier, and Diamanti}}]{JKL:natphoton13}
\bibinfo{author}{\bibfnamefont{P.}~\bibnamefont{Jouguet}},
  \bibinfo{author}{\bibfnamefont{S.}~\bibnamefont{Kunz-Jacques}},
  \bibinfo{author}{\bibfnamefont{A.}~\bibnamefont{Leverrier}},
  \bibinfo{author}{\bibfnamefont{P.}~\bibnamefont{Grangier}}, \bibnamefont{and}
  \bibinfo{author}{\bibfnamefont{E.}~\bibnamefont{Diamanti}},
  \bibinfo{journal}{Nature Photon.} \textbf{\bibinfo{volume}{7}},
  \bibinfo{pages}{378} (\bibinfo{year}{2013}).

\bibitem[{\citenamefont{Lam and Ralph}(2013)}]{LR:natphoton13}
\bibinfo{author}{\bibfnamefont{P.-K.} \bibnamefont{Lam}} \bibnamefont{and}
  \bibinfo{author}{\bibfnamefont{T.}~\bibnamefont{Ralph}},
  \bibinfo{journal}{Nature Photon.} \textbf{\bibinfo{volume}{7}},
  \bibinfo{pages}{350} (\bibinfo{year}{2013}).

\bibitem[{\citenamefont{Korzh et~al.}(2015)\citenamefont{Korzh, Lim, Houlmann,
  Gisin, Li, Nolan, Sanguinetti, Thew, and Zbinden}}]{KLH:natphoton15}
\bibinfo{author}{\bibfnamefont{B.}~\bibnamefont{Korzh}},
  \bibinfo{author}{\bibfnamefont{C.~C.~W.} \bibnamefont{Lim}},
  \bibinfo{author}{\bibfnamefont{R.}~\bibnamefont{Houlmann}},
  \bibinfo{author}{\bibfnamefont{N.}~\bibnamefont{Gisin}},
  \bibinfo{author}{\bibfnamefont{M.~J.} \bibnamefont{Li}},
  \bibinfo{author}{\bibfnamefont{D.}~\bibnamefont{Nolan}},
  \bibinfo{author}{\bibfnamefont{B.}~\bibnamefont{Sanguinetti}},
  \bibinfo{author}{\bibfnamefont{R.}~\bibnamefont{Thew}}, \bibnamefont{and}
  \bibinfo{author}{\bibfnamefont{H.}~\bibnamefont{Zbinden}},
  \bibinfo{journal}{Nature Photon.} \textbf{\bibinfo{volume}{9}},
  \bibinfo{pages}{163} (\bibinfo{year}{2015}).

\bibitem[{\citenamefont{Huang et~al.}(2015{\natexlab{a}})\citenamefont{Huang,
  Lin, Wang, Liu, Fang, Peng, Huang, and Zeng}}]{HLW:opex15}
\bibinfo{author}{\bibfnamefont{D.}~\bibnamefont{Huang}},
  \bibinfo{author}{\bibfnamefont{D.}~\bibnamefont{Lin}},
  \bibinfo{author}{\bibfnamefont{C.}~\bibnamefont{Wang}},
  \bibinfo{author}{\bibfnamefont{W.}~\bibnamefont{Liu}},
  \bibinfo{author}{\bibfnamefont{S.}~\bibnamefont{Fang}},
  \bibinfo{author}{\bibfnamefont{J.}~\bibnamefont{Peng}},
  \bibinfo{author}{\bibfnamefont{P.}~\bibnamefont{Huang}}, \bibnamefont{and}
  \bibinfo{author}{\bibfnamefont{G.}~\bibnamefont{Zeng}},
  \bibinfo{journal}{Opt. Express} \textbf{\bibinfo{volume}{23}},
  \bibinfo{pages}{17511} (\bibinfo{year}{2015}{\natexlab{a}}).

\bibitem[{\citenamefont{Lim et~al.}(2014{\natexlab{a}})\citenamefont{Lim,
  Curty, Walenta, Xu, and Zbinden}}]{LCW:pra14}
\bibinfo{author}{\bibfnamefont{C.~C.~W.} \bibnamefont{Lim}},
  \bibinfo{author}{\bibfnamefont{M.}~\bibnamefont{Curty}},
  \bibinfo{author}{\bibfnamefont{N.}~\bibnamefont{Walenta}},
  \bibinfo{author}{\bibfnamefont{F.}~\bibnamefont{Xu}}, \bibnamefont{and}
  \bibinfo{author}{\bibfnamefont{H.}~\bibnamefont{Zbinden}},
  \bibinfo{journal}{Phys. Rev. A} \textbf{\bibinfo{volume}{89}},
  \bibinfo{pages}{022307} (\bibinfo{year}{2014}{\natexlab{a}}).

\bibitem[{\citenamefont{Lucamarini
  et~al.}(2015{\natexlab{a}})\citenamefont{Lucamarini, Dynes, Fr\"ohlich, Yuan,
  and Shields}}]{LDF:JSTQE15}
\bibinfo{author}{\bibfnamefont{M.}~\bibnamefont{Lucamarini}},
  \bibinfo{author}{\bibfnamefont{J.}~\bibnamefont{Dynes}},
  \bibinfo{author}{\bibfnamefont{B.}~\bibnamefont{Fr\"ohlich}},
  \bibinfo{author}{\bibfnamefont{Z.}~\bibnamefont{Yuan}}, \bibnamefont{and}
  \bibinfo{author}{\bibfnamefont{A.}~\bibnamefont{Shields}},
  \bibinfo{journal}{IEEE J. Sel. Topics Quantum Electron}
  \textbf{\bibinfo{volume}{21}}, \bibinfo{pages}{6601408}
  (\bibinfo{year}{2015}{\natexlab{a}}).

\bibitem[{\citenamefont{Moroder et~al.}(2012)\citenamefont{Moroder, Curty, Lim,
  Thinh, Zbinden, and Gisin}}]{MCL:prl12}
\bibinfo{author}{\bibfnamefont{T.}~\bibnamefont{Moroder}},
  \bibinfo{author}{\bibfnamefont{M.}~\bibnamefont{Curty}},
  \bibinfo{author}{\bibfnamefont{C.~C.~W.} \bibnamefont{Lim}},
  \bibinfo{author}{\bibfnamefont{L.~P.} \bibnamefont{Thinh}},
  \bibinfo{author}{\bibfnamefont{H.}~\bibnamefont{Zbinden}}, \bibnamefont{and}
  \bibinfo{author}{\bibfnamefont{N.}~\bibnamefont{Gisin}},
  \bibinfo{journal}{Phys. Rev. Lett.} \textbf{\bibinfo{volume}{109}},
  \bibinfo{pages}{260501} (\bibinfo{year}{2012}).

\bibitem[{\citenamefont{Leverrier}(2015)}]{Lev:prl15}
\bibinfo{author}{\bibfnamefont{A.}~\bibnamefont{Leverrier}},
  \bibinfo{journal}{Phys. Rev. Lett.} \textbf{\bibinfo{volume}{114}},
  \bibinfo{pages}{070501} (\bibinfo{year}{2015}).

\bibitem[{\citenamefont{Takeoka et~al.}(2014)\citenamefont{Takeoka, Guha, and
  Wilde}}]{TGW:natcomm14}
\bibinfo{author}{\bibfnamefont{M.}~\bibnamefont{Takeoka}},
  \bibinfo{author}{\bibfnamefont{S.}~\bibnamefont{Guha}}, \bibnamefont{and}
  \bibinfo{author}{\bibfnamefont{M.~M.} \bibnamefont{Wilde}},
  \bibinfo{journal}{Nature Commun.} \textbf{\bibinfo{volume}{5}},
  \bibinfo{pages}{5235} (\bibinfo{year}{2014}).

\bibitem[{\citenamefont{Pirandola
  et~al.}(2015{\natexlab{a}})\citenamefont{Pirandola, Laurenza, Ottaviani, and
  Banchi}}]{PLO:arxiv15}
\bibinfo{author}{\bibfnamefont{S.}~\bibnamefont{Pirandola}},
  \bibinfo{author}{\bibfnamefont{R.}~\bibnamefont{Laurenza}},
  \bibinfo{author}{\bibfnamefont{C.}~\bibnamefont{Ottaviani}},
  \bibnamefont{and} \bibinfo{author}{\bibfnamefont{L.}~\bibnamefont{Banchi}},
  \bibinfo{journal}{Arxiv preprint arXiv:1510.08863 [quant-ph]}
  (\bibinfo{year}{2015}{\natexlab{a}}).

\bibitem[{\citenamefont{Winzer}(2015)}]{Winzer:OPN15}
\bibinfo{author}{\bibfnamefont{P.~J.} \bibnamefont{Winzer}},
  \bibinfo{journal}{Optics and Photonics News} \textbf{\bibinfo{volume}{26}},
  \bibinfo{pages}{28} (\bibinfo{year}{2015}).

\bibitem[{\citenamefont{Huang et~al.}(2014)\citenamefont{Huang, Tanaka, Ip,
  Huang, Qian, Zhang, Zhang, Ji, Diordjevic, Wang et~al.}}]{HTI:jlt14}
\bibinfo{author}{\bibfnamefont{M.~F.} \bibnamefont{Huang}},
  \bibinfo{author}{\bibfnamefont{A.}~\bibnamefont{Tanaka}},
  \bibinfo{author}{\bibfnamefont{E.}~\bibnamefont{Ip}},
  \bibinfo{author}{\bibfnamefont{Y.~K.} \bibnamefont{Huang}},
  \bibinfo{author}{\bibfnamefont{D.}~\bibnamefont{Qian}},
  \bibinfo{author}{\bibfnamefont{Y.}~\bibnamefont{Zhang}},
  \bibinfo{author}{\bibfnamefont{S.}~\bibnamefont{Zhang}},
  \bibinfo{author}{\bibfnamefont{P.~N.} \bibnamefont{Ji}},
  \bibinfo{author}{\bibfnamefont{I.~V.} \bibnamefont{Diordjevic}},
  \bibinfo{author}{\bibfnamefont{T.}~\bibnamefont{Wang}}, \bibnamefont{et~al.},
  \bibinfo{journal}{J. Lightwave Technology} \textbf{\bibinfo{volume}{32}},
  \bibinfo{pages}{776} (\bibinfo{year}{2014}).

\bibitem[{\citenamefont{Pernice et~al.}(2012)\citenamefont{Pernice, Schuck,
  Minaeva, Li, Goltsman, Sergienko, and Tang}}]{PSM:natcomm12}
\bibinfo{author}{\bibfnamefont{W.~H.~P.} \bibnamefont{Pernice}},
  \bibinfo{author}{\bibfnamefont{C.}~\bibnamefont{Schuck}},
  \bibinfo{author}{\bibfnamefont{O.}~\bibnamefont{Minaeva}},
  \bibinfo{author}{\bibfnamefont{M.}~\bibnamefont{Li}},
  \bibinfo{author}{\bibfnamefont{G.~N.} \bibnamefont{Goltsman}},
  \bibinfo{author}{\bibfnamefont{A.~V.} \bibnamefont{Sergienko}},
  \bibnamefont{and} \bibinfo{author}{\bibfnamefont{H.~X.} \bibnamefont{Tang}},
  \bibinfo{journal}{Nature Commun.} \textbf{\bibinfo{volume}{3}},
  \bibinfo{pages}{1325} (\bibinfo{year}{2012}).

\bibitem[{\citenamefont{Marsili et~al.}(2013)\citenamefont{Marsili, Verma,
  Stern, Harrington, Lita, Gerrits, Vayshenker, Baek, Shaw, Mirin
  et~al.}}]{MVS:natphoton13}
\bibinfo{author}{\bibfnamefont{F.}~\bibnamefont{Marsili}},
  \bibinfo{author}{\bibfnamefont{V.~B.} \bibnamefont{Verma}},
  \bibinfo{author}{\bibfnamefont{J.~A.} \bibnamefont{Stern}},
  \bibinfo{author}{\bibfnamefont{S.}~\bibnamefont{Harrington}},
  \bibinfo{author}{\bibfnamefont{A.~E.} \bibnamefont{Lita}},
  \bibinfo{author}{\bibfnamefont{T.}~\bibnamefont{Gerrits}},
  \bibinfo{author}{\bibfnamefont{I.}~\bibnamefont{Vayshenker}},
  \bibinfo{author}{\bibfnamefont{B.}~\bibnamefont{Baek}},
  \bibinfo{author}{\bibfnamefont{M.~D.} \bibnamefont{Shaw}},
  \bibinfo{author}{\bibfnamefont{R.~P.} \bibnamefont{Mirin}},
  \bibnamefont{et~al.}, \bibinfo{journal}{Nature Photon.}
  \textbf{\bibinfo{volume}{7}}, \bibinfo{pages}{210} (\bibinfo{year}{2013}).

\bibitem[{\citenamefont{Comandar et~al.}(2015)\citenamefont{Comandar,
  Fr\"ohlich, Dynes, Lucamarini, Sharpe, Yuan, Penty, and Shields}}]{CFD:jap15}
\bibinfo{author}{\bibfnamefont{L.~C.} \bibnamefont{Comandar}},
  \bibinfo{author}{\bibfnamefont{B.}~\bibnamefont{Fr\"ohlich}},
  \bibinfo{author}{\bibfnamefont{J.~F.} \bibnamefont{Dynes}},
  \bibinfo{author}{\bibfnamefont{M.}~\bibnamefont{Lucamarini}},
  \bibinfo{author}{\bibfnamefont{A.~W.} \bibnamefont{Sharpe}},
  \bibinfo{author}{\bibfnamefont{Z.~L.} \bibnamefont{Yuan}},
  \bibinfo{author}{\bibfnamefont{R.~V.} \bibnamefont{Penty}}, \bibnamefont{and}
  \bibinfo{author}{\bibfnamefont{A.~J.} \bibnamefont{Shields}},
  \bibinfo{journal}{J. Appl. Phys.} \textbf{\bibinfo{volume}{117}},
  \bibinfo{pages}{083109} (\bibinfo{year}{2015}).

\bibitem[{SNS()}]{SNSPD}
\bibinfo{note}{\lowercase{www.scontel.ru}, www.singlequantum.com,
  www.idquantique.com, www.photonspot.com}.

\bibitem[{\citenamefont{Bahrani et~al.}(2015)\citenamefont{Bahrani, Razavi, and
  Salehi}}]{BRS:jlt15}
\bibinfo{author}{\bibfnamefont{S.}~\bibnamefont{Bahrani}},
  \bibinfo{author}{\bibfnamefont{M.}~\bibnamefont{Razavi}}, \bibnamefont{and}
  \bibinfo{author}{\bibfnamefont{J.~A.} \bibnamefont{Salehi}},
  \bibinfo{journal}{J. Lightwave Technology} \textbf{\bibinfo{volume}{33}},
  \bibinfo{pages}{4687} (\bibinfo{year}{2015}).

\bibitem[{\citenamefont{Dynes et~al.}(2016)\citenamefont{Dynes, Kindness, Tam,
  Plews, Sharpe, Lucamarini, Fr\"ohlich, Yuan, Penty, and
  Shields}}]{DKT:opex16}
\bibinfo{author}{\bibfnamefont{J.~F.} \bibnamefont{Dynes}},
  \bibinfo{author}{\bibfnamefont{S.~J.} \bibnamefont{Kindness}},
  \bibinfo{author}{\bibfnamefont{S.~W.-B.} \bibnamefont{Tam}},
  \bibinfo{author}{\bibfnamefont{A.}~\bibnamefont{Plews}},
  \bibinfo{author}{\bibfnamefont{A.~W.} \bibnamefont{Sharpe}},
  \bibinfo{author}{\bibfnamefont{M.}~\bibnamefont{Lucamarini}},
  \bibinfo{author}{\bibfnamefont{B.}~\bibnamefont{Fr\"ohlich}},
  \bibinfo{author}{\bibfnamefont{Z.~L.} \bibnamefont{Yuan}},
  \bibinfo{author}{\bibfnamefont{R.~V.} \bibnamefont{Penty}}, \bibnamefont{and}
  \bibinfo{author}{\bibfnamefont{A.~J.} \bibnamefont{Shields}},
  \bibinfo{journal}{Opt. Express} \textbf{\bibinfo{volume}{24}},
  \bibinfo{pages}{8081} (\bibinfo{year}{2016}).

\bibitem[{\citenamefont{Qi et~al.}(2015)\citenamefont{Qi, Lougovski, Pooser,
  Grice, and Bobrek}}]{QLP:prx15}
\bibinfo{author}{\bibfnamefont{B.}~\bibnamefont{Qi}},
  \bibinfo{author}{\bibfnamefont{P.}~\bibnamefont{Lougovski}},
  \bibinfo{author}{\bibfnamefont{R.}~\bibnamefont{Pooser}},
  \bibinfo{author}{\bibfnamefont{W.}~\bibnamefont{Grice}}, \bibnamefont{and}
  \bibinfo{author}{\bibfnamefont{M.}~\bibnamefont{Bobrek}},
  \bibinfo{journal}{Phys. Rev. X} \textbf{\bibinfo{volume}{5}},
  \bibinfo{pages}{041009} (\bibinfo{year}{2015}).

\bibitem[{\citenamefont{Soh et~al.}(2015)\citenamefont{Soh, Brif, Coles,
  L\"utkenhaus, Camacho, Urayama, and Sarovar}}]{SBC:prx15}
\bibinfo{author}{\bibfnamefont{D.~B.~S.} \bibnamefont{Soh}},
  \bibinfo{author}{\bibfnamefont{C.}~\bibnamefont{Brif}},
  \bibinfo{author}{\bibfnamefont{P.~J.} \bibnamefont{Coles}},
  \bibinfo{author}{\bibfnamefont{N.}~\bibnamefont{L\"utkenhaus}},
  \bibinfo{author}{\bibfnamefont{R.~M.} \bibnamefont{Camacho}},
  \bibinfo{author}{\bibfnamefont{J.}~\bibnamefont{Urayama}}, \bibnamefont{and}
  \bibinfo{author}{\bibfnamefont{M.}~\bibnamefont{Sarovar}},
  \bibinfo{journal}{Phys. Rev. X} \textbf{\bibinfo{volume}{5}},
  \bibinfo{pages}{041010} (\bibinfo{year}{2015}).

\bibitem[{\citenamefont{Huang et~al.}(2015{\natexlab{b}})\citenamefont{Huang,
  Huang, Lin, Wang, and Zeng}}]{HHL:ol15}
\bibinfo{author}{\bibfnamefont{D.}~\bibnamefont{Huang}},
  \bibinfo{author}{\bibfnamefont{P.}~\bibnamefont{Huang}},
  \bibinfo{author}{\bibfnamefont{D.}~\bibnamefont{Lin}},
  \bibinfo{author}{\bibfnamefont{C.}~\bibnamefont{Wang}}, \bibnamefont{and}
  \bibinfo{author}{\bibfnamefont{G.}~\bibnamefont{Zeng}},
  \bibinfo{journal}{Opt. Lett.} \textbf{\bibinfo{volume}{40}},
  \bibinfo{pages}{3695} (\bibinfo{year}{2015}{\natexlab{b}}).

\bibitem[{\citenamefont{Comandar et~al.}(2014)\citenamefont{Comandar,
  Fr\"ohlich, Lucamarini, Patel, Sharpe, Dynes, Yuan, Penty, and
  Shields}}]{CFL:apl14}
\bibinfo{author}{\bibfnamefont{L.~C.} \bibnamefont{Comandar}},
  \bibinfo{author}{\bibfnamefont{B.}~\bibnamefont{Fr\"ohlich}},
  \bibinfo{author}{\bibfnamefont{M.}~\bibnamefont{Lucamarini}},
  \bibinfo{author}{\bibfnamefont{K.~A.} \bibnamefont{Patel}},
  \bibinfo{author}{\bibfnamefont{A.~W.} \bibnamefont{Sharpe}},
  \bibinfo{author}{\bibfnamefont{J.~F.} \bibnamefont{Dynes}},
  \bibinfo{author}{\bibfnamefont{Z.~L.} \bibnamefont{Yuan}},
  \bibinfo{author}{\bibfnamefont{R.~V.} \bibnamefont{Penty}}, \bibnamefont{and}
  \bibinfo{author}{\bibfnamefont{A.~J.} \bibnamefont{Shields}},
  \bibinfo{journal}{Appl. Phys. Lett.} \textbf{\bibinfo{volume}{104}},
  \bibinfo{pages}{021101} (\bibinfo{year}{2014}).

\bibitem[{\citenamefont{Fr\"ohlich et~al.}(2015)\citenamefont{Fr\"ohlich,
  Dynes, Lucamarini, Sharpe, Tam, Yuan, and Shields}}]{FDL:scirep15}
\bibinfo{author}{\bibfnamefont{B.}~\bibnamefont{Fr\"ohlich}},
  \bibinfo{author}{\bibfnamefont{J.~F.} \bibnamefont{Dynes}},
  \bibinfo{author}{\bibfnamefont{M.}~\bibnamefont{Lucamarini}},
  \bibinfo{author}{\bibfnamefont{A.~W.} \bibnamefont{Sharpe}},
  \bibinfo{author}{\bibfnamefont{S.~W.~B.} \bibnamefont{Tam}},
  \bibinfo{author}{\bibfnamefont{Z.~L.} \bibnamefont{Yuan}}, \bibnamefont{and}
  \bibinfo{author}{\bibfnamefont{A.~J.} \bibnamefont{Shields}},
  \bibinfo{journal}{Sci. Rep.} \textbf{\bibinfo{volume}{5}},
  \bibinfo{pages}{18121} (\bibinfo{year}{2015}).

\bibitem[{\citenamefont{Shibaba et~al.}(2014)\citenamefont{Shibaba, Honjo, and
  Shimizu}}]{SHS:ol14}
\bibinfo{author}{\bibfnamefont{H.}~\bibnamefont{Shibaba}},
  \bibinfo{author}{\bibfnamefont{T.}~\bibnamefont{Honjo}}, \bibnamefont{and}
  \bibinfo{author}{\bibfnamefont{K.}~\bibnamefont{Shimizu}},
  \bibinfo{journal}{Opt. Lett.} \textbf{\bibinfo{volume}{39}},
  \bibinfo{pages}{5078} (\bibinfo{year}{2014}).

\bibitem[{\citenamefont{Jouguet et~al.}(2014)\citenamefont{Jouguet, Elkouss,
  and Kunz-Jacques}}]{JEK:pra14}
\bibinfo{author}{\bibfnamefont{P.}~\bibnamefont{Jouguet}},
  \bibinfo{author}{\bibfnamefont{D.}~\bibnamefont{Elkouss}}, \bibnamefont{and}
  \bibinfo{author}{\bibfnamefont{S.}~\bibnamefont{Kunz-Jacques}},
  \bibinfo{journal}{Phys. Rev. A} \textbf{\bibinfo{volume}{90}},
  \bibinfo{pages}{042329} (\bibinfo{year}{2014}).

\bibitem[{\citenamefont{Patel et~al.}(2014)\citenamefont{Patel, Dynes,
  Lucamarini, Choi, Sharpe, Yuan, Penty, and Shields}}]{PDL:apl14}
\bibinfo{author}{\bibfnamefont{K.~A.} \bibnamefont{Patel}},
  \bibinfo{author}{\bibfnamefont{J.~F.} \bibnamefont{Dynes}},
  \bibinfo{author}{\bibfnamefont{M.}~\bibnamefont{Lucamarini}},
  \bibinfo{author}{\bibfnamefont{I.}~\bibnamefont{Choi}},
  \bibinfo{author}{\bibfnamefont{A.~W.} \bibnamefont{Sharpe}},
  \bibinfo{author}{\bibfnamefont{Z.~L.} \bibnamefont{Yuan}},
  \bibinfo{author}{\bibfnamefont{R.~V.} \bibnamefont{Penty}}, \bibnamefont{and}
  \bibinfo{author}{\bibfnamefont{A.~J.} \bibnamefont{Shields}},
  \bibinfo{journal}{Appl. Phys. Lett.} \textbf{\bibinfo{volume}{104}},
  \bibinfo{pages}{051123} (\bibinfo{year}{2014}).

\bibitem[{\citenamefont{Choi et~al.}(2014)\citenamefont{Choi, Zhou, Dynes,
  Yuan, Klar, Sharpe, Plews, Lucamarini, Radig, Neubert et~al.}}]{CZD:opex14}
\bibinfo{author}{\bibfnamefont{I.}~\bibnamefont{Choi}},
  \bibinfo{author}{\bibfnamefont{Y.~R.} \bibnamefont{Zhou}},
  \bibinfo{author}{\bibfnamefont{J.~F.} \bibnamefont{Dynes}},
  \bibinfo{author}{\bibfnamefont{Z.}~\bibnamefont{Yuan}},
  \bibinfo{author}{\bibfnamefont{A.}~\bibnamefont{Klar}},
  \bibinfo{author}{\bibfnamefont{A.}~\bibnamefont{Sharpe}},
  \bibinfo{author}{\bibfnamefont{A.}~\bibnamefont{Plews}},
  \bibinfo{author}{\bibfnamefont{M.}~\bibnamefont{Lucamarini}},
  \bibinfo{author}{\bibfnamefont{C.}~\bibnamefont{Radig}},
  \bibinfo{author}{\bibfnamefont{J.}~\bibnamefont{Neubert}},
  \bibnamefont{et~al.}, \bibinfo{journal}{Opt. Express}
  \textbf{\bibinfo{volume}{22}}, \bibinfo{pages}{23121} (\bibinfo{year}{2014}).

\bibitem[{\citenamefont{Qi et~al.}(2010)\citenamefont{Qi, Zhu, Qian, and
  Lo}}]{QZQ:njp10}
\bibinfo{author}{\bibfnamefont{B.}~\bibnamefont{Qi}},
  \bibinfo{author}{\bibfnamefont{W.}~\bibnamefont{Zhu}},
  \bibinfo{author}{\bibfnamefont{L.}~\bibnamefont{Qian}}, \bibnamefont{and}
  \bibinfo{author}{\bibfnamefont{H.-K.} \bibnamefont{Lo}},
  \bibinfo{journal}{New J. Phys.} \textbf{\bibinfo{volume}{12}},
  \bibinfo{pages}{103042} (\bibinfo{year}{2010}).

\bibitem[{\citenamefont{Kumar et~al.}(2015)\citenamefont{Kumar, Qin, and
  All\'eaume}}]{KQA:njp15}
\bibinfo{author}{\bibfnamefont{R.}~\bibnamefont{Kumar}},
  \bibinfo{author}{\bibfnamefont{H.}~\bibnamefont{Qin}}, \bibnamefont{and}
  \bibinfo{author}{\bibfnamefont{R.}~\bibnamefont{All\'eaume}},
  \bibinfo{journal}{New J. Phys.} \textbf{\bibinfo{volume}{17}},
  \bibinfo{pages}{043027} (\bibinfo{year}{2015}).

\bibitem[{\citenamefont{Fr\"ohlich et~al.}(2013)\citenamefont{Fr\"ohlich,
  Dynes, Lucamarini, Sharpe, Yuan, and Shields}}]{FDL:nature13}
\bibinfo{author}{\bibfnamefont{B.}~\bibnamefont{Fr\"ohlich}},
  \bibinfo{author}{\bibfnamefont{J.~F.} \bibnamefont{Dynes}},
  \bibinfo{author}{\bibfnamefont{M.}~\bibnamefont{Lucamarini}},
  \bibinfo{author}{\bibfnamefont{A.~W.} \bibnamefont{Sharpe}},
  \bibinfo{author}{\bibfnamefont{Z.~L.} \bibnamefont{Yuan}}, \bibnamefont{and}
  \bibinfo{author}{\bibfnamefont{A.~J.} \bibnamefont{Shields}},
  \bibinfo{journal}{Nature} \textbf{\bibinfo{volume}{501}}, \bibinfo{pages}{69}
  (\bibinfo{year}{2013}).

\bibitem[{\citenamefont{Hughes et~al.}(2013)\citenamefont{Hughes, Nordholt,
  McCabe, Newell, Peterson, and Somma}}]{HNM:arxiv13}
\bibinfo{author}{\bibfnamefont{R.~J.} \bibnamefont{Hughes}},
  \bibinfo{author}{\bibfnamefont{J.~E.} \bibnamefont{Nordholt}},
  \bibinfo{author}{\bibfnamefont{K.~P.} \bibnamefont{McCabe}},
  \bibinfo{author}{\bibfnamefont{R.~T.} \bibnamefont{Newell}},
  \bibinfo{author}{\bibfnamefont{C.~G.} \bibnamefont{Peterson}},
  \bibnamefont{and} \bibinfo{author}{\bibfnamefont{R.~D.} \bibnamefont{Somma}},
  \bibinfo{journal}{Arxiv preprint arXiv:1305.0305 [quant-ph]}
  (\bibinfo{year}{2013}).

\bibitem[{\citenamefont{Lim et~al.}(2014{\natexlab{b}})\citenamefont{Lim, Song,
  Fang, Li, Tu, Duan, Chen, Tern, and Liow}}]{LSF:JSTQE14}
\bibinfo{author}{\bibfnamefont{A.~E.-J.} \bibnamefont{Lim}},
  \bibinfo{author}{\bibfnamefont{J.}~\bibnamefont{Song}},
  \bibinfo{author}{\bibfnamefont{Q.}~\bibnamefont{Fang}},
  \bibinfo{author}{\bibfnamefont{C.}~\bibnamefont{Li}},
  \bibinfo{author}{\bibfnamefont{X.}~\bibnamefont{Tu}},
  \bibinfo{author}{\bibfnamefont{N.}~\bibnamefont{Duan}},
  \bibinfo{author}{\bibfnamefont{K.~K.} \bibnamefont{Chen}},
  \bibinfo{author}{\bibfnamefont{R.~P.-C.} \bibnamefont{Tern}},
  \bibnamefont{and} \bibinfo{author}{\bibfnamefont{T.-Y.} \bibnamefont{Liow}},
  \bibinfo{journal}{IEEE J. Sel. Topics Quantum Electron.}
  \textbf{\bibinfo{volume}{20}}, \bibinfo{pages}{405}
  (\bibinfo{year}{2014}{\natexlab{b}}).

\bibitem[{\citenamefont{Smit et~al.}(2014)\citenamefont{Smit, Leijtens,
  Ambrosius, Bente, van~der Tol, Smalbrugge, de~Vries, Geluk, Bolk, van
  Veldhoven et~al.}}]{SLA:SST14}
\bibinfo{author}{\bibfnamefont{M.}~\bibnamefont{Smit}},
  \bibinfo{author}{\bibfnamefont{X.}~\bibnamefont{Leijtens}},
  \bibinfo{author}{\bibfnamefont{H.}~\bibnamefont{Ambrosius}},
  \bibinfo{author}{\bibfnamefont{E.}~\bibnamefont{Bente}},
  \bibinfo{author}{\bibfnamefont{J.}~\bibnamefont{van~der Tol}},
  \bibinfo{author}{\bibfnamefont{B.}~\bibnamefont{Smalbrugge}},
  \bibinfo{author}{\bibfnamefont{T.}~\bibnamefont{de~Vries}},
  \bibinfo{author}{\bibfnamefont{E.-J.} \bibnamefont{Geluk}},
  \bibinfo{author}{\bibfnamefont{J.}~\bibnamefont{Bolk}},
  \bibinfo{author}{\bibfnamefont{R.}~\bibnamefont{van Veldhoven}},
  \bibnamefont{et~al.}, \bibinfo{journal}{Semicond. Sci. Technol.}
  \textbf{\bibinfo{volume}{29}}, \bibinfo{pages}{083001}
  (\bibinfo{year}{2014}).

\bibitem[{\citenamefont{Zhang et~al.}(2014{\natexlab{a}})\citenamefont{Zhang,
  Aungskunsiri, Mart\'in-L\'opez, Wabnig, Lobino, Nock, Munns, Bonneau, Jiang,
  Li et~al.}}]{ZAM:prl14}
\bibinfo{author}{\bibfnamefont{P.}~\bibnamefont{Zhang}},
  \bibinfo{author}{\bibfnamefont{K.}~\bibnamefont{Aungskunsiri}},
  \bibinfo{author}{\bibfnamefont{E.}~\bibnamefont{Mart\'in-L\'opez}},
  \bibinfo{author}{\bibfnamefont{J.}~\bibnamefont{Wabnig}},
  \bibinfo{author}{\bibfnamefont{M.}~\bibnamefont{Lobino}},
  \bibinfo{author}{\bibfnamefont{R.~W.} \bibnamefont{Nock}},
  \bibinfo{author}{\bibfnamefont{J.}~\bibnamefont{Munns}},
  \bibinfo{author}{\bibfnamefont{D.}~\bibnamefont{Bonneau}},
  \bibinfo{author}{\bibfnamefont{P.}~\bibnamefont{Jiang}},
  \bibinfo{author}{\bibfnamefont{H.~W.} \bibnamefont{Li}},
  \bibnamefont{et~al.}, \bibinfo{journal}{Phys. Rev. Lett.}
  \textbf{\bibinfo{volume}{112}}, \bibinfo{pages}{130501}
  (\bibinfo{year}{2014}{\natexlab{a}}).

\bibitem[{\citenamefont{Vest et~al.}(2014)\citenamefont{Vest, Rau, Fuchs,
  Corrielli, Weier, Nauerth, Crespi, Osselame, and Weinfurter}}]{VRF:jstqe14}
\bibinfo{author}{\bibfnamefont{G.}~\bibnamefont{Vest}},
  \bibinfo{author}{\bibfnamefont{M.}~\bibnamefont{Rau}},
  \bibinfo{author}{\bibfnamefont{L.}~\bibnamefont{Fuchs}},
  \bibinfo{author}{\bibfnamefont{G.}~\bibnamefont{Corrielli}},
  \bibinfo{author}{\bibfnamefont{H.}~\bibnamefont{Weier}},
  \bibinfo{author}{\bibfnamefont{S.}~\bibnamefont{Nauerth}},
  \bibinfo{author}{\bibfnamefont{A.}~\bibnamefont{Crespi}},
  \bibinfo{author}{\bibfnamefont{R.}~\bibnamefont{Osselame}}, \bibnamefont{and}
  \bibinfo{author}{\bibfnamefont{H.}~\bibnamefont{Weinfurter}},
  \bibinfo{journal}{IEEE J. Sel. Topics Quantum Electron.}
  \textbf{\bibinfo{volume}{21}}, \bibinfo{pages}{6600607}
  (\bibinfo{year}{2014}).

\bibitem[{\citenamefont{Sibson et~al.}(2015)\citenamefont{Sibson, Erven,
  Godfrey, Miki, Yamashita, Fujiwara, Sasaki, Terai, Tanner, Natarajan
  et~al.}}]{SEG:arxiv15}
\bibinfo{author}{\bibfnamefont{P.}~\bibnamefont{Sibson}},
  \bibinfo{author}{\bibfnamefont{C.}~\bibnamefont{Erven}},
  \bibinfo{author}{\bibfnamefont{M.}~\bibnamefont{Godfrey}},
  \bibinfo{author}{\bibfnamefont{S.}~\bibnamefont{Miki}},
  \bibinfo{author}{\bibfnamefont{T.}~\bibnamefont{Yamashita}},
  \bibinfo{author}{\bibfnamefont{M.}~\bibnamefont{Fujiwara}},
  \bibinfo{author}{\bibfnamefont{M.}~\bibnamefont{Sasaki}},
  \bibinfo{author}{\bibfnamefont{H.}~\bibnamefont{Terai}},
  \bibinfo{author}{\bibfnamefont{M.}~\bibnamefont{Tanner}},
  \bibinfo{author}{\bibfnamefont{C.}~\bibnamefont{Natarajan}},
  \bibnamefont{et~al.}, \bibinfo{journal}{Arxiv preprint arXiv:1509.00768
  [quant-ph]}  (\bibinfo{year}{2015}).

\bibitem[{\citenamefont{Takesue et~al.}(2005)\citenamefont{Takesue, Diamanti,
  Honjo, Langrock, Fejer, Inoue, and Yamamoto}}]{TDH:njp05}
\bibinfo{author}{\bibfnamefont{H.}~\bibnamefont{Takesue}},
  \bibinfo{author}{\bibfnamefont{E.}~\bibnamefont{Diamanti}},
  \bibinfo{author}{\bibfnamefont{T.}~\bibnamefont{Honjo}},
  \bibinfo{author}{\bibfnamefont{C.}~\bibnamefont{Langrock}},
  \bibinfo{author}{\bibfnamefont{M.~M.} \bibnamefont{Fejer}},
  \bibinfo{author}{\bibfnamefont{K.}~\bibnamefont{Inoue}}, \bibnamefont{and}
  \bibinfo{author}{\bibfnamefont{Y.}~\bibnamefont{Yamamoto}},
  \bibinfo{journal}{New J. Phys.} \textbf{\bibinfo{volume}{7}},
  \bibinfo{pages}{232} (\bibinfo{year}{2005}).

\bibitem[{\citenamefont{Nambu et~al.}(2008)\citenamefont{Nambu, Yoshino, and
  Tomita}}]{NYT:jmo08}
\bibinfo{author}{\bibfnamefont{Y.}~\bibnamefont{Nambu}},
  \bibinfo{author}{\bibfnamefont{K.}~\bibnamefont{Yoshino}}, \bibnamefont{and}
  \bibinfo{author}{\bibfnamefont{A.}~\bibnamefont{Tomita}},
  \bibinfo{journal}{J. Mod. Opt.} \textbf{\bibinfo{volume}{55}},
  \bibinfo{pages}{1953} (\bibinfo{year}{2008}).

\bibitem[{\citenamefont{Ziebell et~al.}(June 2015)\citenamefont{Ziebell,
  Persechino, Harris, Galland, Marris-Morini, Vivien, Diamanti, and
  Grangier}}]{ZPH:cleoeurope15}
\bibinfo{author}{\bibfnamefont{M.}~\bibnamefont{Ziebell}},
  \bibinfo{author}{\bibfnamefont{M.}~\bibnamefont{Persechino}},
  \bibinfo{author}{\bibfnamefont{N.}~\bibnamefont{Harris}},
  \bibinfo{author}{\bibfnamefont{C.}~\bibnamefont{Galland}},
  \bibinfo{author}{\bibfnamefont{D.}~\bibnamefont{Marris-Morini}},
  \bibinfo{author}{\bibfnamefont{L.}~\bibnamefont{Vivien}},
  \bibinfo{author}{\bibfnamefont{E.}~\bibnamefont{Diamanti}}, \bibnamefont{and}
  \bibinfo{author}{\bibfnamefont{P.}~\bibnamefont{Grangier}}, in
  \emph{\bibinfo{booktitle}{CLEO/Europe - EQEC, Munich, Germany}}
  (\bibinfo{year}{June 2015}).

\bibitem[{\citenamefont{Bechmann-Pasquinucci and Tittel}(2000)}]{BP:pra00}
\bibinfo{author}{\bibfnamefont{H.}~\bibnamefont{Bechmann-Pasquinucci}}
  \bibnamefont{and} \bibinfo{author}{\bibfnamefont{W.}~\bibnamefont{Tittel}},
  \bibinfo{journal}{Phys. Rev. A} \textbf{\bibinfo{volume}{61}},
  \bibinfo{pages}{062308} (\bibinfo{year}{2000}).

\bibitem[{\citenamefont{Bourennane et~al.}(2001)\citenamefont{Bourennane,
  Karlsson, and Bj\"{o}rk}}]{BKB:pra01}
\bibinfo{author}{\bibfnamefont{M.}~\bibnamefont{Bourennane}},
  \bibinfo{author}{\bibfnamefont{A.}~\bibnamefont{Karlsson}}, \bibnamefont{and}
  \bibinfo{author}{\bibfnamefont{G.}~\bibnamefont{Bj\"{o}rk}},
  \bibinfo{journal}{Phys. Rev. A} \textbf{\bibinfo{volume}{64}},
  \bibinfo{pages}{012306} (\bibinfo{year}{2001}).

\bibitem[{\citenamefont{Cerf et~al.}(2002)\citenamefont{Cerf, Bourennane,
  Karlsson, and Gisin}}]{CBK:prl02}
\bibinfo{author}{\bibfnamefont{N.~J.} \bibnamefont{Cerf}},
  \bibinfo{author}{\bibfnamefont{M.}~\bibnamefont{Bourennane}},
  \bibinfo{author}{\bibfnamefont{A.}~\bibnamefont{Karlsson}}, \bibnamefont{and}
  \bibinfo{author}{\bibfnamefont{N.}~\bibnamefont{Gisin}},
  \bibinfo{journal}{Phys. Rev. Lett.} \textbf{\bibinfo{volume}{88}},
  \bibinfo{pages}{127902} (\bibinfo{year}{2002}).

\bibitem[{\citenamefont{Zhang et~al.}(2008)\citenamefont{Zhang, Silberhorn, and
  Walmsley}}]{ZSW:prl08}
\bibinfo{author}{\bibfnamefont{L.}~\bibnamefont{Zhang}},
  \bibinfo{author}{\bibfnamefont{C.}~\bibnamefont{Silberhorn}},
  \bibnamefont{and} \bibinfo{author}{\bibfnamefont{I.~A.}
  \bibnamefont{Walmsley}}, \bibinfo{journal}{Phys. Rev. Lett.}
  \textbf{\bibinfo{volume}{100}}, \bibinfo{pages}{110504}
  (\bibinfo{year}{2008}).

\bibitem[{\citenamefont{Zhang et~al.}(2014{\natexlab{b}})\citenamefont{Zhang,
  Mower, Englund, Wong, and Shapiro}}]{ZME:prl14}
\bibinfo{author}{\bibfnamefont{Z.}~\bibnamefont{Zhang}},
  \bibinfo{author}{\bibfnamefont{J.}~\bibnamefont{Mower}},
  \bibinfo{author}{\bibfnamefont{D.}~\bibnamefont{Englund}},
  \bibinfo{author}{\bibfnamefont{F.~N.~C.} \bibnamefont{Wong}},
  \bibnamefont{and} \bibinfo{author}{\bibfnamefont{J.~H.}
  \bibnamefont{Shapiro}}, \bibinfo{journal}{Phys. Rev. Lett.}
  \textbf{\bibinfo{volume}{112}}, \bibinfo{pages}{120506}
  (\bibinfo{year}{2014}{\natexlab{b}}).

\bibitem[{\citenamefont{Zhong et~al.}(2015)\citenamefont{Zhong, Zhou, Horansky,
  Lee, Verma, Lita, Restelli, Bienfang, Mirin, Gerrits et~al.}}]{ZZH:njp15}
\bibinfo{author}{\bibfnamefont{T.}~\bibnamefont{Zhong}},
  \bibinfo{author}{\bibfnamefont{H.}~\bibnamefont{Zhou}},
  \bibinfo{author}{\bibfnamefont{R.~D.} \bibnamefont{Horansky}},
  \bibinfo{author}{\bibfnamefont{C.}~\bibnamefont{Lee}},
  \bibinfo{author}{\bibfnamefont{V.~B.} \bibnamefont{Verma}},
  \bibinfo{author}{\bibfnamefont{A.~E.} \bibnamefont{Lita}},
  \bibinfo{author}{\bibfnamefont{A.}~\bibnamefont{Restelli}},
  \bibinfo{author}{\bibfnamefont{J.~C.} \bibnamefont{Bienfang}},
  \bibinfo{author}{\bibfnamefont{R.~P.} \bibnamefont{Mirin}},
  \bibinfo{author}{\bibfnamefont{T.}~\bibnamefont{Gerrits}},
  \bibnamefont{et~al.}, \bibinfo{journal}{New J. Phys.}
  \textbf{\bibinfo{volume}{17}}, \bibinfo{pages}{022002}
  (\bibinfo{year}{2015}).

\bibitem[{\citenamefont{Mirhosseini et~al.}(2015)\citenamefont{Mirhosseini,
  Maga\~{n}a Loaiza, O'Sullivan, Rodenburg, Malik, Lavery, Padgett, Gauthier,
  and Boyd}}]{MMO:njp15}
\bibinfo{author}{\bibfnamefont{M.}~\bibnamefont{Mirhosseini}},
  \bibinfo{author}{\bibfnamefont{O.~S.} \bibnamefont{Maga\~{n}a Loaiza}},
  \bibinfo{author}{\bibfnamefont{M.~N.} \bibnamefont{O'Sullivan}},
  \bibinfo{author}{\bibfnamefont{B.}~\bibnamefont{Rodenburg}},
  \bibinfo{author}{\bibfnamefont{M.}~\bibnamefont{Malik}},
  \bibinfo{author}{\bibfnamefont{M.~P.~J.} \bibnamefont{Lavery}},
  \bibinfo{author}{\bibfnamefont{M.~J.} \bibnamefont{Padgett}},
  \bibinfo{author}{\bibfnamefont{D.~J.} \bibnamefont{Gauthier}},
  \bibnamefont{and} \bibinfo{author}{\bibfnamefont{R.~W.} \bibnamefont{Boyd}},
  \bibinfo{journal}{New J. Phys.} \textbf{\bibinfo{volume}{17}},
  \bibinfo{pages}{033033} (\bibinfo{year}{2015}).

\bibitem[{\citenamefont{Etcheverry et~al.}(2013)\citenamefont{Etcheverry,
  Ca\~{n}as, G\'{o}mez, Nogueira, Saavedra, Xavier, and Lima}}]{ECG:scirep13}
\bibinfo{author}{\bibfnamefont{S.}~\bibnamefont{Etcheverry}},
  \bibinfo{author}{\bibfnamefont{G.}~\bibnamefont{Ca\~{n}as}},
  \bibinfo{author}{\bibfnamefont{E.~S.} \bibnamefont{G\'{o}mez}},
  \bibinfo{author}{\bibfnamefont{W.~A.~T.} \bibnamefont{Nogueira}},
  \bibinfo{author}{\bibfnamefont{C.}~\bibnamefont{Saavedra}},
  \bibinfo{author}{\bibfnamefont{G.~B.} \bibnamefont{Xavier}},
  \bibnamefont{and} \bibinfo{author}{\bibfnamefont{G.}~\bibnamefont{Lima}},
  \bibinfo{journal}{Sci. Rep.} \textbf{\bibinfo{volume}{3}},
  \bibinfo{pages}{2316} (\bibinfo{year}{2013}).

\bibitem[{\citenamefont{Sasaki et~al.}(2014)\citenamefont{Sasaki, Yamamoto, and
  Koashi}}]{SWK:nature14}
\bibinfo{author}{\bibfnamefont{T.}~\bibnamefont{Sasaki}},
  \bibinfo{author}{\bibfnamefont{Y.}~\bibnamefont{Yamamoto}}, \bibnamefont{and}
  \bibinfo{author}{\bibfnamefont{M.}~\bibnamefont{Koashi}},
  \bibinfo{journal}{Nature} \textbf{\bibinfo{volume}{509}},
  \bibinfo{pages}{475} (\bibinfo{year}{2014}).

\bibitem[{\citenamefont{Takesue et~al.}(2015)\citenamefont{Takesue, Sasaki,
  Tamaki, and Koashi}}]{TST:natphoton15}
\bibinfo{author}{\bibfnamefont{H.}~\bibnamefont{Takesue}},
  \bibinfo{author}{\bibfnamefont{H.}~\bibnamefont{Sasaki}},
  \bibinfo{author}{\bibfnamefont{K.}~\bibnamefont{Tamaki}}, \bibnamefont{and}
  \bibinfo{author}{\bibfnamefont{M.}~\bibnamefont{Koashi}},
  \bibinfo{journal}{Nature Photon.} \textbf{\bibinfo{volume}{9}},
  \bibinfo{pages}{827 } (\bibinfo{year}{2015}).

\bibitem[{\citenamefont{Guan et~al.}(2015)\citenamefont{Guan, Cao, Liu,
  Shen-Tu, Pelc, Fejer, Peng, Ma, Zhang, and Pan}}]{GCL:prl15}
\bibinfo{author}{\bibfnamefont{J.~Y.} \bibnamefont{Guan}},
  \bibinfo{author}{\bibfnamefont{Z.}~\bibnamefont{Cao}},
  \bibinfo{author}{\bibfnamefont{Y.}~\bibnamefont{Liu}},
  \bibinfo{author}{\bibfnamefont{G.~L.} \bibnamefont{Shen-Tu}},
  \bibinfo{author}{\bibfnamefont{J.~S.} \bibnamefont{Pelc}},
  \bibinfo{author}{\bibfnamefont{M.~M.} \bibnamefont{Fejer}},
  \bibinfo{author}{\bibfnamefont{C.~Z.} \bibnamefont{Peng}},
  \bibinfo{author}{\bibfnamefont{X.}~\bibnamefont{Ma}},
  \bibinfo{author}{\bibfnamefont{Q.}~\bibnamefont{Zhang}}, \bibnamefont{and}
  \bibinfo{author}{\bibfnamefont{J.~W.} \bibnamefont{Pan}},
  \bibinfo{journal}{Phys. Rev. Lett.} \textbf{\bibinfo{volume}{114}},
  \bibinfo{pages}{180502} (\bibinfo{year}{2015}).

\bibitem[{\citenamefont{Wang et~al.}(2015)\citenamefont{Wang, Yin, Chen, He,
  Song, Li, Zhang, Zhou, Guo, and Han}}]{WYC:natphoton15}
\bibinfo{author}{\bibfnamefont{S.}~\bibnamefont{Wang}},
  \bibinfo{author}{\bibfnamefont{Z.~Q.} \bibnamefont{Yin}},
  \bibinfo{author}{\bibfnamefont{W.}~\bibnamefont{Chen}},
  \bibinfo{author}{\bibfnamefont{D.~Y.} \bibnamefont{He}},
  \bibinfo{author}{\bibfnamefont{X.~T.} \bibnamefont{Song}},
  \bibinfo{author}{\bibfnamefont{H.~W.} \bibnamefont{Li}},
  \bibinfo{author}{\bibfnamefont{L.~J.} \bibnamefont{Zhang}},
  \bibinfo{author}{\bibfnamefont{Z.}~\bibnamefont{Zhou}},
  \bibinfo{author}{\bibfnamefont{G.~C.} \bibnamefont{Guo}}, \bibnamefont{and}
  \bibinfo{author}{\bibfnamefont{Z.~F.} \bibnamefont{Han}},
  \bibinfo{journal}{Nature Photon.} \textbf{\bibinfo{volume}{9}},
  \bibinfo{pages}{832} (\bibinfo{year}{2015}),
  \bibinfo{note}{doi:10.1038/nphoton.2015.209}.

\bibitem[{\citenamefont{Li et~al.}(2016)\citenamefont{Li, Cao, Dai, Lin, Zhang,
  Chen, Xu, Guan, Liao, Yin et~al.}}]{LCD:pra16}
\bibinfo{author}{\bibfnamefont{Y.~H.} \bibnamefont{Li}},
  \bibinfo{author}{\bibfnamefont{Y.}~\bibnamefont{Cao}},
  \bibinfo{author}{\bibfnamefont{H.}~\bibnamefont{Dai}},
  \bibinfo{author}{\bibfnamefont{J.}~\bibnamefont{Lin}},
  \bibinfo{author}{\bibfnamefont{Z.}~\bibnamefont{Zhang}},
  \bibinfo{author}{\bibfnamefont{W.}~\bibnamefont{Chen}},
  \bibinfo{author}{\bibfnamefont{Y.}~\bibnamefont{Xu}},
  \bibinfo{author}{\bibfnamefont{J.~Y.} \bibnamefont{Guan}},
  \bibinfo{author}{\bibfnamefont{S.~K.} \bibnamefont{Liao}},
  \bibinfo{author}{\bibfnamefont{J.}~\bibnamefont{Yin}}, \bibnamefont{et~al.},
  \bibinfo{journal}{Phys. Rev. A} \textbf{\bibinfo{volume}{93}},
  \bibinfo{pages}{030302(R)} (\bibinfo{year}{2016}).

\bibitem[{\citenamefont{Mizutani
  et~al.}(2015{\natexlab{a}})\citenamefont{Mizutani, Imoto, and
  Tamaki}}]{MIT:pra15}
\bibinfo{author}{\bibfnamefont{A.}~\bibnamefont{Mizutani}},
  \bibinfo{author}{\bibfnamefont{N.}~\bibnamefont{Imoto}}, \bibnamefont{and}
  \bibinfo{author}{\bibfnamefont{K.}~\bibnamefont{Tamaki}},
  \bibinfo{journal}{Phys. Rev. A} \textbf{\bibinfo{volume}{92}},
  \bibinfo{pages}{060303} (\bibinfo{year}{2015}{\natexlab{a}}).

\bibitem[{\citenamefont{Zhao et~al.}(2008)\citenamefont{Zhao, Fung, Qi, Chen,
  and Lo}}]{ZFQ:pra08}
\bibinfo{author}{\bibfnamefont{Y.}~\bibnamefont{Zhao}},
  \bibinfo{author}{\bibfnamefont{C.-H.~F.} \bibnamefont{Fung}},
  \bibinfo{author}{\bibfnamefont{B.}~\bibnamefont{Qi}},
  \bibinfo{author}{\bibfnamefont{C.}~\bibnamefont{Chen}}, \bibnamefont{and}
  \bibinfo{author}{\bibfnamefont{H.-K.} \bibnamefont{Lo}},
  \bibinfo{journal}{Phys. Rev. A} \textbf{\bibinfo{volume}{78}},
  \bibinfo{pages}{042333} (\bibinfo{year}{2008}).

\bibitem[{\citenamefont{Lydersen et~al.}(2010)\citenamefont{Lydersen, Wiechers,
  Wittmann, Elser, Skaar, and Makarov}}]{LWW:natphoton10}
\bibinfo{author}{\bibfnamefont{L.}~\bibnamefont{Lydersen}},
  \bibinfo{author}{\bibfnamefont{C.}~\bibnamefont{Wiechers}},
  \bibinfo{author}{\bibfnamefont{C.}~\bibnamefont{Wittmann}},
  \bibinfo{author}{\bibfnamefont{D.}~\bibnamefont{Elser}},
  \bibinfo{author}{\bibfnamefont{J.}~\bibnamefont{Skaar}}, \bibnamefont{and}
  \bibinfo{author}{\bibfnamefont{V.}~\bibnamefont{Makarov}},
  \bibinfo{journal}{Nature Photon.} \textbf{\bibinfo{volume}{4}},
  \bibinfo{pages}{686} (\bibinfo{year}{2010}).

\bibitem[{\citenamefont{Xu et~al.}(2010)\citenamefont{Xu, Qi, and
  Lo}}]{XQL:njp10}
\bibinfo{author}{\bibfnamefont{F.}~\bibnamefont{Xu}},
  \bibinfo{author}{\bibfnamefont{B.}~\bibnamefont{Qi}}, \bibnamefont{and}
  \bibinfo{author}{\bibfnamefont{H.-K.} \bibnamefont{Lo}},
  \bibinfo{journal}{New J. Phys.} \textbf{\bibinfo{volume}{12}},
  \bibinfo{pages}{113026} (\bibinfo{year}{2010}).

\bibitem[{\citenamefont{Mayers and Yao}(1998)}]{MY:focs98}
\bibinfo{author}{\bibfnamefont{D.}~\bibnamefont{Mayers}} \bibnamefont{and}
  \bibinfo{author}{\bibfnamefont{A.}~\bibnamefont{Yao}}, in
  \emph{\bibinfo{booktitle}{Foundations of Computer Science, 1998. Proceedings.
  39th Annual Symposium on}} (\bibinfo{organization}{IEEE},
  \bibinfo{year}{1998}), pp. \bibinfo{pages}{503--509}.

\bibitem[{\citenamefont{Ac{\'\i}n et~al.}(2007)\citenamefont{Ac{\'\i}n,
  Brunner, Gisin, Massar, Pironio, and Scarani}}]{ABG:prl07}
\bibinfo{author}{\bibfnamefont{A.}~\bibnamefont{Ac{\'\i}n}},
  \bibinfo{author}{\bibfnamefont{N.}~\bibnamefont{Brunner}},
  \bibinfo{author}{\bibfnamefont{N.}~\bibnamefont{Gisin}},
  \bibinfo{author}{\bibfnamefont{S.}~\bibnamefont{Massar}},
  \bibinfo{author}{\bibfnamefont{S.}~\bibnamefont{Pironio}}, \bibnamefont{and}
  \bibinfo{author}{\bibfnamefont{V.}~\bibnamefont{Scarani}},
  \bibinfo{journal}{Phys. Rev. Lett.} \textbf{\bibinfo{volume}{98}},
  \bibinfo{pages}{230501} (\bibinfo{year}{2007}).

\bibitem[{\citenamefont{Braunstein and Pirandola}(2012)}]{BP:prl12}
\bibinfo{author}{\bibfnamefont{S.~L.} \bibnamefont{Braunstein}}
  \bibnamefont{and}
  \bibinfo{author}{\bibfnamefont{S.}~\bibnamefont{Pirandola}},
  \bibinfo{journal}{Phys. Rev. Lett.} \textbf{\bibinfo{volume}{108}},
  \bibinfo{pages}{130502} (\bibinfo{year}{2012}).

\bibitem[{\citenamefont{Shalm et~al.}(2015)\citenamefont{Shalm, Meyer-Scott,
  Christensen, Bierhorst, Wayne, Stevens, Gerrits, Glancy, Hamel, Allman
  et~al.}}]{SMC:prl15}
\bibinfo{author}{\bibfnamefont{L.~K.} \bibnamefont{Shalm}},
  \bibinfo{author}{\bibfnamefont{E.}~\bibnamefont{Meyer-Scott}},
  \bibinfo{author}{\bibfnamefont{B.~G.} \bibnamefont{Christensen}},
  \bibinfo{author}{\bibfnamefont{P.}~\bibnamefont{Bierhorst}},
  \bibinfo{author}{\bibfnamefont{M.~A.} \bibnamefont{Wayne}},
  \bibinfo{author}{\bibfnamefont{M.~J.} \bibnamefont{Stevens}},
  \bibinfo{author}{\bibfnamefont{T.}~\bibnamefont{Gerrits}},
  \bibinfo{author}{\bibfnamefont{S.}~\bibnamefont{Glancy}},
  \bibinfo{author}{\bibfnamefont{D.~R.} \bibnamefont{Hamel}},
  \bibinfo{author}{\bibfnamefont{M.~S.} \bibnamefont{Allman}},
  \bibnamefont{et~al.}, \bibinfo{journal}{Phys. Rev. Lett.}
  \textbf{\bibinfo{volume}{115}}, \bibinfo{pages}{250402}
  (\bibinfo{year}{2015}).

\bibitem[{\citenamefont{Giustina et~al.}(2015)\citenamefont{Giustina,
  Versteegh, Wengerowsky, Handsteiner, Hochrainer, Phelan, Steinlechner,
  Kofler, Larsson, Abellan et~al.}}]{GVW:prl15}
\bibinfo{author}{\bibfnamefont{M.}~\bibnamefont{Giustina}},
  \bibinfo{author}{\bibfnamefont{M.~A.~M.} \bibnamefont{Versteegh}},
  \bibinfo{author}{\bibfnamefont{S.}~\bibnamefont{Wengerowsky}},
  \bibinfo{author}{\bibfnamefont{J.}~\bibnamefont{Handsteiner}},
  \bibinfo{author}{\bibfnamefont{A.}~\bibnamefont{Hochrainer}},
  \bibinfo{author}{\bibfnamefont{K.}~\bibnamefont{Phelan}},
  \bibinfo{author}{\bibfnamefont{F.}~\bibnamefont{Steinlechner}},
  \bibinfo{author}{\bibfnamefont{J.}~\bibnamefont{Kofler}},
  \bibinfo{author}{\bibfnamefont{J.-A.} \bibnamefont{Larsson}},
  \bibinfo{author}{\bibfnamefont{C.}~\bibnamefont{Abellan}},
  \bibnamefont{et~al.}, \bibinfo{journal}{Phys. Rev. Lett.}
  \textbf{\bibinfo{volume}{115}}, \bibinfo{pages}{250401}
  (\bibinfo{year}{2015}).

\bibitem[{\citenamefont{Biham et~al.}(1996)\citenamefont{Biham, Huttner, and
  Mor}}]{BHM:pra96}
\bibinfo{author}{\bibfnamefont{E.}~\bibnamefont{Biham}},
  \bibinfo{author}{\bibfnamefont{B.}~\bibnamefont{Huttner}}, \bibnamefont{and}
  \bibinfo{author}{\bibfnamefont{T.}~\bibnamefont{Mor}},
  \bibinfo{journal}{Phys. Rev. A} \textbf{\bibinfo{volume}{54}},
  \bibinfo{pages}{2651} (\bibinfo{year}{1996}).

\bibitem[{\citenamefont{Inamori}(2002)}]{Ina:algo02}
\bibinfo{author}{\bibfnamefont{H.}~\bibnamefont{Inamori}},
  \bibinfo{journal}{Algorithmica} \textbf{\bibinfo{volume}{34}},
  \bibinfo{pages}{340} (\bibinfo{year}{2002}).

\bibitem[{\citenamefont{Curty et~al.}(2014)\citenamefont{Curty, Xu, Cui, Lim,
  Tamaki, and Lo}}]{CXC:natcomm14}
\bibinfo{author}{\bibfnamefont{M.}~\bibnamefont{Curty}},
  \bibinfo{author}{\bibfnamefont{F.}~\bibnamefont{Xu}},
  \bibinfo{author}{\bibfnamefont{W.}~\bibnamefont{Cui}},
  \bibinfo{author}{\bibfnamefont{C.~C.~W.} \bibnamefont{Lim}},
  \bibinfo{author}{\bibfnamefont{K.}~\bibnamefont{Tamaki}}, \bibnamefont{and}
  \bibinfo{author}{\bibfnamefont{H.-K.} \bibnamefont{Lo}},
  \bibinfo{journal}{Nature Commun.} \textbf{\bibinfo{volume}{5}},
  \bibinfo{pages}{3732} (\bibinfo{year}{2014}).

\bibitem[{\citenamefont{Lim et~al.}(2013)\citenamefont{Lim, Portmann,
  Tomamichel, Renner, and Gisin}}]{LPT:prx13}
\bibinfo{author}{\bibfnamefont{C.~C.~W.} \bibnamefont{Lim}},
  \bibinfo{author}{\bibfnamefont{C.}~\bibnamefont{Portmann}},
  \bibinfo{author}{\bibfnamefont{M.}~\bibnamefont{Tomamichel}},
  \bibinfo{author}{\bibfnamefont{R.}~\bibnamefont{Renner}}, \bibnamefont{and}
  \bibinfo{author}{\bibfnamefont{N.}~\bibnamefont{Gisin}},
  \bibinfo{journal}{Phys. Rev. X} \textbf{\bibinfo{volume}{3}},
  \bibinfo{pages}{031006} (\bibinfo{year}{2013}).

\bibitem[{\citenamefont{Tang et~al.}(2014{\natexlab{a}})\citenamefont{Tang,
  Yin, Chen, Liu, Zhang, Jiang, Zhang, Wang, You, Guan et~al.}}]{TYC:prl14}
\bibinfo{author}{\bibfnamefont{Y.-L.} \bibnamefont{Tang}},
  \bibinfo{author}{\bibfnamefont{H.-L.} \bibnamefont{Yin}},
  \bibinfo{author}{\bibfnamefont{S.-J.} \bibnamefont{Chen}},
  \bibinfo{author}{\bibfnamefont{Y.}~\bibnamefont{Liu}},
  \bibinfo{author}{\bibfnamefont{W.-J.} \bibnamefont{Zhang}},
  \bibinfo{author}{\bibfnamefont{X.}~\bibnamefont{Jiang}},
  \bibinfo{author}{\bibfnamefont{L.}~\bibnamefont{Zhang}},
  \bibinfo{author}{\bibfnamefont{J.}~\bibnamefont{Wang}},
  \bibinfo{author}{\bibfnamefont{L.-X.} \bibnamefont{You}},
  \bibinfo{author}{\bibfnamefont{J.-Y.} \bibnamefont{Guan}},
  \bibnamefont{et~al.}, \bibinfo{journal}{Phys. Rev. Lett.}
  \textbf{\bibinfo{volume}{113}}, \bibinfo{pages}{190501}
  (\bibinfo{year}{2014}{\natexlab{a}}).

\bibitem[{\citenamefont{Tang et~al.}(2014{\natexlab{b}})\citenamefont{Tang,
  Yin, Chen, Liu, Zhang, Jiang, Zhang, Wang, You, Guan et~al.}}]{TYC:jstqe14}
\bibinfo{author}{\bibfnamefont{Y.-L.} \bibnamefont{Tang}},
  \bibinfo{author}{\bibfnamefont{H.-L.} \bibnamefont{Yin}},
  \bibinfo{author}{\bibfnamefont{S.-J.} \bibnamefont{Chen}},
  \bibinfo{author}{\bibfnamefont{Y.}~\bibnamefont{Liu}},
  \bibinfo{author}{\bibfnamefont{W.-J.} \bibnamefont{Zhang}},
  \bibinfo{author}{\bibfnamefont{X.}~\bibnamefont{Jiang}},
  \bibinfo{author}{\bibfnamefont{L.}~\bibnamefont{Zhang}},
  \bibinfo{author}{\bibfnamefont{J.}~\bibnamefont{Wang}},
  \bibinfo{author}{\bibfnamefont{L.-X.} \bibnamefont{You}},
  \bibinfo{author}{\bibfnamefont{J.-Y.} \bibnamefont{Guan}},
  \bibnamefont{et~al.}, \bibinfo{journal}{IEEE J. Sel. T. Quantum Electron.}
  \textbf{\bibinfo{volume}{21}}, \bibinfo{pages}{6600407}
  (\bibinfo{year}{2014}{\natexlab{b}}).

\bibitem[{\citenamefont{Valivarthi et~al.}(2015)\citenamefont{Valivarthi,
  Lucio-Martinez, Chan, Rubenok, John, Korchinski, Duffin, Marsili, Verma, Shaw
  et~al.}}]{VLC:jmo15}
\bibinfo{author}{\bibfnamefont{R.}~\bibnamefont{Valivarthi}},
  \bibinfo{author}{\bibfnamefont{I.}~\bibnamefont{Lucio-Martinez}},
  \bibinfo{author}{\bibfnamefont{P.}~\bibnamefont{Chan}},
  \bibinfo{author}{\bibfnamefont{A.}~\bibnamefont{Rubenok}},
  \bibinfo{author}{\bibfnamefont{C.}~\bibnamefont{John}},
  \bibinfo{author}{\bibfnamefont{D.}~\bibnamefont{Korchinski}},
  \bibinfo{author}{\bibfnamefont{C.}~\bibnamefont{Duffin}},
  \bibinfo{author}{\bibfnamefont{F.}~\bibnamefont{Marsili}},
  \bibinfo{author}{\bibfnamefont{V.}~\bibnamefont{Verma}},
  \bibinfo{author}{\bibfnamefont{M.~D.} \bibnamefont{Shaw}},
  \bibnamefont{et~al.}, \bibinfo{journal}{J. Mod. Opt.}
  \textbf{\bibinfo{volume}{62}}, \bibinfo{pages}{1141} (\bibinfo{year}{2015}).

\bibitem[{\citenamefont{Tang et~al.}(2015)\citenamefont{Tang, Yin, Zhao, Liu,
  Sun, Huang, Zhang, Chen, Zhang, You et~al.}}]{TYZ:prx15}
\bibinfo{author}{\bibfnamefont{Y.-L.} \bibnamefont{Tang}},
  \bibinfo{author}{\bibfnamefont{H.-L.} \bibnamefont{Yin}},
  \bibinfo{author}{\bibfnamefont{Q.}~\bibnamefont{Zhao}},
  \bibinfo{author}{\bibfnamefont{H.}~\bibnamefont{Liu}},
  \bibinfo{author}{\bibfnamefont{X.-X.} \bibnamefont{Sun}},
  \bibinfo{author}{\bibfnamefont{M.-Q.} \bibnamefont{Huang}},
  \bibinfo{author}{\bibfnamefont{W.-J.} \bibnamefont{Zhang}},
  \bibinfo{author}{\bibfnamefont{S.-J.} \bibnamefont{Chen}},
  \bibinfo{author}{\bibfnamefont{L.}~\bibnamefont{Zhang}},
  \bibinfo{author}{\bibfnamefont{L.-X.} \bibnamefont{You}},
  \bibnamefont{et~al.}, \bibinfo{journal}{Phys. Rev. X}
  \textbf{\bibinfo{volume}{6}}, \bibinfo{pages}{011024} (\bibinfo{year}{2015}).

\bibitem[{\citenamefont{Comandar et~al.}(2016)\citenamefont{Comandar,
  Lucamarini, Fr\"ohlich, Dynes, Sharpe, Tam, Yuan, Penty, and
  Shields}}]{CLF:natphoton16}
\bibinfo{author}{\bibfnamefont{L.~C.} \bibnamefont{Comandar}},
  \bibinfo{author}{\bibfnamefont{M.}~\bibnamefont{Lucamarini}},
  \bibinfo{author}{\bibfnamefont{B.}~\bibnamefont{Fr\"ohlich}},
  \bibinfo{author}{\bibfnamefont{J.~F.} \bibnamefont{Dynes}},
  \bibinfo{author}{\bibfnamefont{A.~W.} \bibnamefont{Sharpe}},
  \bibinfo{author}{\bibfnamefont{S.}~\bibnamefont{Tam}},
  \bibinfo{author}{\bibfnamefont{Z.~L.} \bibnamefont{Yuan}},
  \bibinfo{author}{\bibfnamefont{R.~V.} \bibnamefont{Penty}}, \bibnamefont{and}
  \bibinfo{author}{\bibfnamefont{A.~J.} \bibnamefont{Shields}},
  \bibinfo{journal}{Nature Photon.} \textbf{\bibinfo{volume}{10}},
  \bibinfo{pages}{312} (\bibinfo{year}{2016}).

\bibitem[{\citenamefont{Xu et~al.}(2015{\natexlab{a}})\citenamefont{Xu, Curty,
  Qi, Qian, and Lo}}]{XCQ:natphoton15}
\bibinfo{author}{\bibfnamefont{F.}~\bibnamefont{Xu}},
  \bibinfo{author}{\bibfnamefont{M.}~\bibnamefont{Curty}},
  \bibinfo{author}{\bibfnamefont{B.}~\bibnamefont{Qi}},
  \bibinfo{author}{\bibfnamefont{L.}~\bibnamefont{Qian}}, \bibnamefont{and}
  \bibinfo{author}{\bibfnamefont{H.-K.} \bibnamefont{Lo}},
  \bibinfo{journal}{Nature Photon.} \textbf{\bibinfo{volume}{9}},
  \bibinfo{pages}{772} (\bibinfo{year}{2015}{\natexlab{a}}).

\bibitem[{\citenamefont{Yuan et~al.}(2014{\natexlab{a}})\citenamefont{Yuan,
  Lucamarini, Dynes, Fr\"ohlich, Ward, and Shields}}]{YLD:prapp14}
\bibinfo{author}{\bibfnamefont{Z.-L.} \bibnamefont{Yuan}},
  \bibinfo{author}{\bibfnamefont{M.}~\bibnamefont{Lucamarini}},
  \bibinfo{author}{\bibfnamefont{J.~F.} \bibnamefont{Dynes}},
  \bibinfo{author}{\bibfnamefont{B.}~\bibnamefont{Fr\"ohlich}},
  \bibinfo{author}{\bibfnamefont{M.~B.} \bibnamefont{Ward}}, \bibnamefont{and}
  \bibinfo{author}{\bibfnamefont{A.~J.} \bibnamefont{Shields}},
  \bibinfo{journal}{Phys. Rev. Applied} \textbf{\bibinfo{volume}{2}},
  \bibinfo{pages}{064006} (\bibinfo{year}{2014}{\natexlab{a}}).

\bibitem[{\citenamefont{Tamaki et~al.}(2012)\citenamefont{Tamaki, Lo, Fung, and
  Qi}}]{TLF:pra12}
\bibinfo{author}{\bibfnamefont{K.}~\bibnamefont{Tamaki}},
  \bibinfo{author}{\bibfnamefont{H.-K.} \bibnamefont{Lo}},
  \bibinfo{author}{\bibfnamefont{C.-H.~F.} \bibnamefont{Fung}},
  \bibnamefont{and} \bibinfo{author}{\bibfnamefont{B.}~\bibnamefont{Qi}},
  \bibinfo{journal}{Phys. Rev. A} \textbf{\bibinfo{volume}{85}},
  \bibinfo{pages}{042307} (\bibinfo{year}{2012}).

\bibitem[{not({\natexlab{b}})}]{noteMDI}
\bibinfo{note}{In MDI-QKD, the secure key rate $R$ scales as $T_A *\eta * T_B
  *\eta$, where $T_A$ is the channel transmission from Alice to the measurement
  device, $T_B$ is the channel transmission from Bob to the measurement device,
  and $\eta$ is the single-photon detection efficiency (assuming that all
  detectors have the same efficiency). The overall transmission of the whole
  channel (from Alice to Bob) is $T=T_A*T_B$, hence the key rate $R$ of MDI-QKD
  scales as $T * \eta^2$. This means that the key rate of MDI-QKD scales
  linearly with the whole channel transmittance (same as the case of
  conventional QKD and DDI-QKD), but quadratically with the detector
  efficiency.}

\bibitem[{\citenamefont{Gonz\'{a}lez et~al.}(2015)\citenamefont{Gonz\'{a}lez,
  Reb\'{o}n, Ferreira~da Silva, Figueroa, Saavedra, Curty, Lima, Xavier, and
  Nogueira}}]{GRF:pra15}
\bibinfo{author}{\bibfnamefont{P.}~\bibnamefont{Gonz\'{a}lez}},
  \bibinfo{author}{\bibfnamefont{L.}~\bibnamefont{Reb\'{o}n}},
  \bibinfo{author}{\bibfnamefont{T.}~\bibnamefont{Ferreira~da Silva}},
  \bibinfo{author}{\bibfnamefont{M.}~\bibnamefont{Figueroa}},
  \bibinfo{author}{\bibfnamefont{C.}~\bibnamefont{Saavedra}},
  \bibinfo{author}{\bibfnamefont{M.}~\bibnamefont{Curty}},
  \bibinfo{author}{\bibfnamefont{G.}~\bibnamefont{Lima}},
  \bibinfo{author}{\bibfnamefont{G.~B.} \bibnamefont{Xavier}},
  \bibnamefont{and} \bibinfo{author}{\bibfnamefont{W.~A.~T.}
  \bibnamefont{Nogueira}}, \bibinfo{journal}{Phys. Rev. A}
  \textbf{\bibinfo{volume}{92}}, \bibinfo{pages}{022337}
  (\bibinfo{year}{2015}).

\bibitem[{\citenamefont{Lim et~al.}(2014{\natexlab{c}})\citenamefont{Lim,
  Korzh, Martin, Bussi\`{e}res, Thew, and Zbinden}}]{LKM:apl14}
\bibinfo{author}{\bibfnamefont{C.~C.~W.} \bibnamefont{Lim}},
  \bibinfo{author}{\bibfnamefont{B.}~\bibnamefont{Korzh}},
  \bibinfo{author}{\bibfnamefont{A.}~\bibnamefont{Martin}},
  \bibinfo{author}{\bibfnamefont{F.}~\bibnamefont{Bussi\`{e}res}},
  \bibinfo{author}{\bibfnamefont{R.}~\bibnamefont{Thew}}, \bibnamefont{and}
  \bibinfo{author}{\bibfnamefont{H.}~\bibnamefont{Zbinden}},
  \bibinfo{journal}{Appl. Phys. Lett.} \textbf{\bibinfo{volume}{105}},
  \bibinfo{pages}{221112} (\bibinfo{year}{2014}{\natexlab{c}}).

\bibitem[{\citenamefont{Cao et~al.}(2014)\citenamefont{Cao, Zhen, Zheng, Chen,
  Liu, Chen, and Pan}}]{CZZ:arxiv14}
\bibinfo{author}{\bibfnamefont{W.-F.} \bibnamefont{Cao}},
  \bibinfo{author}{\bibfnamefont{Y.-Z.} \bibnamefont{Zhen}},
  \bibinfo{author}{\bibfnamefont{Y.-L.} \bibnamefont{Zheng}},
  \bibinfo{author}{\bibfnamefont{Z.-B.} \bibnamefont{Chen}},
  \bibinfo{author}{\bibfnamefont{N.-L.} \bibnamefont{Liu}},
  \bibinfo{author}{\bibfnamefont{K.}~\bibnamefont{Chen}}, \bibnamefont{and}
  \bibinfo{author}{\bibfnamefont{J.-W.} \bibnamefont{Pan}},
  \bibinfo{journal}{Arxiv preprint arXiv:1410.2928v1 [quant-ph]}
  (\bibinfo{year}{2014}).

\bibitem[{\citenamefont{Kim}(2003)}]{Kim:pra03}
\bibinfo{author}{\bibfnamefont{Y.-H.} \bibnamefont{Kim}},
  \bibinfo{journal}{Phys. Rev. A} \textbf{\bibinfo{volume}{67}},
  \bibinfo{pages}{040301(R)} (\bibinfo{year}{2003}).

\bibitem[{\citenamefont{Gisin et~al.}(2006)\citenamefont{Gisin, Fasel, Kraus,
  Zbinden, and Ribordy}}]{GFK:pra06}
\bibinfo{author}{\bibfnamefont{N.}~\bibnamefont{Gisin}},
  \bibinfo{author}{\bibfnamefont{S.}~\bibnamefont{Fasel}},
  \bibinfo{author}{\bibfnamefont{B.}~\bibnamefont{Kraus}},
  \bibinfo{author}{\bibfnamefont{H.}~\bibnamefont{Zbinden}}, \bibnamefont{and}
  \bibinfo{author}{\bibfnamefont{G.}~\bibnamefont{Ribordy}},
  \bibinfo{journal}{Phys. Rev. A} \textbf{\bibinfo{volume}{73}},
  \bibinfo{pages}{022320} (\bibinfo{year}{2006}).

\bibitem[{\citenamefont{Qi}(2015)}]{Qi:pra15}
\bibinfo{author}{\bibfnamefont{B.}~\bibnamefont{Qi}}, \bibinfo{journal}{Phys.
  Rev. A} \textbf{\bibinfo{volume}{91}}, \bibinfo{pages}{020303(R)}
  (\bibinfo{year}{2015}).

\bibitem[{\citenamefont{Liang et~al.}(2015)\citenamefont{Liang, Li, Yin, Chen,
  Wang, An, Guo, and Han}}]{LLY:pra15}
\bibinfo{author}{\bibfnamefont{W.-Y.} \bibnamefont{Liang}},
  \bibinfo{author}{\bibfnamefont{M.}~\bibnamefont{Li}},
  \bibinfo{author}{\bibfnamefont{Z.-Q.} \bibnamefont{Yin}},
  \bibinfo{author}{\bibfnamefont{W.}~\bibnamefont{Chen}},
  \bibinfo{author}{\bibfnamefont{S.}~\bibnamefont{Wang}},
  \bibinfo{author}{\bibfnamefont{X.-B.} \bibnamefont{An}},
  \bibinfo{author}{\bibfnamefont{G.-C.} \bibnamefont{Guo}}, \bibnamefont{and}
  \bibinfo{author}{\bibfnamefont{Z.-F.} \bibnamefont{Han}},
  \bibinfo{journal}{Phys. Rev. A} \textbf{\bibinfo{volume}{92}},
  \bibinfo{pages}{012319} (\bibinfo{year}{2015}).

\bibitem[{\citenamefont{Pirandola
  et~al.}(2015{\natexlab{b}})\citenamefont{Pirandola, Ottaviani, Spedalieri,
  Weedbrook, Braunstein, Lloyd, Gehring, Jacobsen, and
  Andersen}}]{POS:natphoton15}
\bibinfo{author}{\bibfnamefont{S.}~\bibnamefont{Pirandola}},
  \bibinfo{author}{\bibfnamefont{C.}~\bibnamefont{Ottaviani}},
  \bibinfo{author}{\bibfnamefont{G.}~\bibnamefont{Spedalieri}},
  \bibinfo{author}{\bibfnamefont{C.}~\bibnamefont{Weedbrook}},
  \bibinfo{author}{\bibfnamefont{S.~L.} \bibnamefont{Braunstein}},
  \bibinfo{author}{\bibfnamefont{S.}~\bibnamefont{Lloyd}},
  \bibinfo{author}{\bibfnamefont{T.}~\bibnamefont{Gehring}},
  \bibinfo{author}{\bibfnamefont{C.~S.} \bibnamefont{Jacobsen}},
  \bibnamefont{and} \bibinfo{author}{\bibfnamefont{U.~L.}
  \bibnamefont{Andersen}}, \bibinfo{journal}{Nature Photon.}
  \textbf{\bibinfo{volume}{9}}, \bibinfo{pages}{397}
  (\bibinfo{year}{2015}{\natexlab{b}}).

\bibitem[{\citenamefont{Li et~al.}(2014)\citenamefont{Li, Zhang, Xu, Peng, and
  Guo}}]{LZX:pra14}
\bibinfo{author}{\bibfnamefont{Z.}~\bibnamefont{Li}},
  \bibinfo{author}{\bibfnamefont{Y.-C.} \bibnamefont{Zhang}},
  \bibinfo{author}{\bibfnamefont{F.}~\bibnamefont{Xu}},
  \bibinfo{author}{\bibfnamefont{X.}~\bibnamefont{Peng}}, \bibnamefont{and}
  \bibinfo{author}{\bibfnamefont{H.}~\bibnamefont{Guo}},
  \bibinfo{journal}{Phys. Rev. A} \textbf{\bibinfo{volume}{89}},
  \bibinfo{pages}{052301} (\bibinfo{year}{2014}).

\bibitem[{\citenamefont{Ma et~al.}(2014)\citenamefont{Ma, Sun, Jiang, Gui, and
  Liang}}]{MSJ:pra14}
\bibinfo{author}{\bibfnamefont{X.-C.} \bibnamefont{Ma}},
  \bibinfo{author}{\bibfnamefont{S.-H.} \bibnamefont{Sun}},
  \bibinfo{author}{\bibfnamefont{M.-S.} \bibnamefont{Jiang}},
  \bibinfo{author}{\bibfnamefont{M.}~\bibnamefont{Gui}}, \bibnamefont{and}
  \bibinfo{author}{\bibfnamefont{L.-M.} \bibnamefont{Liang}},
  \bibinfo{journal}{Phys. Rev. A} \textbf{\bibinfo{volume}{89}},
  \bibinfo{pages}{042335} (\bibinfo{year}{2014}).

\bibitem[{\citenamefont{Gehring et~al.}(2015)\citenamefont{Gehring,
  H\"{a}ndchen, Duhme, Furrer, Franz, Pacher, Werner, and
  Schnabel}}]{GHD:natcomm15}
\bibinfo{author}{\bibfnamefont{T.}~\bibnamefont{Gehring}},
  \bibinfo{author}{\bibfnamefont{V.}~\bibnamefont{H\"{a}ndchen}},
  \bibinfo{author}{\bibfnamefont{J.}~\bibnamefont{Duhme}},
  \bibinfo{author}{\bibfnamefont{F.}~\bibnamefont{Furrer}},
  \bibinfo{author}{\bibfnamefont{T.}~\bibnamefont{Franz}},
  \bibinfo{author}{\bibfnamefont{C.}~\bibnamefont{Pacher}},
  \bibinfo{author}{\bibfnamefont{R.~F.} \bibnamefont{Werner}},
  \bibnamefont{and} \bibinfo{author}{\bibfnamefont{R.}~\bibnamefont{Schnabel}},
  \bibinfo{journal}{Nature Commun.} \textbf{\bibinfo{volume}{6}},
  \bibinfo{pages}{8795} (\bibinfo{year}{2015}).

\bibitem[{\citenamefont{Pirandola
  et~al.}(2015{\natexlab{c}})\citenamefont{Pirandola, Ottaviani, Spedalieri,
  Weedbrook, Braunstein, Lloyd, Gehring, Jacobsen, and
  Andersen}}]{POSW:natphoton15}
\bibinfo{author}{\bibfnamefont{S.}~\bibnamefont{Pirandola}},
  \bibinfo{author}{\bibfnamefont{C.}~\bibnamefont{Ottaviani}},
  \bibinfo{author}{\bibfnamefont{G.}~\bibnamefont{Spedalieri}},
  \bibinfo{author}{\bibfnamefont{C.}~\bibnamefont{Weedbrook}},
  \bibinfo{author}{\bibfnamefont{S.~L.} \bibnamefont{Braunstein}},
  \bibinfo{author}{\bibfnamefont{S.}~\bibnamefont{Lloyd}},
  \bibinfo{author}{\bibfnamefont{T.}~\bibnamefont{Gehring}},
  \bibinfo{author}{\bibfnamefont{C.~S.} \bibnamefont{Jacobsen}},
  \bibnamefont{and} \bibinfo{author}{\bibfnamefont{U.~L.}
  \bibnamefont{Andersen}}, \bibinfo{journal}{Nature Photon.}
  \textbf{\bibinfo{volume}{9}}, \bibinfo{pages}{773}
  (\bibinfo{year}{2015}{\natexlab{c}}).

\bibitem[{\citenamefont{Yin et~al.}(2013)\citenamefont{Yin, Fung, Ma, Zhang,
  Li, Chen, Wang, Guo, and Han}}]{YFM:pra13}
\bibinfo{author}{\bibfnamefont{Z.-Q.} \bibnamefont{Yin}},
  \bibinfo{author}{\bibfnamefont{C.-H.~F.} \bibnamefont{Fung}},
  \bibinfo{author}{\bibfnamefont{X.}~\bibnamefont{Ma}},
  \bibinfo{author}{\bibfnamefont{C.-M.} \bibnamefont{Zhang}},
  \bibinfo{author}{\bibfnamefont{H.-W.} \bibnamefont{Li}},
  \bibinfo{author}{\bibfnamefont{W.}~\bibnamefont{Chen}},
  \bibinfo{author}{\bibfnamefont{S.}~\bibnamefont{Wang}},
  \bibinfo{author}{\bibfnamefont{G.}~\bibnamefont{Guo}}, \bibnamefont{and}
  \bibinfo{author}{\bibfnamefont{Z.~F.} \bibnamefont{Han}},
  \bibinfo{journal}{Phys. Rev. A} \textbf{\bibinfo{volume}{88}},
  \bibinfo{pages}{062322} (\bibinfo{year}{2013}).

\bibitem[{\citenamefont{Yin et~al.}(2014)\citenamefont{Yin, Fung, Ma, Zhang,
  Li, Chen, Wang, Guo, and Han}}]{YFM:pra14}
\bibinfo{author}{\bibfnamefont{Z.-Q.} \bibnamefont{Yin}},
  \bibinfo{author}{\bibfnamefont{C.-H.~F.} \bibnamefont{Fung}},
  \bibinfo{author}{\bibfnamefont{X.}~\bibnamefont{Ma}},
  \bibinfo{author}{\bibfnamefont{C.-M.} \bibnamefont{Zhang}},
  \bibinfo{author}{\bibfnamefont{H.-W.} \bibnamefont{Li}},
  \bibinfo{author}{\bibfnamefont{W.}~\bibnamefont{Chen}},
  \bibinfo{author}{\bibfnamefont{S.}~\bibnamefont{Wang}},
  \bibinfo{author}{\bibfnamefont{G.~C.} \bibnamefont{Guo}}, \bibnamefont{and}
  \bibinfo{author}{\bibfnamefont{Z.~F.} \bibnamefont{Han}},
  \bibinfo{journal}{Phys. Rev. A} \textbf{\bibinfo{volume}{90}},
  \bibinfo{pages}{052319} (\bibinfo{year}{2014}).

\bibitem[{\citenamefont{Barnett et~al.}(1993)\citenamefont{Barnett, Huttner,
  and Phoenix}}]{BHP:jmo93}
\bibinfo{author}{\bibfnamefont{S.~M.} \bibnamefont{Barnett}},
  \bibinfo{author}{\bibfnamefont{B.}~\bibnamefont{Huttner}}, \bibnamefont{and}
  \bibinfo{author}{\bibfnamefont{S.}~\bibnamefont{Phoenix}},
  \bibinfo{journal}{J. Mod. Opt.} \textbf{\bibinfo{volume}{40}},
  \bibinfo{pages}{2501} (\bibinfo{year}{1993}).

\bibitem[{\citenamefont{Du\v{s}ek et~al.}(2000)\citenamefont{Du\v{s}ek, Jahma,
  and L\"{u}tkenhaus}}]{DJL:pra00}
\bibinfo{author}{\bibfnamefont{M.}~\bibnamefont{Du\v{s}ek}},
  \bibinfo{author}{\bibfnamefont{M.}~\bibnamefont{Jahma}}, \bibnamefont{and}
  \bibinfo{author}{\bibfnamefont{N.}~\bibnamefont{L\"{u}tkenhaus}},
  \bibinfo{journal}{Phys. Rev. A} \textbf{\bibinfo{volume}{62}},
  \bibinfo{pages}{022306} (\bibinfo{year}{2000}).

\bibitem[{\citenamefont{Xu et~al.}(2015{\natexlab{b}})\citenamefont{Xu, Wei,
  Sajeed, Kaiser, Sun, Tang, Qian, Makarov, and Lo}}]{XWS:pra15}
\bibinfo{author}{\bibfnamefont{F.}~\bibnamefont{Xu}},
  \bibinfo{author}{\bibfnamefont{K.}~\bibnamefont{Wei}},
  \bibinfo{author}{\bibfnamefont{S.}~\bibnamefont{Sajeed}},
  \bibinfo{author}{\bibfnamefont{S.}~\bibnamefont{Kaiser}},
  \bibinfo{author}{\bibfnamefont{S.}~\bibnamefont{Sun}},
  \bibinfo{author}{\bibfnamefont{Z.}~\bibnamefont{Tang}},
  \bibinfo{author}{\bibfnamefont{L.}~\bibnamefont{Qian}},
  \bibinfo{author}{\bibfnamefont{V.}~\bibnamefont{Makarov}}, \bibnamefont{and}
  \bibinfo{author}{\bibfnamefont{H.-K.} \bibnamefont{Lo}},
  \bibinfo{journal}{Phys. Rev. A} \textbf{\bibinfo{volume}{92}},
  \bibinfo{pages}{032305} (\bibinfo{year}{2015}{\natexlab{b}}).

\bibitem[{\citenamefont{Mizutani
  et~al.}(2015{\natexlab{b}})\citenamefont{Mizutani, Curty, Lim, Imoto, and
  Tamaki}}]{MCL:njp15}
\bibinfo{author}{\bibfnamefont{A.}~\bibnamefont{Mizutani}},
  \bibinfo{author}{\bibfnamefont{M.}~\bibnamefont{Curty}},
  \bibinfo{author}{\bibfnamefont{C.~C.~W.} \bibnamefont{Lim}},
  \bibinfo{author}{\bibfnamefont{N.}~\bibnamefont{Imoto}}, \bibnamefont{and}
  \bibinfo{author}{\bibfnamefont{K.}~\bibnamefont{Tamaki}},
  \bibinfo{journal}{New J. Phys.} \textbf{\bibinfo{volume}{17}},
  \bibinfo{pages}{093011} (\bibinfo{year}{2015}{\natexlab{b}}).

\bibitem[{\citenamefont{Cao et~al.}(2015)\citenamefont{Cao, Zhang, Lo, and
  Ma}}]{CZL:njp15}
\bibinfo{author}{\bibfnamefont{Z.}~\bibnamefont{Cao}},
  \bibinfo{author}{\bibfnamefont{Z.}~\bibnamefont{Zhang}},
  \bibinfo{author}{\bibfnamefont{H.-K.} \bibnamefont{Lo}}, \bibnamefont{and}
  \bibinfo{author}{\bibfnamefont{X.}~\bibnamefont{Ma}}, \bibinfo{journal}{New
  J. Phys.} \textbf{\bibinfo{volume}{17}}, \bibinfo{pages}{053014}
  (\bibinfo{year}{2015}).

\bibitem[{\citenamefont{Yuan et~al.}(2014{\natexlab{b}})\citenamefont{Yuan,
  Lucamarini, Dynes, Fr{\"o}hlich, Plews, and Shields}}]{YLD:apl14}
\bibinfo{author}{\bibfnamefont{Z.~L.} \bibnamefont{Yuan}},
  \bibinfo{author}{\bibfnamefont{M.}~\bibnamefont{Lucamarini}},
  \bibinfo{author}{\bibfnamefont{J.~F.} \bibnamefont{Dynes}},
  \bibinfo{author}{\bibfnamefont{B.}~\bibnamefont{Fr{\"o}hlich}},
  \bibinfo{author}{\bibfnamefont{A.}~\bibnamefont{Plews}}, \bibnamefont{and}
  \bibinfo{author}{\bibfnamefont{A.~J.} \bibnamefont{Shields}},
  \bibinfo{journal}{Appl. Phys. Lett.} \textbf{\bibinfo{volume}{104}},
  \bibinfo{pages}{261112} (\bibinfo{year}{2014}{\natexlab{b}}).

\bibitem[{\citenamefont{Jain et~al.}(2014)\citenamefont{Jain, Anisimova, Khan,
  Makarov, Marquardt, and Leuchs}}]{JAK:njp14}
\bibinfo{author}{\bibfnamefont{N.}~\bibnamefont{Jain}},
  \bibinfo{author}{\bibfnamefont{E.}~\bibnamefont{Anisimova}},
  \bibinfo{author}{\bibfnamefont{I.}~\bibnamefont{Khan}},
  \bibinfo{author}{\bibfnamefont{V.}~\bibnamefont{Makarov}},
  \bibinfo{author}{\bibfnamefont{C.}~\bibnamefont{Marquardt}},
  \bibnamefont{and} \bibinfo{author}{\bibfnamefont{G.}~\bibnamefont{Leuchs}},
  \bibinfo{journal}{New J. Phys.} \textbf{\bibinfo{volume}{16}},
  \bibinfo{pages}{123030} (\bibinfo{year}{2014}).

\bibitem[{\citenamefont{Jain et~al.}(2015)\citenamefont{Jain, Stiller, Khan,
  Makarov, Marquardt, and Leuchs}}]{JSK:jstqe15}
\bibinfo{author}{\bibfnamefont{N.}~\bibnamefont{Jain}},
  \bibinfo{author}{\bibfnamefont{B.}~\bibnamefont{Stiller}},
  \bibinfo{author}{\bibfnamefont{I.}~\bibnamefont{Khan}},
  \bibinfo{author}{\bibfnamefont{V.}~\bibnamefont{Makarov}},
  \bibinfo{author}{\bibfnamefont{C.}~\bibnamefont{Marquardt}},
  \bibnamefont{and} \bibinfo{author}{\bibfnamefont{G.}~\bibnamefont{Leuchs}},
  \bibinfo{journal}{IEEE J. Sel. Topics Quantum Electron.}
  \textbf{\bibinfo{volume}{21}}, \bibinfo{pages}{6600710}
  (\bibinfo{year}{2015}).

\bibitem[{\citenamefont{Stiller et~al.}(2015)\citenamefont{Stiller, Khan, Jain,
  Jouguet, Kunz-Jacques, Diamanti, Marquardt, and Leuchs}}]{SKJ:cleo15}
\bibinfo{author}{\bibfnamefont{B.}~\bibnamefont{Stiller}},
  \bibinfo{author}{\bibfnamefont{I.}~\bibnamefont{Khan}},
  \bibinfo{author}{\bibfnamefont{N.}~\bibnamefont{Jain}},
  \bibinfo{author}{\bibfnamefont{P.}~\bibnamefont{Jouguet}},
  \bibinfo{author}{\bibfnamefont{S.}~\bibnamefont{Kunz-Jacques}},
  \bibinfo{author}{\bibfnamefont{E.}~\bibnamefont{Diamanti}},
  \bibinfo{author}{\bibfnamefont{C.}~\bibnamefont{Marquardt}},
  \bibnamefont{and} \bibinfo{author}{\bibfnamefont{G.}~\bibnamefont{Leuchs}},
  in \emph{\bibinfo{booktitle}{CLEO: 2015, OSA Technical Digest (online)}},
  edited by \bibinfo{editor}{\bibfnamefont{S.}~\bibnamefont{Goldwasser}}
  (\bibinfo{publisher}{Optical Society of America}, \bibinfo{year}{2015}),
  \bibinfo{note}{paper FF1A.7.}

\bibitem[{\citenamefont{Lucamarini
  et~al.}(2015{\natexlab{b}})\citenamefont{Lucamarini, Choi, Ward, Dynes, Yuan,
  and Shields}}]{LCW:prx15}
\bibinfo{author}{\bibfnamefont{M.}~\bibnamefont{Lucamarini}},
  \bibinfo{author}{\bibfnamefont{I.}~\bibnamefont{Choi}},
  \bibinfo{author}{\bibfnamefont{M.~B.} \bibnamefont{Ward}},
  \bibinfo{author}{\bibfnamefont{J.~F.} \bibnamefont{Dynes}},
  \bibinfo{author}{\bibfnamefont{Z.}~\bibnamefont{Yuan}}, \bibnamefont{and}
  \bibinfo{author}{\bibfnamefont{A.~J.} \bibnamefont{Shields}},
  \bibinfo{journal}{Phys. Rev. X} \textbf{\bibinfo{volume}{5}},
  \bibinfo{pages}{031030} (\bibinfo{year}{2015}{\natexlab{b}}).

\bibitem[{\citenamefont{Tang et~al.}(2016)\citenamefont{Tang, Wei, Bedroya,
  Qian, and Lo}}]{TWB:pra16}
\bibinfo{author}{\bibfnamefont{Z.}~\bibnamefont{Tang}},
  \bibinfo{author}{\bibfnamefont{K.}~\bibnamefont{Wei}},
  \bibinfo{author}{\bibfnamefont{O.}~\bibnamefont{Bedroya}},
  \bibinfo{author}{\bibfnamefont{L.}~\bibnamefont{Qian}}, \bibnamefont{and}
  \bibinfo{author}{\bibfnamefont{H.-K.} \bibnamefont{Lo}},
  \bibinfo{journal}{Phys. Rev. A} \textbf{\bibinfo{volume}{93}},
  \bibinfo{pages}{042308} (\bibinfo{year}{2016}).

\bibitem[{\citenamefont{Kocher}(1996)}]{Kocher:crypto96}
\bibinfo{author}{\bibfnamefont{P.~C.} \bibnamefont{Kocher}}, in
  \emph{\bibinfo{booktitle}{Advances in Cryptology - CRYPTO 1996}}
  (\bibinfo{organization}{Springer}, \bibinfo{year}{1996}), pp.
  \bibinfo{pages}{104--113}.

\bibitem[{\citenamefont{Kocher et~al.}(1999)\citenamefont{Kocher, Jaffe, and
  Jun}}]{KJJ:crypto99}
\bibinfo{author}{\bibfnamefont{P.}~\bibnamefont{Kocher}},
  \bibinfo{author}{\bibfnamefont{J.}~\bibnamefont{Jaffe}}, \bibnamefont{and}
  \bibinfo{author}{\bibfnamefont{B.}~\bibnamefont{Jun}}, in
  \emph{\bibinfo{booktitle}{Advances in Cryptology - CRYPTO 1999}}
  (\bibinfo{organization}{Springer}, \bibinfo{year}{1999}), pp.
  \bibinfo{pages}{388--397}.

\bibitem[{\citenamefont{Genkin et~al.}(2014)\citenamefont{Genkin, Shamir, and
  Tromer}}]{GST:crypto14}
\bibinfo{author}{\bibfnamefont{D.}~\bibnamefont{Genkin}},
  \bibinfo{author}{\bibfnamefont{A.}~\bibnamefont{Shamir}}, \bibnamefont{and}
  \bibinfo{author}{\bibfnamefont{E.}~\bibnamefont{Tromer}}, in
  \emph{\bibinfo{booktitle}{Advances in Cryptology - CRYPTO 2014}}
  (\bibinfo{publisher}{Springer}, \bibinfo{year}{2014}), pp.
  \bibinfo{pages}{444--461}.

\bibitem[{\citenamefont{Sasaki et~al.}(2011)\citenamefont{Sasaki, Fujiwara,
  Ishizuka, Klaus, Wakui, Takeoka, Tanaka, Yoshino, Nambu, Takahashi
  et~al.}}]{SFI:oe11}
\bibinfo{author}{\bibfnamefont{M.}~\bibnamefont{Sasaki}},
  \bibinfo{author}{\bibfnamefont{M.}~\bibnamefont{Fujiwara}},
  \bibinfo{author}{\bibfnamefont{H.}~\bibnamefont{Ishizuka}},
  \bibinfo{author}{\bibfnamefont{W.}~\bibnamefont{Klaus}},
  \bibinfo{author}{\bibfnamefont{K.}~\bibnamefont{Wakui}},
  \bibinfo{author}{\bibfnamefont{M.}~\bibnamefont{Takeoka}},
  \bibinfo{author}{\bibfnamefont{A.}~\bibnamefont{Tanaka}},
  \bibinfo{author}{\bibfnamefont{K.}~\bibnamefont{Yoshino}},
  \bibinfo{author}{\bibfnamefont{Y.}~\bibnamefont{Nambu}},
  \bibinfo{author}{\bibfnamefont{S.}~\bibnamefont{Takahashi}},
  \bibnamefont{et~al.}, \bibinfo{journal}{Opt. Express}
  \textbf{\bibinfo{volume}{19}}, \bibinfo{pages}{10387} (\bibinfo{year}{2011}).

\bibitem[{\citenamefont{Northup and Blatt}(2014)}]{NB:natphoton14}
\bibinfo{author}{\bibfnamefont{T.~E.} \bibnamefont{Northup}} \bibnamefont{and}
  \bibinfo{author}{\bibfnamefont{R.}~\bibnamefont{Blatt}},
  \bibinfo{journal}{Nature Photon.} \textbf{\bibinfo{volume}{8}},
  \bibinfo{pages}{356} (\bibinfo{year}{2014}).

\bibitem[{\citenamefont{Bussi\`eres et~al.}(2014)\citenamefont{Bussi\`eres,
  Clausen, Tiranov, Korzh, Verma, Nam, Marsili, Ferrier, Goldner, Herrmann
  et~al.}}]{BCT:natphoton14}
\bibinfo{author}{\bibfnamefont{F.}~\bibnamefont{Bussi\`eres}},
  \bibinfo{author}{\bibfnamefont{C.}~\bibnamefont{Clausen}},
  \bibinfo{author}{\bibfnamefont{A.}~\bibnamefont{Tiranov}},
  \bibinfo{author}{\bibfnamefont{B.}~\bibnamefont{Korzh}},
  \bibinfo{author}{\bibfnamefont{V.~B.} \bibnamefont{Verma}},
  \bibinfo{author}{\bibfnamefont{S.~W.} \bibnamefont{Nam}},
  \bibinfo{author}{\bibfnamefont{F.}~\bibnamefont{Marsili}},
  \bibinfo{author}{\bibfnamefont{A.}~\bibnamefont{Ferrier}},
  \bibinfo{author}{\bibfnamefont{P.}~\bibnamefont{Goldner}},
  \bibinfo{author}{\bibfnamefont{H.}~\bibnamefont{Herrmann}},
  \bibnamefont{et~al.}, \bibinfo{journal}{Nature Photon.}
  \textbf{\bibinfo{volume}{8}}, \bibinfo{pages}{775} (\bibinfo{year}{2014}).

\bibitem[{\citenamefont{Munro et~al.}(2012)\citenamefont{Munro, Stephens,
  Devitt, Harrison, and Nemoto}}]{MSD:natphoton12}
\bibinfo{author}{\bibfnamefont{W.~J.} \bibnamefont{Munro}},
  \bibinfo{author}{\bibfnamefont{A.~M.} \bibnamefont{Stephens}},
  \bibinfo{author}{\bibfnamefont{S.~J.} \bibnamefont{Devitt}},
  \bibinfo{author}{\bibfnamefont{K.~A.} \bibnamefont{Harrison}},
  \bibnamefont{and} \bibinfo{author}{\bibfnamefont{K.}~\bibnamefont{Nemoto}},
  \bibinfo{journal}{Nature Photon.} \textbf{\bibinfo{volume}{6}},
  \bibinfo{pages}{777} (\bibinfo{year}{2012}).

\bibitem[{\citenamefont{Azuma et~al.}(2015)\citenamefont{Azuma, Tamaki, and
  Lo}}]{ATL:natcomm15}
\bibinfo{author}{\bibfnamefont{K.}~\bibnamefont{Azuma}},
  \bibinfo{author}{\bibfnamefont{K.}~\bibnamefont{Tamaki}}, \bibnamefont{and}
  \bibinfo{author}{\bibfnamefont{H.-K.} \bibnamefont{Lo}},
  \bibinfo{journal}{Nature Commun.} \textbf{\bibinfo{volume}{6}},
  \bibinfo{pages}{6787} (\bibinfo{year}{2015}).

\bibitem[{\citenamefont{Buttler et~al.}(2000)\citenamefont{Buttler, Hughes,
  Lamoreaux, Morgan, Nordholt, and Peterson}}]{BHL:prl00}
\bibinfo{author}{\bibfnamefont{W.~T.} \bibnamefont{Buttler}},
  \bibinfo{author}{\bibfnamefont{R.~J.} \bibnamefont{Hughes}},
  \bibinfo{author}{\bibfnamefont{S.~K.} \bibnamefont{Lamoreaux}},
  \bibinfo{author}{\bibfnamefont{G.~L.} \bibnamefont{Morgan}},
  \bibinfo{author}{\bibfnamefont{J.~E.} \bibnamefont{Nordholt}},
  \bibnamefont{and} \bibinfo{author}{\bibfnamefont{C.~G.}
  \bibnamefont{Peterson}}, \bibinfo{journal}{Phys. Rev. Lett.}
  \textbf{\bibinfo{volume}{84}}, \bibinfo{pages}{5652} (\bibinfo{year}{2000}).

\bibitem[{\citenamefont{Nauerth et~al.}(2013)\citenamefont{Nauerth, Moll, Rau,
  Fuchs, Horwath, Frick, and Weinfurter}}]{NMR:natphoton13}
\bibinfo{author}{\bibfnamefont{S.}~\bibnamefont{Nauerth}},
  \bibinfo{author}{\bibfnamefont{F.}~\bibnamefont{Moll}},
  \bibinfo{author}{\bibfnamefont{M.}~\bibnamefont{Rau}},
  \bibinfo{author}{\bibfnamefont{C.}~\bibnamefont{Fuchs}},
  \bibinfo{author}{\bibfnamefont{J.}~\bibnamefont{Horwath}},
  \bibinfo{author}{\bibfnamefont{S.}~\bibnamefont{Frick}}, \bibnamefont{and}
  \bibinfo{author}{\bibfnamefont{H.}~\bibnamefont{Weinfurter}},
  \bibinfo{journal}{Nature Photon.} \textbf{\bibinfo{volume}{7}},
  \bibinfo{pages}{382} (\bibinfo{year}{2013}).

\bibitem[{\citenamefont{Wang et~al.}(2013)\citenamefont{Wang, Yang, Liao,
  Zhang, Shen, Hu, Wu, Yang, Jiang, Tang et~al.}}]{WYL:natphoton13}
\bibinfo{author}{\bibfnamefont{J.-Y.} \bibnamefont{Wang}},
  \bibinfo{author}{\bibfnamefont{B.}~\bibnamefont{Yang}},
  \bibinfo{author}{\bibfnamefont{S.-K.} \bibnamefont{Liao}},
  \bibinfo{author}{\bibfnamefont{L.}~\bibnamefont{Zhang}},
  \bibinfo{author}{\bibfnamefont{Q.}~\bibnamefont{Shen}},
  \bibinfo{author}{\bibfnamefont{X.-F.} \bibnamefont{Hu}},
  \bibinfo{author}{\bibfnamefont{J.-C.} \bibnamefont{Wu}},
  \bibinfo{author}{\bibfnamefont{S.-J.} \bibnamefont{Yang}},
  \bibinfo{author}{\bibfnamefont{H.}~\bibnamefont{Jiang}},
  \bibinfo{author}{\bibfnamefont{Y.-L.} \bibnamefont{Tang}},
  \bibnamefont{et~al.}, \bibinfo{journal}{Nature Photon.}
  \textbf{\bibinfo{volume}{7}}, \bibinfo{pages}{387} (\bibinfo{year}{2013}).

\bibitem[{\citenamefont{Vallone et~al.}(2015)\citenamefont{Vallone, Bacco,
  Dequal, Gaiarin, Luceri, Bianco, and Villoresi}}]{VBD:prl15}
\bibinfo{author}{\bibfnamefont{G.}~\bibnamefont{Vallone}},
  \bibinfo{author}{\bibfnamefont{D.}~\bibnamefont{Bacco}},
  \bibinfo{author}{\bibfnamefont{D.}~\bibnamefont{Dequal}},
  \bibinfo{author}{\bibfnamefont{S.}~\bibnamefont{Gaiarin}},
  \bibinfo{author}{\bibfnamefont{V.}~\bibnamefont{Luceri}},
  \bibinfo{author}{\bibfnamefont{G.}~\bibnamefont{Bianco}}, \bibnamefont{and}
  \bibinfo{author}{\bibfnamefont{P.}~\bibnamefont{Villoresi}},
  \bibinfo{journal}{Phys. Rev. Lett.} \textbf{\bibinfo{volume}{115}},
  \bibinfo{pages}{040502} (\bibinfo{year}{2015}).

\bibitem[{\citenamefont{Meyers}(2015)}]{Meyers:fso15}
\bibinfo{author}{\bibfnamefont{R.~E.} \bibnamefont{Meyers}}, in
  \emph{\bibinfo{booktitle}{Advanced Free Space Optics (FSO)}}
  (\bibinfo{publisher}{Springer}, \bibinfo{year}{2015}), pp.
  \bibinfo{pages}{343--387}.

\bibitem[{\citenamefont{Elser et~al.}(October 2015)\citenamefont{Elser,
  G\"unthner, Khan, Stiller, Marquardt, Leuchs, Saucke, Tr\"ondle, Heine, Seel
  et~al.}}]{EGK:icsos15}
\bibinfo{author}{\bibfnamefont{D.}~\bibnamefont{Elser}},
  \bibinfo{author}{\bibfnamefont{K.}~\bibnamefont{G\"unthner}},
  \bibinfo{author}{\bibfnamefont{I.}~\bibnamefont{Khan}},
  \bibinfo{author}{\bibfnamefont{B.}~\bibnamefont{Stiller}},
  \bibinfo{author}{\bibfnamefont{C.}~\bibnamefont{Marquardt}},
  \bibinfo{author}{\bibfnamefont{G.}~\bibnamefont{Leuchs}},
  \bibinfo{author}{\bibfnamefont{K.}~\bibnamefont{Saucke}},
  \bibinfo{author}{\bibfnamefont{D.}~\bibnamefont{Tr\"ondle}},
  \bibinfo{author}{\bibfnamefont{F.}~\bibnamefont{Heine}},
  \bibinfo{author}{\bibfnamefont{S.}~\bibnamefont{Seel}}, \bibnamefont{et~al.},
  in \emph{\bibinfo{booktitle}{IEEE ICSOS 2015, New Orleans, USA}}
  (\bibinfo{year}{October 2015}).

\bibitem[{\citenamefont{Bourgoin et~al.}(2015)\citenamefont{Bourgoin, Higgins,
  Gigov, Holloway, Pugh, Kaiser, Cranmer, and Jennewein}}]{BHZ:opex15}
\bibinfo{author}{\bibfnamefont{J.-P.} \bibnamefont{Bourgoin}},
  \bibinfo{author}{\bibfnamefont{B.~L.} \bibnamefont{Higgins}},
  \bibinfo{author}{\bibfnamefont{N.}~\bibnamefont{Gigov}},
  \bibinfo{author}{\bibfnamefont{C.}~\bibnamefont{Holloway}},
  \bibinfo{author}{\bibfnamefont{C.~J.} \bibnamefont{Pugh}},
  \bibinfo{author}{\bibfnamefont{S.}~\bibnamefont{Kaiser}},
  \bibinfo{author}{\bibfnamefont{M.}~\bibnamefont{Cranmer}}, \bibnamefont{and}
  \bibinfo{author}{\bibfnamefont{T.}~\bibnamefont{Jennewein}},
  \bibinfo{journal}{Opt. Express} \textbf{\bibinfo{volume}{23}},
  \bibinfo{pages}{33437} (\bibinfo{year}{2015}).

\bibitem[{\citenamefont{Heim et~al.}(2014)\citenamefont{Heim, Peuntinger,
  Killoran, Khan, Wittmann, Marquardt, and Leuchs}}]{HPK:njp14}
\bibinfo{author}{\bibfnamefont{B.}~\bibnamefont{Heim}},
  \bibinfo{author}{\bibfnamefont{C.}~\bibnamefont{Peuntinger}},
  \bibinfo{author}{\bibfnamefont{N.}~\bibnamefont{Killoran}},
  \bibinfo{author}{\bibfnamefont{I.}~\bibnamefont{Khan}},
  \bibinfo{author}{\bibfnamefont{C.}~\bibnamefont{Wittmann}},
  \bibinfo{author}{\bibfnamefont{C.}~\bibnamefont{Marquardt}},
  \bibnamefont{and} \bibinfo{author}{\bibfnamefont{G.}~\bibnamefont{Leuchs}},
  \bibinfo{journal}{New J. Phys.} \textbf{\bibinfo{volume}{16}},
  \bibinfo{pages}{113018} (\bibinfo{year}{2014}).

\bibitem[{\citenamefont{Weedbrook et~al.}(2010)\citenamefont{Weedbrook,
  Pirandola, Lloyd, and Ralph}}]{WPL:prl10}
\bibinfo{author}{\bibfnamefont{C.}~\bibnamefont{Weedbrook}},
  \bibinfo{author}{\bibfnamefont{S.}~\bibnamefont{Pirandola}},
  \bibinfo{author}{\bibfnamefont{S.}~\bibnamefont{Lloyd}}, \bibnamefont{and}
  \bibinfo{author}{\bibfnamefont{T.~C.} \bibnamefont{Ralph}},
  \bibinfo{journal}{Phys. Rev. Lett.} \textbf{\bibinfo{volume}{105}},
  \bibinfo{pages}{110501} (\bibinfo{year}{2010}).

\bibitem[{\citenamefont{Weedbrook et~al.}(2012)\citenamefont{Weedbrook,
  Pirandola, and Ralph}}]{WPR:pra12}
\bibinfo{author}{\bibfnamefont{C.}~\bibnamefont{Weedbrook}},
  \bibinfo{author}{\bibfnamefont{S.}~\bibnamefont{Pirandola}},
  \bibnamefont{and} \bibinfo{author}{\bibfnamefont{T.~C.} \bibnamefont{Ralph}},
  \bibinfo{journal}{Phys. Rev. A} \textbf{\bibinfo{volume}{86}},
  \bibinfo{pages}{022318} (\bibinfo{year}{2012}).

\bibitem[{\citenamefont{Mayers}(1997)}]{Mayers:prl97}
\bibinfo{author}{\bibfnamefont{D.}~\bibnamefont{Mayers}},
  \bibinfo{journal}{Phys. Rev. Lett.} \textbf{\bibinfo{volume}{78}},
  \bibinfo{pages}{3414} (\bibinfo{year}{1997}).

\bibitem[{\citenamefont{Lo and Chau}(1997)}]{LC:prl97}
\bibinfo{author}{\bibfnamefont{H.-K.} \bibnamefont{Lo}} \bibnamefont{and}
  \bibinfo{author}{\bibfnamefont{H.~F.} \bibnamefont{Chau}},
  \bibinfo{journal}{Phys. Rev. Lett.} \textbf{\bibinfo{volume}{78}},
  \bibinfo{pages}{3410} (\bibinfo{year}{1997}).

\bibitem[{\citenamefont{Lunghi et~al.}(2015)\citenamefont{Lunghi, Kaniewski,
  Bussieres, Houlmann, Tomamichel, Wehner, and Zbinden}}]{LKB:prl15}
\bibinfo{author}{\bibfnamefont{T.}~\bibnamefont{Lunghi}},
  \bibinfo{author}{\bibfnamefont{J.}~\bibnamefont{Kaniewski}},
  \bibinfo{author}{\bibfnamefont{F.}~\bibnamefont{Bussieres}},
  \bibinfo{author}{\bibfnamefont{R.}~\bibnamefont{Houlmann}},
  \bibinfo{author}{\bibfnamefont{M.}~\bibnamefont{Tomamichel}},
  \bibinfo{author}{\bibfnamefont{S.}~\bibnamefont{Wehner}}, \bibnamefont{and}
  \bibinfo{author}{\bibfnamefont{H.}~\bibnamefont{Zbinden}},
  \bibinfo{journal}{Phys. Rev. Lett.} \textbf{\bibinfo{volume}{115}},
  \bibinfo{pages}{030502} (\bibinfo{year}{2015}).

\bibitem[{\citenamefont{Cleve et~al.}(1999)\citenamefont{Cleve, Gottesman, and
  Lo}}]{CGL:prl99}
\bibinfo{author}{\bibfnamefont{R.}~\bibnamefont{Cleve}},
  \bibinfo{author}{\bibfnamefont{D.}~\bibnamefont{Gottesman}},
  \bibnamefont{and} \bibinfo{author}{\bibfnamefont{H.-K.} \bibnamefont{Lo}},
  \bibinfo{journal}{Phys. Rev. Lett.} \textbf{\bibinfo{volume}{83}},
  \bibinfo{pages}{648} (\bibinfo{year}{1999}).

\bibitem[{\citenamefont{Hillery et~al.}(1999)\citenamefont{Hillery, Bu\v{z}ek,
  and Berthiaume}}]{HBB:pra99}
\bibinfo{author}{\bibfnamefont{M.}~\bibnamefont{Hillery}},
  \bibinfo{author}{\bibfnamefont{V.}~\bibnamefont{Bu\v{z}ek}},
  \bibnamefont{and}
  \bibinfo{author}{\bibfnamefont{A.}~\bibnamefont{Berthiaume}},
  \bibinfo{journal}{Phys. Rev. A} \textbf{\bibinfo{volume}{59}},
  \bibinfo{pages}{1829} (\bibinfo{year}{1999}).

\bibitem[{\citenamefont{Bell et~al.}(2014)\citenamefont{Bell, Markham,
  Herrera-Marti, Marin, Wadsworth, Rarity, and Tame}}]{BMH:natcommun14}
\bibinfo{author}{\bibfnamefont{B.~A.} \bibnamefont{Bell}},
  \bibinfo{author}{\bibfnamefont{D.}~\bibnamefont{Markham}},
  \bibinfo{author}{\bibfnamefont{D.~A.} \bibnamefont{Herrera-Marti}},
  \bibinfo{author}{\bibfnamefont{A.}~\bibnamefont{Marin}},
  \bibinfo{author}{\bibfnamefont{W.~J.} \bibnamefont{Wadsworth}},
  \bibinfo{author}{\bibfnamefont{J.~G.} \bibnamefont{Rarity}},
  \bibnamefont{and} \bibinfo{author}{\bibfnamefont{M.~S.} \bibnamefont{Tame}},
  \bibinfo{journal}{Nature Commun.} \textbf{\bibinfo{volume}{5}},
  \bibinfo{pages}{5480} (\bibinfo{year}{2014}).

\bibitem[{\citenamefont{Berlin et~al.}(2011)\citenamefont{Berlin, Brassard,
  Bussieres, Godbout, Slater, and Tittel}}]{BBB:natcomm11}
\bibinfo{author}{\bibfnamefont{G.}~\bibnamefont{Berlin}},
  \bibinfo{author}{\bibfnamefont{G.}~\bibnamefont{Brassard}},
  \bibinfo{author}{\bibfnamefont{F.}~\bibnamefont{Bussieres}},
  \bibinfo{author}{\bibfnamefont{N.}~\bibnamefont{Godbout}},
  \bibinfo{author}{\bibfnamefont{J.~A.} \bibnamefont{Slater}},
  \bibnamefont{and} \bibinfo{author}{\bibfnamefont{W.}~\bibnamefont{Tittel}},
  \bibinfo{journal}{Nature Commun.} \textbf{\bibinfo{volume}{2}},
  \bibinfo{pages}{561} (\bibinfo{year}{2011}).

\bibitem[{\citenamefont{Pappa et~al.}(2014)\citenamefont{Pappa, Jouguet,
  Lawson, Chailloux, Legr\'e, Trinkler, Kerenidis, and
  Diamanti}}]{PJL:natcommun14}
\bibinfo{author}{\bibfnamefont{A.}~\bibnamefont{Pappa}},
  \bibinfo{author}{\bibfnamefont{P.}~\bibnamefont{Jouguet}},
  \bibinfo{author}{\bibfnamefont{T.}~\bibnamefont{Lawson}},
  \bibinfo{author}{\bibfnamefont{A.}~\bibnamefont{Chailloux}},
  \bibinfo{author}{\bibfnamefont{M.}~\bibnamefont{Legr\'e}},
  \bibinfo{author}{\bibfnamefont{P.}~\bibnamefont{Trinkler}},
  \bibinfo{author}{\bibfnamefont{I.}~\bibnamefont{Kerenidis}},
  \bibnamefont{and} \bibinfo{author}{\bibfnamefont{E.}~\bibnamefont{Diamanti}},
  \bibinfo{journal}{Nature Commun.} \textbf{\bibinfo{volume}{5}},
  \bibinfo{pages}{3717} (\bibinfo{year}{2014}).

\bibitem[{\citenamefont{Buhrman et~al.}(2001)\citenamefont{Buhrman, Cleve,
  Watrous, and Wolf}}]{BCW:prl01}
\bibinfo{author}{\bibfnamefont{H.}~\bibnamefont{Buhrman}},
  \bibinfo{author}{\bibfnamefont{R.}~\bibnamefont{Cleve}},
  \bibinfo{author}{\bibfnamefont{J.}~\bibnamefont{Watrous}}, \bibnamefont{and}
  \bibinfo{author}{\bibfnamefont{R.~D.} \bibnamefont{Wolf}},
  \bibinfo{journal}{Phys. Rev. Lett.} \textbf{\bibinfo{volume}{87}},
  \bibinfo{pages}{167902} (\bibinfo{year}{2001}).

\bibitem[{\citenamefont{Xu et~al.}(2015{\natexlab{c}})\citenamefont{Xu,
  Arrazola, Wei, Wang, Palacios-Avila, Feng, Sajeed, L\"utkenhaus, and
  Lo}}]{XAW:natcommun15}
\bibinfo{author}{\bibfnamefont{F.}~\bibnamefont{Xu}},
  \bibinfo{author}{\bibfnamefont{J.~M.} \bibnamefont{Arrazola}},
  \bibinfo{author}{\bibfnamefont{K.}~\bibnamefont{Wei}},
  \bibinfo{author}{\bibfnamefont{W.}~\bibnamefont{Wang}},
  \bibinfo{author}{\bibfnamefont{P.}~\bibnamefont{Palacios-Avila}},
  \bibinfo{author}{\bibfnamefont{C.}~\bibnamefont{Feng}},
  \bibinfo{author}{\bibfnamefont{S.}~\bibnamefont{Sajeed}},
  \bibinfo{author}{\bibfnamefont{N.}~\bibnamefont{L\"utkenhaus}},
  \bibnamefont{and} \bibinfo{author}{\bibfnamefont{H.-K.} \bibnamefont{Lo}},
  \bibinfo{journal}{Nature Commun.} \textbf{\bibinfo{volume}{6}},
  \bibinfo{pages}{8735} (\bibinfo{year}{2015}{\natexlab{c}}).

\bibitem[{\citenamefont{Gottesman and Chuang}(2001)}]{GC:arxiv01}
\bibinfo{author}{\bibfnamefont{D.}~\bibnamefont{Gottesman}} \bibnamefont{and}
  \bibinfo{author}{\bibfnamefont{I.}~\bibnamefont{Chuang}},
  \bibinfo{journal}{arXiv preprint quant-ph/0105032}  (\bibinfo{year}{2001}).

\bibitem[{\citenamefont{Donaldson et~al.}(2016)\citenamefont{Donaldson,
  Collins, Kleczkowska, Amiri, Wallden, Dunjko, Jeffers, Andersson, and
  Buller}}]{DCK:pra16}
\bibinfo{author}{\bibfnamefont{R.~J.} \bibnamefont{Donaldson}},
  \bibinfo{author}{\bibfnamefont{R.~J.} \bibnamefont{Collins}},
  \bibinfo{author}{\bibfnamefont{K.}~\bibnamefont{Kleczkowska}},
  \bibinfo{author}{\bibfnamefont{R.}~\bibnamefont{Amiri}},
  \bibinfo{author}{\bibfnamefont{P.}~\bibnamefont{Wallden}},
  \bibinfo{author}{\bibfnamefont{V.}~\bibnamefont{Dunjko}},
  \bibinfo{author}{\bibfnamefont{J.}~\bibnamefont{Jeffers}},
  \bibinfo{author}{\bibfnamefont{E.}~\bibnamefont{Andersson}},
  \bibnamefont{and} \bibinfo{author}{\bibfnamefont{G.~S.}
  \bibnamefont{Buller}}, \bibinfo{journal}{Phys. Rev. A}
  \textbf{\bibinfo{volume}{93}}, \bibinfo{pages}{012329}
  (\bibinfo{year}{2016}).

\bibitem[{\citenamefont{Broadbent et~al.}(2009)\citenamefont{Broadbent,
  Fitzsimons, and Kashefi}}]{BFK:focs09}
\bibinfo{author}{\bibfnamefont{A.}~\bibnamefont{Broadbent}},
  \bibinfo{author}{\bibfnamefont{J.}~\bibnamefont{Fitzsimons}},
  \bibnamefont{and} \bibinfo{author}{\bibfnamefont{E.}~\bibnamefont{Kashefi}},
  in \emph{\bibinfo{booktitle}{Foundations of Computer Science, 2009.
  Proceedings. 50th Annual Symposium on}} (\bibinfo{organization}{IEEE},
  \bibinfo{year}{2009}), pp. \bibinfo{pages}{517--526}.

\bibitem[{\citenamefont{Barz et~al.}(2012)\citenamefont{Barz, Kashefi,
  Broadbent, Fitzsimons, Zeilinger, and Walther}}]{BKB:science12}
\bibinfo{author}{\bibfnamefont{S.}~\bibnamefont{Barz}},
  \bibinfo{author}{\bibfnamefont{E.}~\bibnamefont{Kashefi}},
  \bibinfo{author}{\bibfnamefont{A.}~\bibnamefont{Broadbent}},
  \bibinfo{author}{\bibfnamefont{J.~F.} \bibnamefont{Fitzsimons}},
  \bibinfo{author}{\bibfnamefont{A.}~\bibnamefont{Zeilinger}},
  \bibnamefont{and} \bibinfo{author}{\bibfnamefont{P.}~\bibnamefont{Walther}},
  \bibinfo{journal}{Science} \textbf{\bibinfo{volume}{335}},
  \bibinfo{pages}{303} (\bibinfo{year}{2012}).

\bibitem[{\citenamefont{Lau and Lo}(2011)}]{LL:pra11}
\bibinfo{author}{\bibfnamefont{H.-K.} \bibnamefont{Lau}} \bibnamefont{and}
  \bibinfo{author}{\bibfnamefont{H.-K.} \bibnamefont{Lo}},
  \bibinfo{journal}{Phys. Rev. A} \textbf{\bibinfo{volume}{83}},
  \bibinfo{pages}{012322} (\bibinfo{year}{2011}).

\bibitem[{\citenamefont{Buhrman et~al.}(2014)\citenamefont{Buhrman, Chandran,
  Fehr, Gelles, Goyal, Ostrovsky, and Schaffner}}]{BCF:siam14}
\bibinfo{author}{\bibfnamefont{H.}~\bibnamefont{Buhrman}},
  \bibinfo{author}{\bibfnamefont{N.}~\bibnamefont{Chandran}},
  \bibinfo{author}{\bibfnamefont{S.}~\bibnamefont{Fehr}},
  \bibinfo{author}{\bibfnamefont{R.}~\bibnamefont{Gelles}},
  \bibinfo{author}{\bibfnamefont{V.}~\bibnamefont{Goyal}},
  \bibinfo{author}{\bibfnamefont{R.}~\bibnamefont{Ostrovsky}},
  \bibnamefont{and}
  \bibinfo{author}{\bibfnamefont{C.}~\bibnamefont{Schaffner}},
  \bibinfo{journal}{SIAM Journal on Computing} \textbf{\bibinfo{volume}{43}},
  \bibinfo{pages}{150} (\bibinfo{year}{2014}).

\bibitem[{\citenamefont{Chakraborty and Leverrier}(2015)}]{CL:pra15}
\bibinfo{author}{\bibfnamefont{K.}~\bibnamefont{Chakraborty}} \bibnamefont{and}
  \bibinfo{author}{\bibfnamefont{A.}~\bibnamefont{Leverrier}},
  \bibinfo{journal}{Phys. Rev. A} \textbf{\bibinfo{volume}{92}},
  \bibinfo{pages}{052304} (\bibinfo{year}{2015}).

\bibitem[{\citenamefont{Wehner et~al.}(2010)\citenamefont{Wehner, Curty,
  Schaffner, and Lo}}]{WCS:pra10}
\bibinfo{author}{\bibfnamefont{S.}~\bibnamefont{Wehner}},
  \bibinfo{author}{\bibfnamefont{M.}~\bibnamefont{Curty}},
  \bibinfo{author}{\bibfnamefont{C.}~\bibnamefont{Schaffner}},
  \bibnamefont{and} \bibinfo{author}{\bibfnamefont{H.-K.} \bibnamefont{Lo}},
  \bibinfo{journal}{Phys. Rev. A} \textbf{\bibinfo{volume}{81}},
  \bibinfo{pages}{052336} (\bibinfo{year}{2010}).

\end{thebibliography}
\bibliographystyle{apsrev}

\end{document}